\documentclass[12pt]{article} 

\usepackage[utf8]{inputenc}
\usepackage[T1]{fontenc} 
\usepackage{lmodern}
\usepackage[english]{babel}           

\usepackage{cancel}
\usepackage[pdftex]{graphicx}
\usepackage{epsfig}
\usepackage{graphicx}
\usepackage{comment}
\usepackage{latexsym}
\usepackage{hyperref}
\usepackage{amsmath}
\allowdisplaybreaks 
\usepackage{mathrsfs}
\usepackage[usenames, dvipsnames]{color}
\usepackage{amsbsy}
\usepackage{amssymb}
\usepackage{amsthm}
\usepackage{amsfonts}
\usepackage{cite}
\usepackage{enumitem}
\usepackage{xcolor}
\usepackage{color}
\usepackage{mathtools}
\usepackage{caption} 
\usepackage{subcaption} 
\usepackage{euscript} 

\usepackage{float}

\usepackage{diagbox}

\usepackage[title]{appendix}

\usepackage{tabularx}

\newcommand{\be}[0]{\begin{equation}}
\newcommand{\ee}[0]{\end{equation}}
\newcommand{\dis}{\displaystyle}
\renewcommand{\thefootnote}{\fnsymbol{footnote}}

\newcommand{\Eq}[1]{Eq.~\eqref{#1}}
\newcommand{\Eqs}[1]{Eqs.~\eqref{#1}}
\newcommand{\Reff}[1]{Ref.~\cite{#1}}
\newcommand{\Reffs}[1]{Refs.~\cite{#1}}
\newcommand{\Sect}[1]{Sect.~\ref{#1}}
\newcommand{\Sects}[1]{Sects.~\ref{#1}}
\newcommand{\Appendix}[1]{Appendix~\ref{#1}}
\newcommand{\Fig}[1]{Fig.~\ref{#1}}

\newcommand{\X}{Z}

\newcommand{\R}{\mathbb{R}}
\newcommand{\Z}{\mathbb{Z}}
\renewcommand{\natural}{\mathbb{N}}

\renewcommand{\O}{{\cal O}}
\renewcommand{\Re}{{\rm Re}\,}
\renewcommand{\Im}{{\rm Im}\,}
\newcommand{\sign}{{\rm sign}\,}

\newcommand{\tr}{\textrm{tr}}
\newcommand{\Str}{\textrm{Str}\,}

\newcommand{\diag}{{\rm diag}\,}

\newcommand{\rmv}{{\rm v}}

\newcommand{\ie}{{\em i.e.} }

\newcommand{\via}{{\it via} }

\newcommand{\where}{\mbox{where}}
\newcommand{\with}{\mbox{with}}
\newcommand{\when}{\mbox{when}}
\renewcommand{\and}{\mbox{and}}
\newcommand{\for}{\mbox{for}}

\newcommand{\esps}{\phantom{\!\!\!\overset{|}{a}}}
\newcommand{\esp}{\phantom{\!\!\overset{\displaystyle |}{|}}}
\newcommand{\espD}{\phantom{\!\!\underset{\displaystyle |}{\cdot}}}

\newcommand{\bm}{\boldmath} 

\newcommand{\W}{{\cal W}}
\newcommand{\F}{{\cal F}}
\newcommand{\N}{{\cal N}}

\newcommand{\I}{{\cal I}}
\newcommand{\K}{{\cal K}}
\renewcommand{\P}{{\cal P}}

\newcommand{\A}{{\cal A}}

\newcommand{\T}{{\cal T}}
\newcommand{\C}{{\cal C}}
\newcommand{\M}{{\cal M}}

\newcommand{\cR}{{\cal R}}
\newcommand{\cH}{{\cal H}}
\newcommand{\tl}{{\tilde l}}

\renewcommand{\SS}{Scherk--Schwarz }
\newcommand{\KK}{Kaluza--Klein }
\newcommand{\Teich}{Teichm\"uller }
\newcommand{\Mob}{M\"obius }
\newcommand{\Ms}{M_{\rm s}}

\newcommand{\nF}{n_{\rm F}}
\newcommand{\nB}{n_{\rm B}}

\newcommand{\Vone}{{\cal V}_{\mbox{\scriptsize 1-loop}}}

\newcommand{\taudc}{\tau^{\rm dc}}
\newcommand{\qdc}{q_{\rm dc}}

\newcommand{\nui}{\nu_{\rm int}}
\newcommand{\nue}{\nu_{\rm ext}}

\newcommand{\half}{\frac{1}{2}}

\newcommand\dd{\text{d}}

\def\jac(#1,#2){%
\begin{bsmallmatrix}
#1\cr 
#2\cr
\end{bsmallmatrix}}

\catcode`\@=11
\def\marginnote#1{}
\newcount\hour
\newcount\minute
\newtoks\amorpm
\hour=\time\divide\hour by60 \minute=\time{\multiply\hour by60
\global\advance\minute by-\hour}
\edef\standardtime{{\ifnum\hour<12 \global\amorpm={am}%
        \else\global\amorpm={pm}\advance\hour by-12 \fi
        \ifnum\hour=0 \hour=12 \fi
        \number\hour:\ifnum\minute<10 0\fi\number\minute\the\amorpm}}
\edef\militarytime{\number\hour:\ifnum\minute<10 0\fi\number\minute}
\def\draftlabel#1{{\@bsphack\if@filesw {\let\thepage\relax
   \xdef\@gtempa{\write\@auxout{\string
      \newlabel{#1}{{\@currentlabel}{\thepage}}}}}\@gtempa
   \if@nobreak \ifvmode\nobreak\fi\fi\fi\@esphack}
        \gdef\@eqnlabel{#1}}
\def\@eqnlabel{}
\def\@vacuum{}
\def\draftmarginnote#1{\marginpar{\raggedright\scriptsize\tt#1}}
\def\draft{\oddsidemargin -.2truein
        \def\@oddfoot{\sl preliminary draft \hfil
        \rm\thepage\hfil\sl\today\quad\militarytime}
        \let\@evenfoot\@oddfoot \overfullrule 3pt
        \let\label=\draftlabel
        \let\marginnote=\draftmarginnote
   \def\@eqnnum{(\theequation)\rlap{\kern\marginparsep\tt\@eqnlabel}%
\global\let\@eqnlabel\@vacuum}  }
\def\thebibliography#1{
\vskip 0.5cm \centerline{\bf \Large References}
\list{
[\arabic{enumi}]}{\settowidth\labelwidth{[#1]}
\leftmargin\labelwidth
\advance\leftmargin\labelsep
\usecounter{enumi}}
\def\newblock{\hskip .11em plus .33em minus .07em}
\sloppy\clubpenalty4000\widowpenalty4000
\sfcode`\.=1000\relax}

\renewcommand{\theequation}{\arabic{section}.\arabic{equation}}
\renewcommand{\section}{\setcounter{equation}{0}\@startsection
{section}{1}{0mm}{-\baselineskip}{0.5\baselineskip} {\normalfont\Large\bfseries}}
\renewcommand{\subsection}{\@startsection
{subsection}{2}{0mm}{-\baselineskip}{0.5\baselineskip} {\normalfont\large\bfseries}}
\renewcommand{\subsubsection}{\@startsection
{subsubsection}{3}{0mm}{-\baselineskip}{0.5\baselineskip}
{\normalfont\normalsize\slshape}}

\newcommand{\lattice}{\sum_{\vec{m},\vec{n}}\frac{\Lambda^{(4,4)}_{\vec{m},\vec{n}}}{\left|\eta^{4}\right|^{2}}}
\newcommand{\tetad}{\left|\frac{2\eta}{\vartheta_{2}}\right|^{4}}
\newcommand{\tetaq}{\left|\frac{\eta}{\vartheta_{4}}\right|^{4}}
\newcommand{\tetat}{\left|\frac{\eta}{\vartheta_{3}}\right|^{4}}

\newcommand{\tetado}{\left(\frac{2\eta}{\vartheta_{2}}\right)^{2}}
\newcommand{\tetaqo}{\left(\frac{\eta}{\vartheta_{4}}\right)^{2}}
\newcommand{\tetato}{\left(\frac{\eta}{\vartheta_{3}}\right)^{2}}

\newcommand{\tetadm}{\left(\frac{2\hat{\eta}}{\hat{\vartheta}_{2}}\right)^{2}}

\newcommand{\latticek}{\bigg(\sum_{\vec{m}}\frac{P^{(4)}_{\vec{m}}}{\eta^{4}}+\sum_{\vec{n}}\frac{W^{(4)}_{\vec{n}}}{\eta^{4}}\bigg)}

\newcommand{\vq}{V_{4}}
\newcommand{\oq}{O_{4}}
\newcommand{\sq}{S_{4}}
\newcommand{\cq}{C_{4}}

\def\car(#1,#2,#3,#4,#5,#6,#7,#8){
\def\ArgI{{#1}}
\def\ArgII{{#2}}
\def\ArgIII{{#3}}
\def\ArgIV{{#4}}
\def\ArgV{{#5}}
\def\ArgVI{{#6}}
\def\ArgVII{{#7}}
\def\ArgVIII{{#8}}
}

\def\carRelay(#1,#2,#3){
\left| \ArgI_{4}\ArgII_{4}#1\ArgIII_{4}\ArgIV_{4}#2\ArgV_{4}\ArgVI_{4}#3\ArgVII_{4}\ArgVIII_{4}\right|^{2}
}

\def\caro(#1,#2,#3,#4,#5,#6,#7,#8){
\def\ArgI{{#1}}
\def\ArgII{{#2}}
\def\ArgIII{{#3}}
\def\ArgIV{{#4}}
\def\ArgV{{#5}}
\def\ArgVI{{#6}}
\def\ArgVII{{#7}}
\def\ArgVIII{{#8}}
}

\def\carRelayo(#1,#2,#3){
\left( \ArgI_{4}\ArgII_{4}#1\ArgIII_{4}\ArgIV_{4}#2\ArgV_{4}\ArgVI_{4}#3\ArgVII_{4}\ArgVIII_{4}\right)
}

\def\carm(#1,#2,#3,#4,#5,#6,#7,#8){
\def\ArgI{{#1}}
\def\ArgII{{#2}}
\def\ArgIII{{#3}}
\def\ArgIV{{#4}}
\def\ArgV{{#5}}
\def\ArgVI{{#6}}
\def\ArgVII{{#7}}
\def\ArgVIII{{#8}}
}

\def\carRelaym(#1,#2,#3){
\left( \hat{\ArgI}_{4}\hat{\ArgII}_{4}#1\hat{\ArgIII}_{4}\hat{\ArgIV}_{4}#2\hat{\ArgV}_{4}\hat{\ArgVI}_{4}#3\hat{\ArgVII}_{4}\hat{\ArgVIII}_{4}\right)
}

\def\carq(#1,#2,#3,#4,#5){
\left|#1_{4}#2_{4}#5#3_{4}#4_{4}\right|^{2}
}

\def\cars(#1,#2,#3,#4,#5){
(#1_{4}#2_{4}#5#3_{4}#4_{4})
}

\def\carsm(#1,#2,#3,#4,#5){
(\hat{#1}_{4}\hat{#2}_{4}#5\hat{#3}_{4}\hat{#4}_{4})
}

\def\carsbar(#1,#2,#3,#4,#5){
(\bar{#1}_{4}\bar{#2}_{4}#5\bar{#3}_{4}\bar{#4}_{4})
}

\def\LAMBDA(#1,#2){
\ifnum #2=0
\ifnum #1=0
\Lambda^{(2,2)}_{\vec{m},\vec{n}}
\else
\Lambda^{(2,2)}_{\vec{m}+\half,\vec{n}}
\fi
\else
\ifnum #1=0
\ifnum #2=0
\Lambda_{\vec{m},\vec{n}}
\else
\Lambda^{(2,2)}_{\vec{m},\vec{n}+\vec{a_{\text{S}}}}
\fi
\fi
\fi
\ifnum #1=1
\ifnum #2=1
\Lambda^{(2,2)}_{\vec{m}+\half,\vec{n}+\vec{a_{\text{S}}}}
\fi
\fi
}

\def\LAMBDA(#1,#2){
\frac{\Lambda^{(2,2)}_{#1,#2}}{\left|\eta^{4}\right|^{2}}
}

\def\LAMBDAO(#1,#2){
\frac{\Lambda^{(2,2)}_{#1,#2}}{\eta^{4}}
}

\def\LAMBDAM(#1,#2){
\frac{\Lambda^{(2,2)}_{#1,#2}}{\hat{\eta}^{4}}
}

\newcommand{\overbar}[1]{\mkern 1.5mu\overline{\mkern-1.5mu#1\mkern-1.5mu}\mkern 1.5mu}

\newcommand{\Xcl}{Z_{\text{cl}}}
\newcommand{\Xclbar}{\overbar{Z}_{\text{cl}}}
\newcommand{\Xqu}{Z_{\text{qu}}}
\newcommand{\Xqubar}{\overbar{Z}_{\text{qu}}}

\newcommand{\Scl}{S_{\text{cl}}}

\newcommand{\zu}{z_{1}}
\newcommand{\zd}{z_{2}}

\newcommand{\zud}{z_{12}}

\newcommand{\thetau}{\vartheta_{1}}
\newcommand{\thetauzud}{\vartheta_{1}(\zud)}

\newcommand{\thetan}{\vartheta_{\nu}}

\makeatletter
\newcommand{\vast}{\bBigg@{3.5}}
\makeatother

\begin{document}


\begin{titlepage}
\begin{flushright}
CPHT-RR090.112020, November 2020
\vspace{1.5cm}
\end{flushright}
\begin{centering}
{\bm\bf \Large One-loop masses of Neumann--Dirichlet open strings \\  
\vspace{0.2cm}and boundary-changing vertex operators}

\vspace{7mm}

 {\bf Thibaut Coudarchet and Herv\'e Partouche}

 \vspace{4mm}

{CPHT, CNRS, Ecole Polytechnique, IP Paris, \\F-91128 Palaiseau, France \\ \vspace{2mm}
\textit{thibaut.coudarchet@polytechnique.edu}\\  \textit{herve.partouche@polytechnique.edu}}

\end{centering}
\vspace{0.1cm}
$~$\\
\centerline{\bf\Large Abstract}\\
\vspace{-1cm}

\begin{quote}

\hspace{.6cm} 

We derive the masses acquired at one loop by massless scalars in the Neumann--Dirichlet sector of open strings, when supersymmetry is spontaneously broken. It is done by computing two-point functions of ``boundary-changing vertex operators''  inserted on the boundaries of the annulus and \Mob strip. This requires the evaluation of correlators of ``excited boundary-changing fields,'' which are analogous to excited twist fields for closed strings. We work in the type~IIB orientifold theory compactified on $T^2\times T^4/\Z_2$, where $\N=2$ supersymmetry is broken to $\N=0$ by the \SS mechanism implemented along $T^2$. Even though the full expression of the squared masses is complicated, it reduces to a very simple form when the lowest scale of the background is the supersymmetry breaking scale $M_{3/2}$. We apply our results to analyze in this regime the stability at the quantum level of the moduli fields arising in the Neumann--Dirichlet sector. This completes the study of \Reff{ACP}, where the quantum masses of all other types of moduli arising in the open- or closed-string sectors are derived. Ultimately, we identify all brane configurations that produce backgrounds without tachyons at one loop and yield an effective potential exponentially suppressed, or strictly positive with runaway behavior of $M_{3/2}$.

\end{quote}

\end{titlepage}
\newpage
\setcounter{footnote}{0}
\renewcommand{\thefootnote}{\arabic{footnote}}
 \setlength{\baselineskip}{.7cm} \setlength{\parskip}{.2cm}

\setcounter{section}{0}


\section{Introduction}

Superstring-theory models based on two-dimensional conformal field theories of free fields have the advantage of allowing, at least in principle, string amplitudes to be computed exactly in string tension $\alpha'$ by including all worldsheet instantons. Backgrounds whose internal spaces are $\Z_N$-twist orbifolds of tori are of particular interest since their numbers of spacetime supersymmetries  are reduced in a ``hard way'' compared to the case of  toroidal compactifications. In this framework, twisted states in the closed-string Hilbert space are mandatory for modular invariance to hold, which implies ``twist fields'' to exist in the conformal field theory to create them~\cite{Dixon}. String amplitudes involving external states in the twisted sectors are based on correlation functions of twist fields, which are notoriously difficult to handle. Indeed, the seminal work of \Reff{Atick} presents results only for the case of twist fields creating ground states in the closed-string sector.  

In open-string theory, the consistency of orbifold models also implies the presence of distinct D-brane sectors. For instance, in the type~IIB orientifold on $T^4/\Z_2$~\cite{BianchiSagnotti,GimonPolchinski,GimonPolchinski2}, open strings have either Neumann~(N) or  Dirichlet~(D) boundary conditions in the orbifold directions, and are thus attached to D9- or  D5-branes. In particular, strings with Neumann boundary conditions at one end and Dirichlet conditions at the other end populate the ND sector. In string amplitudes involving external states of this type, a conformal transformation maps the legs of the diagram to vertex operators localized along the worldsheet boundary. The key point is that the nature of an ND-sector state implies that the worldsheet boundary condition changes from Neumann on one side of the vertex to Dirichlet on the other side. Hence, vertex operators creating states in the ND sector involve ``boundary-changing fields''~\cite{Hashimoto:1996he} dressed by other objects encoding the quantum numbers. 

It turns out that twist fields and boundary-changing fields have identical OPE's~\cite{Hashimoto:1996he,Atick}, up to the fact that the former are inserted in the bulk of the worldsheet and the latter on the boundary. Combining this with the method of images which defines surfaces with boundaries as Rienmann surfaces modded by involutions~\cite{Burgess1,Burgess2}, correlation functions of boundary-changing fields can be related to those of twist fields. In the literature, this point of view was applied for computing amplitudes with external states of ND sectors in supersymmetric theories at tree level~\cite{Cvetic-Abel1,Cvetic-Abel2,Cvetic-Abel3,Good} and one loop~\cite{Abel:2004ue,Schofield,markG1,markG2,markG3,Benakli:2008ub}, while other approaches were followed in \Reffs{Frohlich,Anastasopoulos:2011gn,Mattiello}.  

In the present work, we consider the type~IIB orientifold model of \Reffs{BianchiSagnotti,GimonPolchinski,GimonPolchinski2} compactified on $T^2\times T^4/\Z_2$, when $\N=2$ supersymmetry is spontaneously broken to $\N=0$. The implementation of the breaking consists of a string version~\cite{openSS1,openSS2,openSS3,openSS4,openSS5,openSS6,openSS7, openSS8} of the \SS mechanism~\cite{SS1,SS2} along one direction of $T^2$. In this case, the supersymmetry breaking scale, $M_{3/2}$, is a modulus inversely proportional to the size of the compact direction involved in the mechanism. Moreover, the free nature of the bosonic and fermionic fields defining the worldsheet conformal field theory is preserved and the results of \Reff{Atick} apply. An effective potential which depends on all moduli fields is generated by quantum corrections and the question of their stability must be addressed. Assuming the string coupling to be in perturbative regime, loci in moduli space where the one-loop effective potential  is extremal with respect to all moduli fields except $M_{3/2}$ have been determined in \Reff{ACP}, up to exponentially suppressed terms. At these points, the potential reads
\begin{equation}
\Vone=\upsilon (\nF-\nB)M_{3/2}^4 +\O\!\left((\Ms M_{3/2})^{2}\, e^{-\pi \frac{c\Ms }{M_{3/ 2}}}\right),
\label{vd}
\end{equation}
where $\nF$ and $\nB$ are the numbers of massless fermionic and bosonic degrees of freedom present at genus-0.  Moreover, $\upsilon>0$ is a constant, $\Ms\equiv 1/\sqrt{\alpha'}$ is the string scale, and $c\Ms$ is the lowest non-vanishing mass scale other than $M_{3/2}$, $0<c\le 1$. Hence, \Eq{vd} is of interest in all regions in moduli space where $c\Ms$, which is a compactification scale, is greater than $M_{3/2}$. When this is the case and the exponential contributions are neglected, $M_{3/2}$ runs away, except when the background satisfies a Bose/Fermi degeneracy at \mbox{genus-0}, $\nF-\nB= 0$, implying $M_{3/2}$ to be a flat direction (for other theories, see  \Reffs{PreviousPaper,Itoyama:1986ei,Abel:2015oxa,SNS1,SNS2,CoudarchetPartouche,FR,Abel:2017rch,Abel:2017vos,Itoyama1,Itoyama2}). For arbitrary $\nF-\nB$, though, stability of all remaining moduli fields can be analyzed from different points of view.

In \Reff{ACP}, the mass terms of all moduli fields in the NN and DD open-string sectors were derived by direct computation of the potential for arbitrary backgrounds of these scalars. The untwisted closed-string sector contains three types of moduli fields: Firstly, since the internal metric components do not show up in the dominant term of \Eq{vd} (except the combination $M_{3/2}$), they parametrize flat directions up to the suppressed terms. Secondly, heterotic/type~I duality was used to show that the Ramond-Ramond (RR) two-form moduli are also flat directions. Finally, the same conclusion definitely applies to  the dilaton at  one loop. The twisted closed-string sector contains 16 blowing-up modes of $T^4/\Z_2$ among which 2 to 16 are absorbed by anomalous $U(1)$'s, which become massive vector fields thanks to a generalized Green--Schwarz mechanism. 
In this regard, the present work can be seen as a companion paper of \Reff{ACP}, as it provides a derivation of the mass terms generated at one loop by the remaining moduli fields, namely those belonging to the ND+DN open-string sector.  This will be done by computing two-point functions  of boundary-changing vertex operators of massless scalars in the ND+DN sector, on the annulus and M\"obius strip. 

In \Sect{susymod}, we review the description of the type~IIB orientifold model with broken $\N=2$ supersymmetry, which involves  D9- and D5-branes. Alternative T-dual pictures are also introduced for describing the NN- and DD-sector moduli as positions of D3-branes in the internal space. \Sect{ndmass} defines the string amplitudes we are interested in. \Sect{Atick} presents all correlators needed to calculate these amplitudes on the double-cover tori of the annulus and M\"obius strip. In particular, we review the derivation of \Reff{Atick} of the correlation function of twist fields that create ground states in the twisted sectors of closed strings. Following the method introduced in \Reffs{Abel:2004ue,Schofield,markG1,markG2,markG3}, we extend the result to the case of ``excited twist fields'' \ie operators appearing as higher order terms in the OPE of ordinary twist fields. 

In \Sect{NDmasses} we compute the two-point functions of interest. While the formulas can be used to extract the one-loop corrections to the K\"ahler metric and masses of the classically massless scalars of the ND+DN sector, they turn out to be rather cumbersome and obscure. For this reason, we derive in \Sect{alpha'0} a simplified expression of the squared masses at one loop that is valid when $M_{3/2}$ is lower than all other non-vanishing mass scales present in the background, precisely in the spirit of \Eq{vd} which holds in this regime. 

In \Sect{stanaly}, we apply this result to the last two models highlighted in  \Reff{ACP}, which presented all brane configurations that are tachyon free (or potentially tachyon free) at one loop\footnote{When the suppressed terms in \Eq{vd} can be neglected.} and  satisfy $\nF-\nB\ge 0$. The outcome of the two papers is that among the $\O(10^{11})$ non-perturbatively consistent brane configurations, there exist 2 tachyon free setups with $\nF-\nB=0$, and 5 with $\nF-\nB>0$. A third configuration with $\nF-\nB=0$ and ND+DN-sector moduli is tachyon free at one loop, up to 2 blowing-up modes of $T^4/\Z_2$ for which we have not computed the quantum mass terms.

Finally, our conclusions can be found in \Sect{conclusion}, while technical points are reported in three appendices. 


\section{\bm The $\N=2\to \N=0$ open-string model}
\label{susymod}

 In this section, we review the open-string  model considered in \Reff{ACP,Coudarchet:2020ozn}, which realizes at tree level the spontaneous breaking of $\N=2$ supersymmetry in  four-dimensional Minkowski spacetime. Our goals are to fix our notations,  list the  massless spectrum at genus-0, and specify the moduli fields whose masses will be computed at one loop in the sections to come. 


\subsection{The supersymmetric parent model}
\label{BSGP}

At the supersymmetric level, our starting point is the type~IIB orientifold model constructed in six dimensions by  Bianchi and Sagnotti~\cite{BianchiSagnotti}, as well as by Gimon and Polchinski \cite{GimonPolchinski,GimonPolchinski2}. Compactified down to four dimensions, the full gravitational background becomes  
\be
\mathbb{R}^{1,3}\times T^{2}\times {T^4\over \Z_2} \, ,
\label{back}
\ee
whose coordinates will be labeled by Greek, primed Latin and unprimed Latin indices
\be
\begin{aligned}
&\mbox{spacetime:} && X^\mu\, ,  &&~~\mu\in\{0,\dots,3\}\,,\\
&\mbox{two-torus:} &&X^{I'}\, , && ~~I'\in\{4,5\}\,,\\
&\mbox{four-torus:} &&X^{I}\, , &&~~ I\in\{6,\dots,9\}\, ,
\end{aligned}
\ee
and where the $\Z_2$-orbifold generator is defined as 
\be
g:\quad (X^{6},X^{7},X^{8},X^{9})\longrightarrow(-X^{6},-X^{7},-X^{8},-X^{9})\, .
\ee
The background also contains orientifold planes and D-branes. First of all, there is an O9-plane and 32 D9-branes spanned along all spatial directions. Second, there is an O5-plane localized at each of the 16 fixed points of $T^4/\Z_2$, and 32 D5-branes transverse to $T^4/\Z_2$. Open strings with one end attached to a D9-brane have Neumann boundary conditions in all spacetime coordinates, while those stuck to a D5-brane have Dirichlet boundary conditions along the directions of  $T^4/\Z_2$  (and Neumann along $\mathbb{R}^{1,3}\times T^{2}$).  


\paragraph{\em Moduli fields: } 


\begin{itemize}
\item On the   worldvolumes of the 32 D5-branes, the gauge bosons can develop vacuum expectation values (vev's) along $T^2$, which are Wilson lines.  T-dualizing the two-torus, the 32 D5-branes  become D3-branes whose positions along $\tilde X^{I'}$, the coordinates along the T-dual torus $\tilde T^2$ of metric $\tilde G_{I'J'}\equiv G^{I'J'}$, are nothing but the Wilson-line moduli of the original description~\cite{review-3}. Because in the T-dual picture a D3-brane at $(\tilde X^{I'},X^I)$ is transformed under $\Omega$, the orientifold generator, into an ``orientifold-mirror'' D3-brane located at $(-\tilde X^{I'},-X^I)$~\cite{review-3}, there are 64 fixed points in this description, all supporting one O3-plane.\footnote{In addition, the initial D9-branes become D7-branes.} Moreover, at genus-0, there are only 16 independent positions along $\tilde T^2$, which are  associated with the brane/mirror brane pairs. 

\item The locations of the 32 D3-branes (T-dual to the D5-branes) in $T^4/\Z_2$ are also allowed to vary, provided this is done consistently with the symmetries generated by $g$ and $\Omega$. Indeed, a D3-brane sitting at $(\tilde X^{I'},X^I)$ must be paired with an image brane under $g$ at $(\tilde X^{I'},-X^I)$. Moreover, both admit ``mirror branes'' under $\Omega$, which are located at $(-\tilde X^{I'},-X^I)$ and $(-\tilde X^{I'},X^I)$. Hence, there are at most 8 independent D3-brane positions in $T^4/\Z_2$. Notice that this number is lowered when there are $2$ modulo~4 D3-branes sitting on one of the 64 O3-planes, since such a configuration is still symmetric under $g$ and $\Omega$ but does not allow 2 D3-branes to move in the bulk of $T^4/\Z_2$. In other words, 2 D3-branes have rigid positions in $T^4/\Z_2$.

\item Applying a T-duality on the four-torus of the background~(\ref{back}), D5-branes and  D9-branes are turned into each other. Therefore, all moduli fields described for the D5-branes admit counterparts for the D9-branes. In particular, the D9-brane moduli are mapped into positions of 32 D3-branes in $\tilde T^2\times \tilde T^4/\Z_2$, where $\tilde T^4$ is the dual four-torus with  metric $\tilde G_{IJ}\equiv G^{IJ}$ and coordinates $\tilde X^I$. In this alternative T-dual picture, there are again 64 O3-planes at the fixed points of the inversion $(\tilde X^{I'},\tilde X^I)\to (-\tilde X^{I'},-\tilde X^I)$.\footnote{The initial D5-branes also become D7-branes in this alternative T-dual picture.}

\item In the original picture  involving  D5- and D9-branes, all  open-string moduli described so far correspond to modes realized in the DD and NN sectors. However, open strings stretched between one D5-brane and one D9-brane can also lead to moduli fields. The present paper is devoted to the study of these moduli. To be specific, we will derive the masses they acquire at one loop, when supersymmetry is spontaneously broken and their vev's vanish. When these moduli condense, the backgrounds can be described in terms of brane recombinations or magnetized branes~\cite{recomb1,recomb2,recomb3,recomb4}. 
\end{itemize}


\paragraph{\em Geometric picture: } 

\begin{figure}
\captionsetup[subfigure]{position=t}
\begin{center}
\begin{subfigure}[t]{0.48\textwidth}
\begin{center}
\includegraphics [scale=0.70]{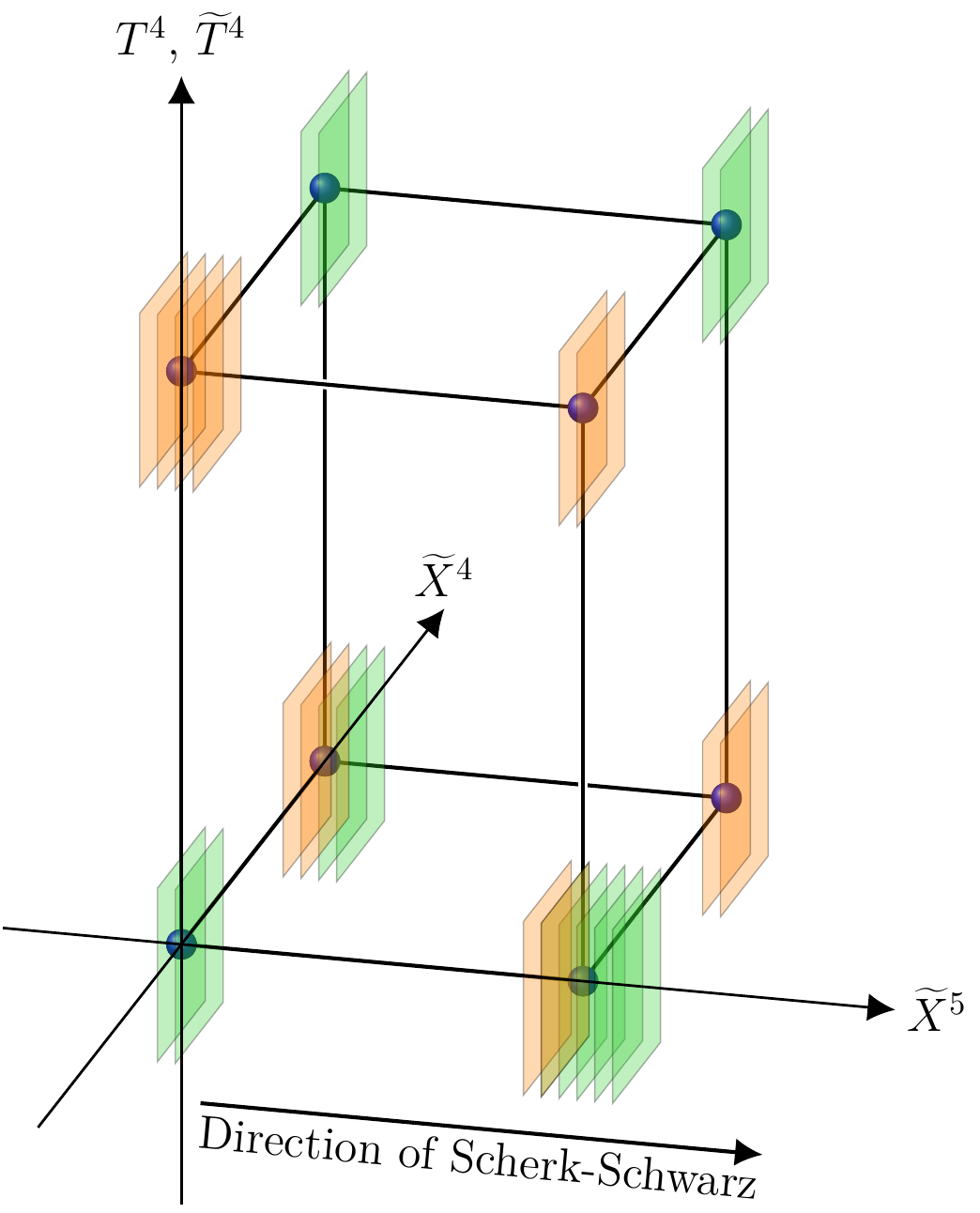}
\end{center}
\caption{\footnotesize Configuration of D3-branes associated with D5-branes (orange) and D9-branes (green) in T-dual pictures. In this example, all D3-branes sit on O3-planes (blue dots).}
\label{D5D9}
\end{subfigure}
\quad
\begin{subfigure}[t]{0.48\textwidth}
\begin{center}
\includegraphics [scale=0.65]{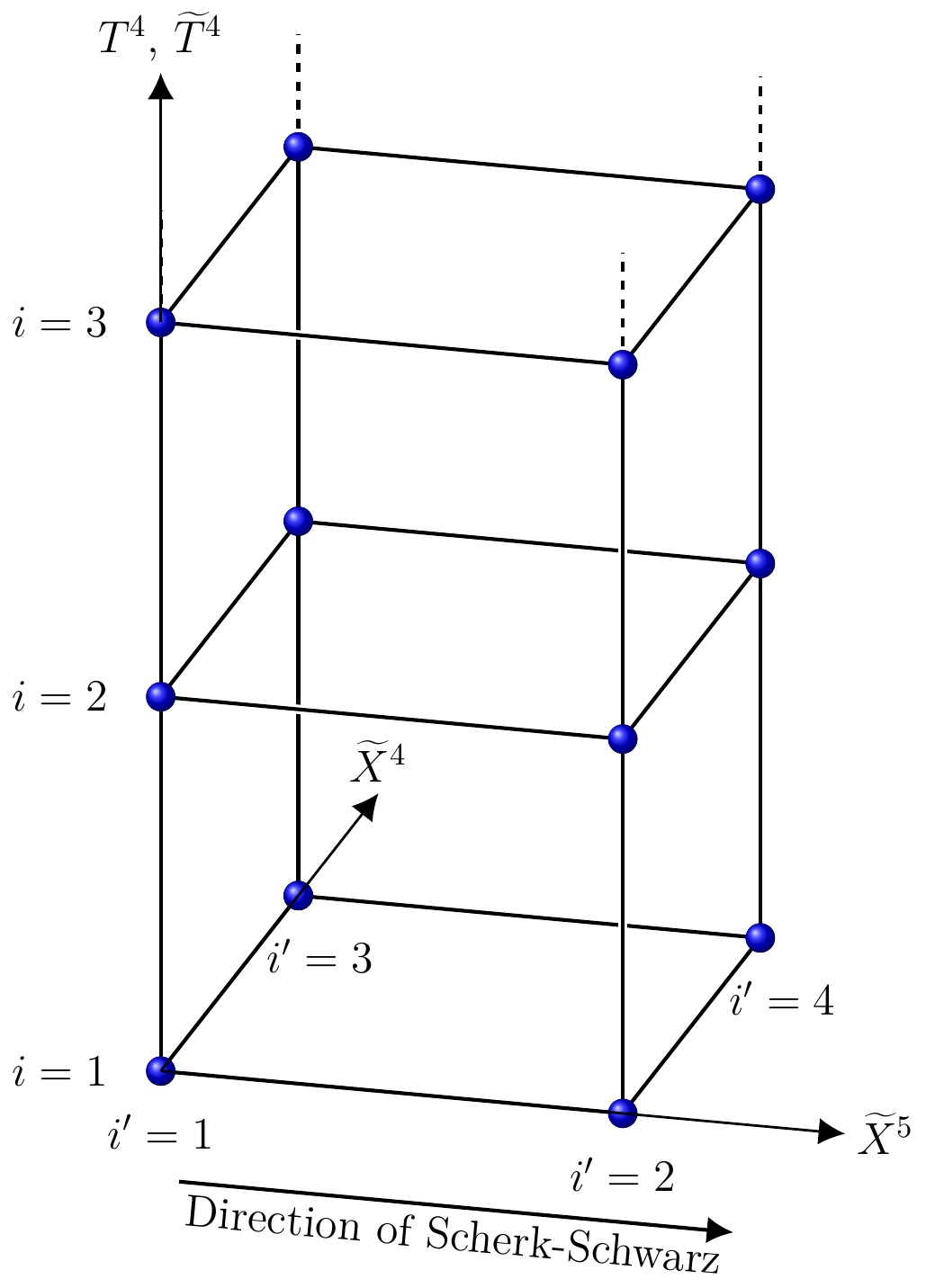}
\end{center}
\caption{\footnotesize Labelling of the fixed points $i'\in\{1,...,4\}$ along the directions of $\tilde T^2$, and schematic labelling of the fixed points $i\in\{1,...,16\}$ along the directions of  $T^4$ or $\tilde T^4$. $i'=1$ or 3 correspond to  points at $\tilde X^5=0$, while $i'=2$ or 4  correspond to  points at $\tilde X^5=\pi$, where $\tilde X^5$ is the coordinate T-dual to the direction along which the Scherk--Schwarz mechanism is implemented.}
\label{corners}
\end{subfigure}
\caption{\footnotesize Description in terms of D3-brane positions of the moduli arising from the NN and DD sectors of the orientifold theory.
}
\label{D5D9D5D9}
\end{center}
\end{figure}

In order to specify a particular set of vev's for the moduli arising from the DD and NN sectors, we will use a  pictorial representation~\cite{ACP}, as shown in \Fig{D5D9}.
We represent the fundamental domain of $\tilde T^2\times T^4/\Z_2$ modded by the involution $(\tilde X^{I'},X^I)\to (-\tilde X^{I'},-X^I)$ by a schematic six-dimensional ``box'', with an O3-plane represented by a dot at each fixed point \ie corner of the box. The moduli in the DD sector correspond to the positions of the 32 D3-branes (drawn in orange) T-dual to the D5-branes. Similarly, we consider a second box corresponding to the fundamental domain of $\tilde T^2\times \tilde T^4/\Z_2$ modded by the involution $(\tilde X^{I'},\tilde X^I)\to (-\tilde X^{I'},-\tilde X^I)$, with the moduli of the NN sector given by the positions of the 32 D3-branes (drawn in green) T-dual to the D9-branes. Finally, we superpose the two boxes, keeping in mind that the resulting picture combines information from two distinct T-dual descriptions of the same theory. 

In the schematic example of \Fig{D5D9}, all D3-branes are located on O3-planes. Indeed, it has been shown in \Reff{ACP} that in presence of supersymmetry breaking (to be introduced in the next subsection) these configurations are of particular interest, since they yield extrema of the one-loop effective potential with respect to the moduli arising from the NN and DD sectors (except for $M_{3/2}$ when $\nF\neq \nB$), up to exponentially suppressed terms (see \Eq{vd}). Therefore, from now on, we will consider background values of the moduli in the DD and NN sectors corresponding to stacks of D3-branes all located on corners of the  six-dimensional boxes. To this end, we label the 64 corners by a double index $ii'$, where $i\in\{1,\dots, 16\}$ refers to the fixed points of $T^4/\Z_2$ (or $\tilde T^4/\Z_2$), and $i'\in\{1,\dots,4\}$ is associated with those in the $\tilde T^2$ directions. Hence, the coordinates of corner $ii'$ are captured by a two-vector $2\pi \vec a_{i'}$ and a four-vector $2\pi \vec a_i$, whose components satisfy
\be
a_{i'}^{I'}\,,~a_i^I\in \left\{0,\half\right\},\quad i'\in\{1,\dots,4\}\, , ~~ i\in\{1,\dots,16\}\, .
\ee
\Fig{corners} shows how the labelling looks like when the fixed points $i\in\{1,\dots, 16\}$ are schematically arranged linearly along a  vertical axis. In these notations, we will denote by $D_{ii'}$ and $N_{ii'}$ the numbers of D3-branes T-dual to the D5-branes and D9-branes that are located at corners $ii'$ of the appropriate boxes. 


\subsection{Spontaneous supersymmetry breaking}
\label{nonsusy}

In quantum field theory, the \SS mechanism amounts to imposing fields to satisfy boundary conditions along compact directions that are compatible with  global symmetries and depend on associated conserved charges~\cite{SS1,SS2}. When charges vary between superpartners, distinct \KK masses arise in lower dimension and supersymmetry is spontaneously broken. The implementation of this mechanism in closed-string theory was developed in \Reffs{SSstring1,SSstring11,SSstring2,SSstring3,SSstring4}, and extended to the open-string framework in \Reffs{openSS1,openSS2,openSS3,openSS4,openSS5,openSS6,openSS7, openSS8}. 

In the model based on the background~(\ref{back}), we make the choice to implement the \SS mechanism along the periodic direction $X^5$ only, and to use the fermionic number $F$ as conserved charge. In practice, $F=0$ for the bosonic degrees of freedom and $F=1$ for the fermionic ones.  Denoting $\vec m'$ the two-vector whose components are the \KK momenta  $(m_4,m_5)\in\Z^2$ along $T^2$, the lattices of zero modes appearing in the one-loop partition function are shifted according to the rules\footnote{In the closed-string sector, this is the only modification in the untwisted sector of the extra generator that implements the \SS breaking  in orbifold  language.} 
\be
\label{shifts}
\begin{aligned}
& \vec m'+F\,\vec a'_{\rm S}&& \quad \mbox{for closed string}\, , \\
&\vec m'+F\,\vec a'_{\rm S}+\vec a_{i'}-\vec a_{j'}&&\quad \mbox{for open string}\, ,
\end{aligned}
\ee
where we have defined 
\be
\vec a'_{\rm S}=\left(0,\half\right) .
\ee
As a result, the two gravitino masses are  
 \be
\label{breakingscale}
M_{3/2}=\frac{\sqrt{G^{55}}}{2}\,\Ms\, ,
\ee
which is the scale of $\N=2\to \N=0$ spontaneous breaking of supersymmetry. 
In the open-string case, the extra shift $\vec a_{i'}-\vec a_{j'}$ arises from the Wilson-line background along $T^2$ of the gauge fields living on the worldvolumes of the D5- and D9-branes. In the D3-brane T-dual pictures, it means that an open string is stretched between D3-branes sitting on corners $ii'$ and $jj'$, regardless of whether they are dual to D5- or D9-branes.\footnote{\label{precision} For the ND and DN sectors, our description in terms of ``stretched strings'' is somewhat abusive since the corners $ii'$ and $jj'$ are to be understood in distinct T-dual descriptions.} Because of the particular role played by the direction $\tilde X^5$, which is T-dual to the \SS direction $X^5$ of the original picture, it is convenient to specify our labelling of the fixed points along the directions of $\tilde T^2$. We will denote by $i'=1$ and 3 those located at $\tilde X^5=0$, and by $i'=2$ and 4 those located at $\tilde X^5=\pi$ (see Fig.~\ref{corners}). 


\paragraph{\em Partition function: } 

The one-loop partition function can be divided into four contributions $Z_\Sigma$, which can be derived from path integrals on worldsheets whose topologies are those of a torus~$(\T)$, Klein bottle~$(\K)$, annulus~$(\A)$ and \Mob strip~$(\M)$. These contributions can also be expressed as supertraces over the modes belonging to the untwisted and twisted closed-string sectors, as well as over those in the NN, DD, ND and DN open-string sectors. For the closed strings, we have  
\be
Z_\T= {1\over \tau_2^{2}}\, \Str \half\, \frac{1+g}{2} \,q^{L_0-\half}\,\bar q^{\tilde L_0-\half}\,,  \quad Z_\K={1\over \tau_2^{2}}\, \Str {\Omega\over 2}\, \frac{1+g}{2} \,q^{L_0-\half}\, \bar q^{\tilde L_0-\half}\,, \quad q=e^{2i\pi \tau}\, ,
\label{Zcl}
\ee
where $\tau$ is the \Teich parameter of the worldsheet torus with real and imaginary parts denoted $\tau_1$ and  $\tau_2>0$, while for the open strings we have 
\be
Z_\A={1\over \tau_2^{2}}\, \Str \half\, \frac{1+g}{2}\, q^{\half(L_0-1)}\, ,\qquad Z_\M= {1\over \tau_2^{2}}\, \Str  {\Omega\over 2}\, \frac{1+g}{2}\,  q^{\half(L_0-1)}\,,\qquad q=e^{-2\pi \tau_2}\, .
\ee
In these formulas, $L_0$, $\tilde L_0$ are the zero-frequency Virasoro operators. 

In order to give explicit expressions of $Z_\A$ and $Z_\M$, we first define four-vectors $\vec m$ and $\vec n$ whose components are the Kaluza--Klein momenta $m_I\in\Z$ and winding numbers $n_I\in\Z$ along the directions of $T^4$. The lattices of zero modes (to be shifted by Wilson lines) of the bosonic coordinates  are then given by 
\be
\begin{aligned}
\sum_{\vec m}P_{\vec{m}}^{(4)}(i\tau_{2})&=\sum_{\vec m}e^{-\pi\tau_2m_{I}G^{IJ}m_{J}}&& \quad \mbox{for the NN sector}\, ,\\
\sum_{\vec n}W_{\vec{n}}^{(4)}(i\tau_{2})&=\sum_{\vec n}e^{-\pi \tau_{2} n_I G_{IJ}n_J}&& \quad \mbox{for the DD sector}\, ,\\
&~\,1&& \quad \mbox{for the ND and DN sectors}\, ,
\end{aligned}
\label{latop}
\ee
while the lattice of momenta along $T^2$ is
\be
\sum_{\vec m'}P_{\vec m'}^{(2)}(i\tau_{2})=\sum_{\vec m'}e^{-\pi\tau_2m_{I'}G^{I'J'}m_{J'}}
\label{latop2}
\ee
in all open-string sectors. 

In  the annulus contribution to the partition function, the actions of the neutral group element 1 and  generator $g$ on the Chan--Paton indices can be represented by matrices acting on each Neumann or Dirichlet sector $ii'$~\cite{GimonPolchinski,GimonPolchinski2},
\be
\begin{aligned}
&\gamma^{ii'}_{{\rm N},1}=I_{N_{ii'}}\,, ~~\quad &&\gamma^{ii'}_{{\rm N},g}=J_{N_{ii'}}\, ,\\
&\gamma^{ii'}_{{\rm D},1}=I_{D_{ii'}}\,, ~~\quad &&\gamma^{ii'}_{{\rm D},g}=J_{D_{ii'}}\, ,
\end{aligned}
\ee 
where $I_k$ is the $k\times k$ identity matrix while for $k$ even 
\be
J_{k}=\begin{pmatrix} 0&I_{k\over 2} \\-I_{k\over 2} & 0\end{pmatrix}.
\ee
Actually, the precise dictionary between the above  matrices and those defined in \Reffs{GimonPolchinski,GimonPolchinski2} can be found in \Appendix{dico}. 
To be specific, by labelling the branes with Greek indices, the actions of $G=1$ or  $g$ are represented in the NN sector as follows:
\be
\forall \alpha\in\{1,\dots, N_{ii'}\}\, , \forall \beta\in\{1,\dots, N_{jj'}\}\, , \quad |\alpha\beta\rangle\to \sum_{\alpha'=1}^{N_{ii'}}\sum_{\beta'=1}^{N_{jj'}}(\gamma^{ii'}_{{\rm N},G})_{\alpha\alpha'}|\alpha'\beta'\rangle ( \gamma^{jj'\, -1}_{{\rm N},G})_{\beta'\beta}\, .
\label{Arule}
\ee 
Similar expressions apply to the DD sector for $G=1$, $g$, as well as  to the ND and DN sectors for $G=1$. There exists only one subtlety in the ND and DN sectors for $G=g$, where one has to multiply all Neumann matrices by signs in the transformation rules, 
\be
\label{repla}
\gamma_{{\rm N},g}^{ii'}~~ \longrightarrow ~~ e^{4i\pi \vec a_i\cdot \vec a_j}\gamma_{{\rm N},g}^{ii'}\, ,
\ee
where the index $j$ refers to the fixed point of $T^4/\Z_2$ where the stack of D5-branes sits. This is explained in \Reff{GimonPolchinski2} and translated into the notations of our paper in \Appendix{dico}. 
Moreover, the worldsheet fermions associated with the directions $X^2,\dots,X^5$ on the one-hand, and those associated with the directions $X^6,\dots,X^9$ on the other hand, yield contributions expressed as characters of the $SO(4)$ affine algebra. The latter are associated with a singlet (O), vectorial~(V) and two spinorial (S and C) conjugacy classes~\cite{characters2, characters1, BLT}. For the annulus partition function, these characters denoted $O_4$, $V_4$, $S_4$, $C_4$ are defined in \Eq{eq:CharacterDef} and evaluated at argument  $i\tau_2/2$. Altogether, one obtains 
\begin{align}
\label{annulus}
&Z_\A={1\over 4}\,{1\over \tau_{2}^{2}} \sum_{\substack{i,i'\\ j,j'}}\nonumber \\
&\bigg\{\bigg[\cars(V,O,O,V,+)\bigg(\tr(\gamma_{{\rm N},1}^{ii'})\tr(\gamma_{{\rm N},1}^{jj'\,-1})\sum_{\vec m}{P^{(4)}_{\vec m+\vec a_i-\vec a_j}\over \eta^4}+\tr(\gamma_{{\rm D},1}^{ii'})\tr(\gamma_{{\rm D},1}^{jj'\,-1})\sum_{\vec n}{W^{(4)}_{\vec n+\vec a_i-\vec a_j}\over \eta^4}\bigg)\nonumber\\
&-\cars(V,O,O,V,-)\,\delta_{ij}\,\big(\tr(\gamma_{{\rm N},g}^{ii'})\tr(\gamma_{{\rm N},g}^{jj'\,-1})+\tr(\gamma_{{\rm D},g}^{ii'})\tr(\gamma_{{\rm D},g}^{jj'\,-1})\big)\tetado\nonumber\\
&+\cars(O,C,V,S,+)\big(\tr(\gamma_{{\rm N},1}^{ii'})\tr(\gamma_{{\rm D},1}^{jj'\,-1})+\tr(\gamma_{{\rm D},1}^{ii'})\tr(\gamma_{{\rm N},1}^{jj'\,-1})\big)\tetaqo\esp\\
&-\cars(O,C,V,S,-)\,e^{4i\pi \vec a_i\cdot \vec a_j}\big(\tr(\gamma_{{\rm N},g}^{ii'})\tr(\gamma_{{\rm D},g}^{jj'\,-1})+\tr(\gamma_{{\rm D},g}^{ii'})\tr(\gamma_{{\rm N},g}^{jj'\,-1})\big)\tetato\bigg]\sum_{\vec{m}'}\frac{P^{(2)}_{\vec{m}'+\vec{a}_{i'}-\vec{a}_{j'}}}{\eta^{4}}\nonumber \\
&-\bigg[\cars(S,S,C,C,+)\bigg(\tr(\gamma_{{\rm N},1}^{ii'})\tr(\gamma_{{\rm N},1}^{jj'\,-1})\sum_{\vec m}{P^{(4)}_{\vec m+\vec a_i-\vec a_j}\over \eta^4}+\tr(\gamma_{{\rm D},1}^{ii'})\tr(\gamma_{{\rm D},1}^{jj'\,-1})\sum_{\vec n}{W^{(4)}_{\vec n+\vec a_i-\vec a_j}\over \eta^4}\bigg)\nonumber\\
&-\cars(C,C,S,S,-)\,\delta_{ij}\,\big(\tr(\gamma_{{\rm N},g}^{ii'})\tr(\gamma_{{\rm N},g}^{jj'\,-1})+\tr(\gamma_{{\rm D},g}^{ii'})\tr(\gamma_{{\rm D},g}^{jj'\,-1})\big)\tetado\nonumber \\
&+\cars(S,O,C,V,+)\big(\tr(\gamma_{{\rm N},1}^{ii'})\tr(\gamma_{{\rm D},1}^{jj'\,-1})+\tr(\gamma_{{\rm D},1}^{ii'})\tr(\gamma_{{\rm N},1}^{jj'\,-1})\big)\tetaqo\nonumber\esp\\
&-\cars(S,O,C,V,-)\, e^{4i\pi \vec a_i\cdot \vec a_j}\big(\tr(\gamma_{{\rm N},g}^{ii'})\tr(\gamma_{{\rm D},g}^{jj'\,-1})+\tr(\gamma_{{\rm D},g}^{ii'})\tr(\gamma_{{\rm N},g}^{jj'\,-1})\big)\tetato\bigg]\sum_{\vec{m}'}\frac{P^{(2)}_{\vec{m}'+\vec{a}'_{\text{S}}+\vec{a}_{i'}-\vec{a}_{j'}}}{\eta^{4}}\bigg\}\,.\nonumber
\end{align}

In the the M\"obius-strip contribution to the partition function, the actions of $\Omega$ and $\Omega g$ on the Chan--Paton indices can be represented by matrices associated with each Neumann or Dirichlet sector $ii'$,
\be
\begin{aligned}
&\gamma^{ii'}_{{\rm N},\Omega}~=I_{N_{ii'}}\,, ~~\quad &&\gamma^{ii'}_{{\rm N},\Omega g}=J_{N_{ii'}}\, ,\\
&\gamma^{ii'}_{{\rm D},\Omega g}=I_{D_{ii'}}\,, ~~\quad &&\gamma^{ii'}_{{\rm D},\Omega}~=J_{D_{ii'}}\, .
\end{aligned}
\ee 
Notice the inverted roles of $\Omega$ and $\Omega g$ in the  Neumann and Dirichlet sectors. The precise actions of $\Omega G$ for $G=1$ or $g$ on the NN sector are~\cite{GimonPolchinski}
\be
\forall \alpha\in\{1,\dots, N_{ii'}\}\, , \forall \beta\in\{1,\dots, N_{jj'}\}\, , ~~~ |\alpha\beta\rangle\to \sum_{\alpha'=1}^{N_{ii'}}\sum_{\beta'=1}^{N_{jj'}}(\gamma^{ii'}_{{\rm N},\Omega G})_{\alpha\alpha'}|\beta'\alpha'\rangle ( \gamma^{jj'\, -1}_{{\rm N},\Omega G})_{\beta'\beta}\, , 
\ee
and similarly for the DD sector. As compared to \Eq{Arule}, note the reversal $\alpha'\beta'\to \beta'\alpha'$ in the transformation rule. As a result, the ND and DN sectors automatically yield vanishing contributions in the defining trace of $Z_\M$.   Moreover, all characters denoted generically as $\hat \chi$ appearing in the M\"obius strip partition function can be defined in terms of their counterparts $\chi$ in the annulus amplitude by the relation~\cite{review-1,review-2} 
\be
\hat{\chi}\Big(\frac{1}{2}+i{\tau_{2}\over 2}\Big)=e^{-i\pi(h-{c\over 24})}\, \chi\Big(\frac{1}{2}+i{\tau_{2}\over 2}\Big)\,,
\label{hatc}
\ee
where $h$ is the weight of the associated primary state and
$c$ the central charge of the Verma module. With these notations, one obtains 
\be
\begin{aligned}
\label{mobius}
Z_\M=&-{1\over 4}\,\sum_{i,i'} \bigg\{\bigg[\carsm(V,O,O,V,+)\bigg(\tr(\gamma_{{\rm N},\Omega}^{ii'\, {\rm T}}\gamma_{{\rm N},\Omega}^{ii'\,-1})\sum_{\vec m}{P^{(4)}_{\vec m}\over \hat \eta^4}+\tr(\gamma_{{\rm D},\Omega g}^{ii'\, {\rm T}}\gamma_{{\rm D},\Omega g}^{ii'\,-1})\sum_{\vec n}{W^{(4)}_{\vec n}\over \hat\eta^4}\bigg)\\
&-\carsm(V,O,O,V,-)\big(\tr(\gamma_{{\rm N},\Omega g}^{ii'\, {\rm T}}\gamma_{{\rm N},\Omega g}^{ii'\,-1})+\tr(\gamma_{{\rm D},\Omega}^{ii'\, {\rm T}}\gamma_{{\rm D},\Omega}^{ii'\,-1})\big)\tetadm\bigg]\sum_{\vec{m}'}\frac{P^{(2)}_{\vec{m}'}}{\hat{\eta}^{4}}\\
&-\bigg[\carsm(C,C,S,S,+)\bigg(\tr(\gamma_{{\rm N},\Omega}^{ii'\, {\rm T}}\gamma_{{\rm N},\Omega}^{ii'\,-1})\sum_{\vec m}{P^{(4)}_{\vec m}\over \hat\eta^4}+\tr(\gamma_{{\rm D},\Omega g}^{ii'\, {\rm T}}\gamma_{{\rm D},\Omega g}^{ii'\,-1})\sum_{\vec n}{W^{(4)}_{\vec n}\over \hat\eta^4}\bigg)\\
&-\carsm(C,C,S,S,-)\big(\tr(\gamma_{{\rm N},\Omega g}^{ii'\, {\rm T}}\gamma_{{\rm N},\Omega g}^{ii'\,-1})+\tr(\gamma_{{\rm D},\Omega}^{ii'\, {\rm T}}\gamma_{{\rm D},\Omega}^{ii'\,-1})\big)\tetadm\bigg]\sum_{\vec{m}'}\frac{P^{(2)}_{\vec{m}'+\vec{a}'_{\text{S}}}}{\hat{\eta}^{4}}\bigg\}\,,
\end{aligned}
\ee  
where the arguments of all hatted characters are $(1+i\tau_2)/2$, and  the superscript $\rm T$ stands for the transposition of the matrix to which it applies.

For completeness, the closed-string sector contributions to the partition function $Z_\T$ and $Z_\K$ are displayed in \Appendix{ztk}. 


\paragraph{\em Spectrum: } The classical massless spectrum can be read from the partition function. To this end, it is useful to evaluate the traces over the Chan--Paton indices in the open string sector, which yields
\be
\begin{aligned}
N_{ii'}&\equiv n_{ii'}+\bar n_{ii'}&&=\tr \gamma^{ii'}_{{\rm N},1}=\tr \gamma^{ii'\, -1}_{{\rm N},1}=\tr(\gamma_{{\rm N},\Omega}^{ii'\, T}\gamma_{{\rm N},\Omega}^{ii'\,-1})=\tr(\gamma_{{\rm N},\Omega g}^{ii'\, T}\gamma_{{\rm N},\Omega g}^{ii'\,-1})\, , \\
0&\equiv i (n_{ii'}-\bar n_{ii'})&&=\tr \gamma^{ii'}_{{\rm N},g}=-\tr \gamma^{ii'\, -1}_{{\rm N},g}\, ,\\
D_{ii'}&\equiv d_{ii'}+\bar d_{ii'}&&=\tr \gamma^{ii'}_{{\rm D},1}=\tr \gamma^{ii'\, -1}_{{\rm D},1}=\tr(\gamma_{{\rm D},\Omega g}^{ii'\, T}\gamma_{{\rm D},\Omega g}^{ii'\,-1})=\tr(\gamma_{{\rm D},\Omega}^{ii'\, T}\gamma_{{\rm D},\Omega}^{ii'\,-1})\, , \\
0&\equiv i (d_{ii'}-\bar d_{ii'})&&=\tr \gamma^{ii'}_{{\rm D},g}=-\tr \gamma^{ii'\, -1}_{{\rm D},g}\, ,
\end{aligned}
\ee
where we use the fact that the matrix $J_k$ for $k$ even has equal number of eigenvalues $i$ and~$-i$. 

\begin{figure}[H]
\captionsetup[subfigure]{position=t}
\begin{center}
\begin{subfigure}[t]{0.48\textwidth}
\begin{center}
\includegraphics [trim=0cm 16cm 10cm 0cm,clip,scale=0.70]{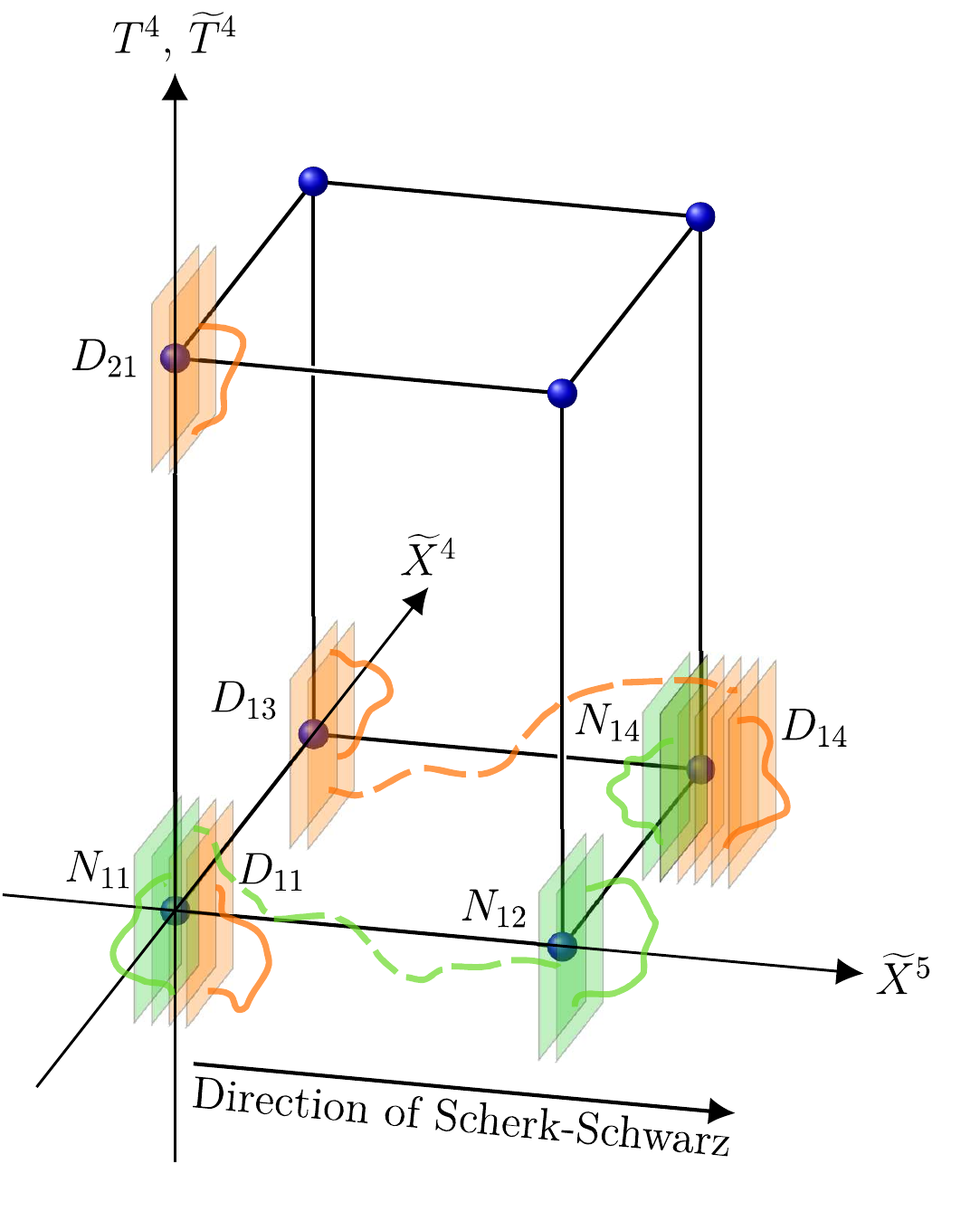}
\end{center}
\caption{\footnotesize Bosonic states in the NN and DD sectors  are massless when their ends are attached to the same stack of branes. 
By contrast, fermionic states in the NN and DD sectors  are massless when they are realized as strings stretched between corners of the six-dimensional boxes that are facing each other along the \mbox{T-dual} Scherk--Schwarz direction.}
\label{massless_NNDD}
\end{subfigure}
\quad
\begin{subfigure}[t]{0.48\textwidth}
\begin{center}
\includegraphics [trim=0cm 16cm 10cm 0cm,clip,scale=0.70]{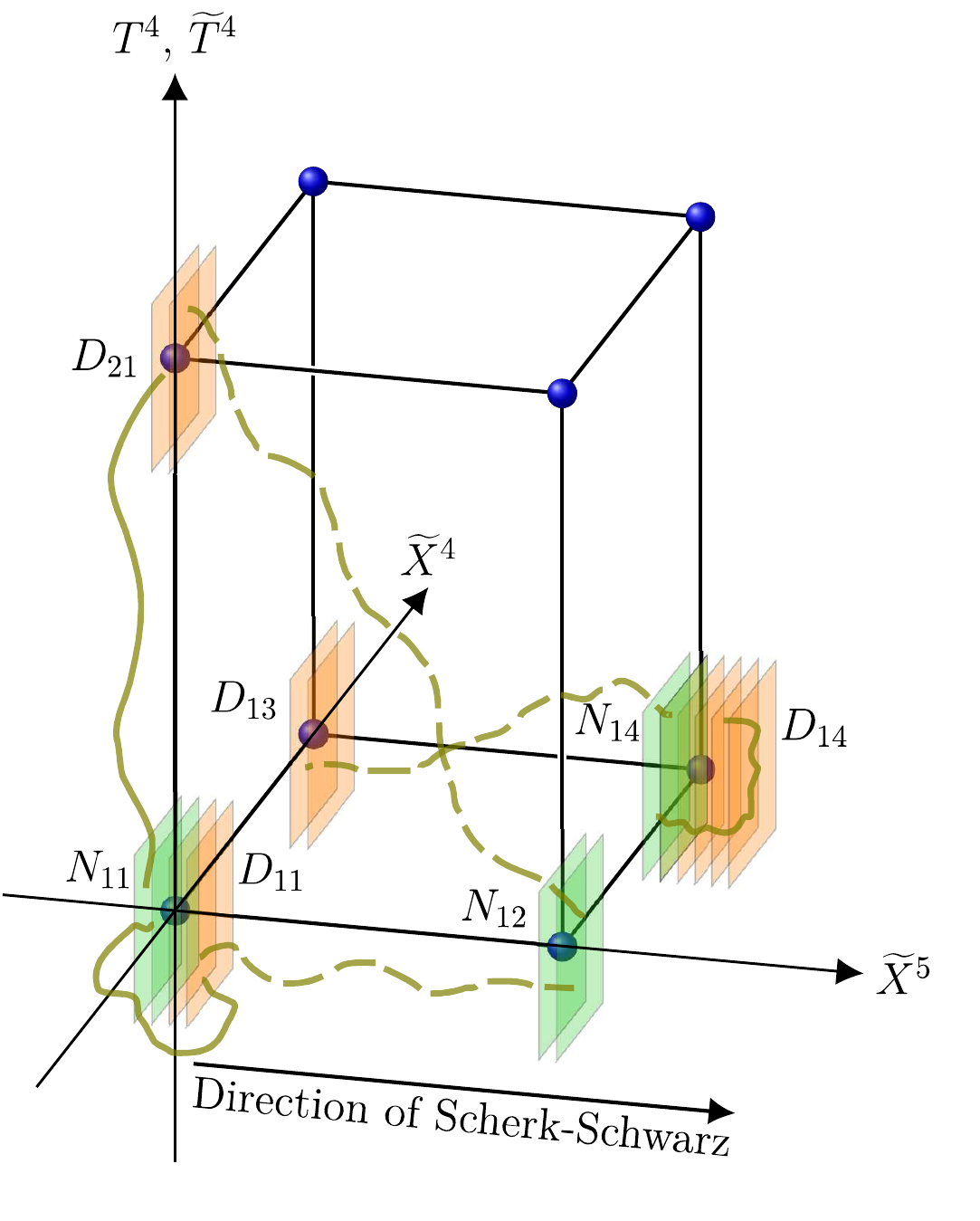}
\end{center}
\caption{\footnotesize Massless bosonic states in the ND+DN sector are symbolized as strings attached to stacks of D3-branes T-dual to D9-branes and D5-branes that are located at corners having the same  coordinates $\tilde X^4$ and $\tilde X^5$. For the massless fermionic states in the ND+DN sector, the corners have same coordinate $\tilde X^4$ and distinct coordinate~$\tilde X^5$. }
\label{massless_ND}
\end{subfigure}
\caption{\footnotesize Visualization of the massless open-string states in the D3-brane pictures. The scalars are depicted as solid lines and the fermions as dashed lines.}
\label{massless_picture}
\end{center}
\end{figure}

From $Z_\A+Z_\M$, one finds that the massless bosonic degrees of freedom are the low-lying modes of the combinations of characters  
\be
\begin{aligned}
\label{massless_bosons}
&
{1\over \eta^8}\sum_{i,i'}\bigg\{\vq\oq\left[n_{ii'}\bar{n}_{ii'}+d_{ii'}\bar{d}_{ii'}\right]\\
&+\oq\vq\left[\frac{n_{ii'}(n_{ii'}-1)}{2}+\frac{\bar{n}_{ii'}(\bar{n}_{ii'}-1)}{2}+\frac{d_{ii'}(d_{ii'}-1)}{2}+\frac{\bar{d}_{ii'}(\bar{d}_{ii'}-1)}{2}\right]\\
&+\oq\cq\sum_{j}\bigg[\frac{1-e^{4i\pi\vec{a}_{i}\cdot\vec{a}_{j}}}{2}\left(n_{ii'}d_{ji'}+\bar{n}_{ii'}\bar{d}_{ji'}\right)+\frac{1+e^{4i\pi\vec{a}_{i}\cdot\vec{a}_{j}}}{2}\left(n_{ii'}\bar{d}_{ji'}+\bar{n}_{ii'}d_{ji'}\right)\bigg]\bigg\}\, .
\end{aligned}
\ee
In the products of $SO(4)$ characters, the first is telling us whether the states belong to vectorial or singlet representations of the six-dimensional Lorentz group. Hence, the first line corresponds to the bosonic parts of an $\N=2$ vector multiplet in the adjoint representation of the open-string gauge group 
\be
\prod_{ii' / n_{ii'}   \neq 0} U(n_{ii'})  ~~\times \prod_{ jj' / d_{jj'}   \neq 0} U(d_{jj'})\, ,\quad \where \quad \sum_{ii'}n_{ii'}=\sum_{ii'}d_{ii'}=16\, ,
\ee
while the second line corresponds to the bosonic parts of one hypermultiplet in the antisymmetric $\oplus$ 
$\overline{\text{antisymmetric}}$ representation of each unitary factor. All of these states, which arise in the NN and DD sectors, are visualized in \Fig{massless_NNDD} as strings drawn in green (NN) or orange (DD) solid  lines with both ends attached to the same stacks of D3-branes. On the contrary, the third line in (\ref{massless_bosons}), which  is associated with the ND + DN sector, corresponds to the bosonic part of one hypermultiplet in the fundamental $\otimes$ fundamental representation of each $U(n_{ii'})\times U(d_{ji'})$ if $e^{4i\pi\vec a_i\cdot \vec a_j}=-1$, and  in the fundamental $\otimes$ $\overline{\text{fundamental}}$ of this group product if $e^{4i\pi\vec a_i\cdot \vec a_j}=1$.\footnote{Each product of characters $O_4C_4$ yields 2 massless degrees of freedom which can be combined into complex scalars.} They are depicted in \Fig{massless_ND} as khaki strings drawn in solid lines stretched between corners $ii'$ and $ji'$, \ie fixed points with identical positions in $\tilde T^2$. As already mentioned in Footnote~\ref{precision}, even if convenient, the visualization in terms of stretched strings is abusive in this case since  corners $ii'$ and $ji'$ are understood in distinct T-dual descriptions. Notice that the moduli whose masses we want to calculate in the present work are among these scalars.

To proceed the same way for the  fermions, it is convenient to define a new double-primed index $i''\in\{1,2\}$ and write $i'=2i''$ or $2i''-1$. The massless fermionic degrees of freedom extracted from $Z_\A+Z_\M$ are then identified as the low-lying modes of the following characters,   
\be
\begin{aligned}
\label{massless_fermions}
&
{1\over \eta^8}\sum_{i,i''}\bigg\{\cq\cq\left[n_{i,2i''-1}\bar{n}_{i,2i''}+\bar{n}_{i,2i''-1}n_{i,2i''}+d_{i,2i''-1}\bar{d}_{i,2i''}+\bar{d}_{i,2i''-1}d_{i,2i''}\right]\\
&+\sq\sq\left[n_{i,2i''-1}n_{i,2i''}+\bar{n}_{i,2i''-1}\bar{n}_{i,2i''}+d_{i,2i''-1}d_{i,2i''}+\bar{d}_{i,2i''-1}\bar{d}_{i,2i''}\right]\\
&+\sq\oq\sum_{j}\bigg[\frac{1-e^{4i\pi\vec{a}_{i}\cdot\vec{a}_{j}}}{2}\left(n_{i,2i''-1}d_{j,2i''}+\bar{n}_{i,2i''-1}\bar{d}_{j,2i''}+n_{i,2i''}d_{j,2i''-1}+\bar{n}_{i,2i''}\bar{d}_{j,2i''-1}\right)\\
&\hspace{1cm}+\frac{1+e^{4i\pi\vec{a}_{i}\cdot\vec{a}_{j}}}{2}\left(n_{i,2i''-1}\bar{d}_{j,2i''}+\bar{n}_{i,2i''-1}d_{j,2i''}+n_{i,2i''}\bar{d}_{j,2i''-1}+\bar{n}_{i,2i''}d_{j,2i''-1}\right)\bigg]\bigg\}\,.
\end{aligned}
\ee
They all correspond to fermionic parts of hypermultiplets in  fundamental $\otimes$ fundamental or  fundamental $\otimes$ $\overline{\text{fundamental}}$ representations of pairs of unitary groups supported by stacks of D3-branes (in the T-dual pictures) located at corners with distinct coordinates along the \SS direction $\tilde X^5$ (and possibly distinct positions in $T^4/\Z_2$ or $\tilde T^4/\Z_2$ in the ND+DN sector). They appear as strings drawn in dashed lines in \Fig{massless_picture}: Green and orange for the NN and DD sectors in \Fig{massless_NNDD}, and khaki for the ND+DN sector in   \Fig{massless_ND}. Actually, massless fermions are realized as string stretched along the $\tilde X^5$ direction, translating the fact that the shifts of $m'_5$ arising from the Wilson lines and the \SS mechanism compensate each other (see \Eq{shifts}).

Because the closed-string spectrum is neutral with respect to the gauge group generated by the open strings, it is independent of the deformations $\vec a_i$ and $\vec a_{i'}$. As a result, all fermions initially massless in the parent supersymmetric model of \Sect{BSGP} acquire tree-level masses equal to $M_{3/2}$ thanks to the \SS mechanism. At the massless level, we are left with bosons, which are  easily listed from a six-dimensional point of view. The untwisted sector contains the components $(G+C)_{\hat \mu\hat\nu}$, $\hat\mu,\hat \nu\in\{2,\dots, 5\}$, and the internal components $(G+C)_{IJ}$, which yield $(6-2)\times (6-2)+4\times 4$ degrees of freedom. Moreover, there are $4\times 16$ real scalars arising from the twisted hypermultiplets.  


\section{Two-point functions of massless ND and DN states}
\label{ndmass}

In \Reff{ACP} the masses at one loop of the open-string moduli arising from the NN and DD sectors were derived by using the background field method. However, in the case of the moduli in the ND+DN sector, the partition function for arbitrary vev's of these scalars is not known and this approach cannot be applied. Therefore, we will derive in \Sects{NDmasses}  and~\ref{alpha'0} the one-loop masses of all classically massless scalars in the bifundamental representations of unitary groups supported by D9- and D5-branes by computing two-point correlation functions with external states in the massless ND and DN bosonic sectors. This will be done by applying techniques first introduced in classical open-string theories in \Reffs{Cvetic-Abel1,Cvetic-Abel2,Cvetic-Abel3}, and at one loop in \Reffs{Abel:2004ue,Schofield,markG1,markG2,markG3}. For now, we define the relevant vertex operators and open-string amplitudes. 


\subsection{Vertex operators and amplitudes}
\label{veramp}

In the T-dual pictures, let us consider two corners $i_0i_0'$ and $j_0i_0'$ on which are located $N_{i_0i_0'}\ge 2$ and $D_{j_0i_0'}\ge 2$ D3-branes T-dual to D9-branes and D5-branes, respectively. As seen in the third line of \Eq{massless_bosons}, the open strings ``stretched'' between these stacks give rise to $2n_{i_0i_0'}d_{j_0j_0'}$ massless complex scalars (depicted as solid  strings in \Fig{massless_ND}). In the initial description in terms of D9- and D5-branes, we are interested in correlation functions of vertex operators in ghost pictures $p$ and $-p$ of the form 
\be
\sum_{\alpha_0=1}^{N_{i_0i_0'}}\sum_{\beta_0=1}^{D_{j_0i_0'}}\left\langle V^{\alpha_0\beta_0}_p(z_1,k,\epsilon)V^{\beta_0\alpha_0}_{-p}(z_2,-k,-\epsilon)\right\rangle^\Sigma\,,
\label{ampi}
\ee
where $z_1$, $z_2$ are insertion points on the boundary of a worldsheet  whose topology is either that of the  annulus  or M\"obius strip, $\Sigma\in\{\A,\M\}$, and 
\be
\begin{aligned}
V_{-1}^{\alpha_0\beta_0}(\zu,k,\epsilon)&=\lambda_{\alpha_0\beta_0}\,e^{-\phi}\,e^{ik\cdot X}\, e^{\epsilon{i\over2}(H_{3}-H_{4})}\, \sigma^{3}\sigma^{4}(\zu)\,,\\
V_{-1}^{\beta_0\alpha_0}(\zd,-k,-\epsilon)&=\lambda^{\rm T}_{\beta_0\alpha_0}\,e^{-\phi}\,e^{-ik\cdot X}\,e^{-\epsilon{i\over 2}(H_{3}-H_{4})}\, \sigma^{3}\sigma^{4}(\zd)\,.
\end{aligned}
\label{vop}
\ee
In the above definitions, we use the following  notations:
\begin{itemize}
\item $k^\mu$ is the external momentum satisfying on-shell the condition $k^\mu k_\mu=0$.

\item $\phi(z)$ is the ghost field encountered in the bosonization of the superconformal ghosts~\cite{FMS}. 

\item $\lambda$ is the matrix 
\be
\lambda=\begin{pmatrix} \Lambda_1 & \Lambda_2\\  -\Lambda_2 & \Lambda_1\end{pmatrix} , 
\ee
where $\Lambda_1$, $\Lambda_2$ are arbitrary $n_{i_0i_0'}\times d_{j_0i_0'}$ real matrices~\cite{GimonPolchinski}. It labels the states that transform as the {\bm $(n_{i_0i_0'},d_{j_0i_0'})\oplus (\bar n_{i_0i_0'},\bar d_{j_0i_0'})$} or {\bm $(n_{i_0i_0'},\bar d_{j_0i_0'})\oplus (\bar n_{i_0i_0'}, d_{j_0i_0'})$} bifundamental representation of $U(n_{i_0i_0'})\times  U(d_{j_0i_0'})$.

\item From now until  \Sect{alpha'0}, we restrict our analysis to the case where the internal metric is diagonal, 
\be
G_{I'J'}\equiv \delta_{I'J'}\, {R_{I'}^2\over \alpha'}\, , \quad~~G_{IJ}\equiv \delta_{IJ}\, {R_{I}^2\over \alpha'}\, ,
\ee
for some radii $R_{I'}$, $R_I$. In this case, the formalism of \Reff{Atick} applies without having to generalize it.  
Denoting $\psi^\mu(z)$, $\psi^{I'}(z)$, $\psi^I(z)$ the Grassmann fields superpartners of the bosonic-coordinate fields $X^\mu(z)$, $X^{I'}(z)$, $X^I(z)$, we define a new basis of degrees of freedom
\be
\label{Zu}
\begin{aligned}
&{\X}^u\equiv \frac{ X^{2u}+i X^{2u+1}}{\sqrt{2}}\, , &&\overbar{\X}^u\equiv \frac{ X^{2u}-iX^{2u+1}}{\sqrt{2}}\, ,\\
&{\Psi}^u\equiv  \frac{ \psi^{2u}+i\psi^{2u+1}}{\sqrt{2}}\equiv e^{iH_u}\, , &&\overbar{\Psi}^u\equiv \frac{\psi^{2u}-i\psi^{2u+1}}{\sqrt{2}}\equiv e^{-iH_u}\, ,\quad u\in\{0,\dots,4\}\,,
\end{aligned}
\ee
where $H_u$ are scalars introduced to bosonize the fermionic fields.\footnote{These definitions apply to a Euclidean spacetime. In the Lorentzian case, replace $(X^0, \psi^0)\to i(X^0, \psi^0)$.} 

\item The characters $O_4C_4$ tell us that the scalars we are interested in are organized as singlet from a six-dimensional point of view, and spinors of the  $T^4/\Z_2$ orbifold space. The operators $e^{\pm\epsilon i( H_3-H_4)}$ are therefore spin fields, which means that the coefficients of $H_3$, $H_4$ in the exponentials are  the weights of the dimension-two spinorial representation of negative chirality of $SO(4)$, which are $\epsilon (\half,-\half)$, $\epsilon\in\{-1,+1\}$.\footnote{Characters $O_4S_4$ would yield states in the spinorial representation of positive chirality of $SO(4)$, whose weights are $\epsilon (\half,\half)$.} 

\item $\sigma^u$, $u\in\{3,4\}$, are so-called ``boundary-changing fields'' associated with the complex directions $Z^u$~\cite{Hashimoto:1996he}.  
\end{itemize}

\begin{figure}[H]
\begin{center}
\raisebox{-0.5\height}{\includegraphics[scale=0.60]{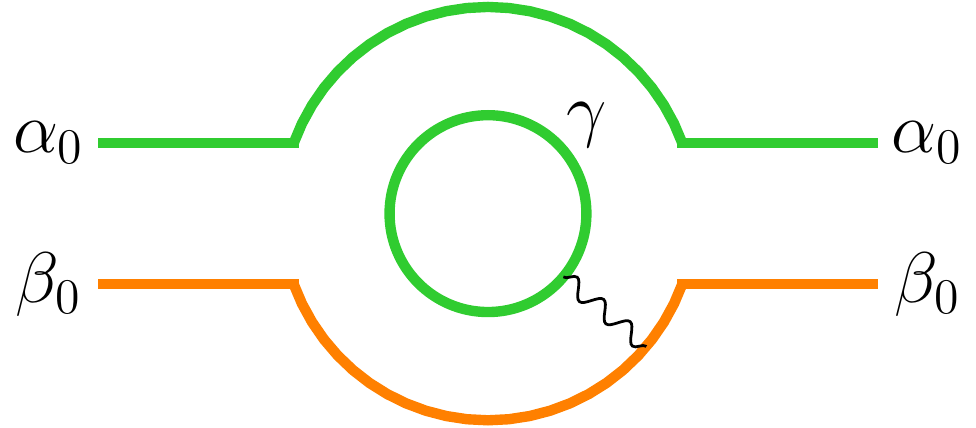}}
\quad~
\raisebox{-0.5\height}{\includegraphics[scale=0.60]{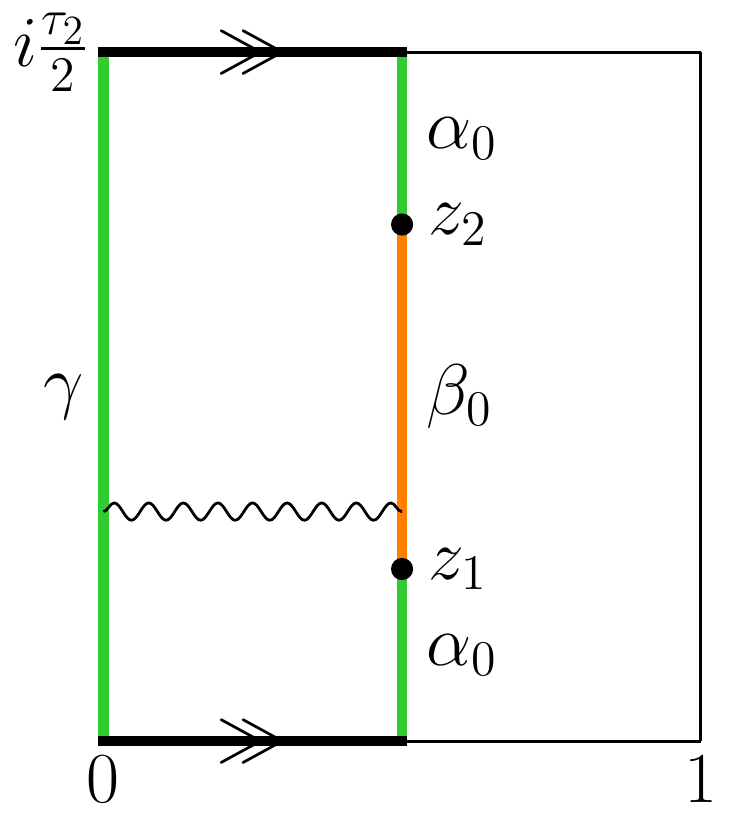}}
\quad~
\raisebox{-0.5\height}{\includegraphics[scale=0.60]{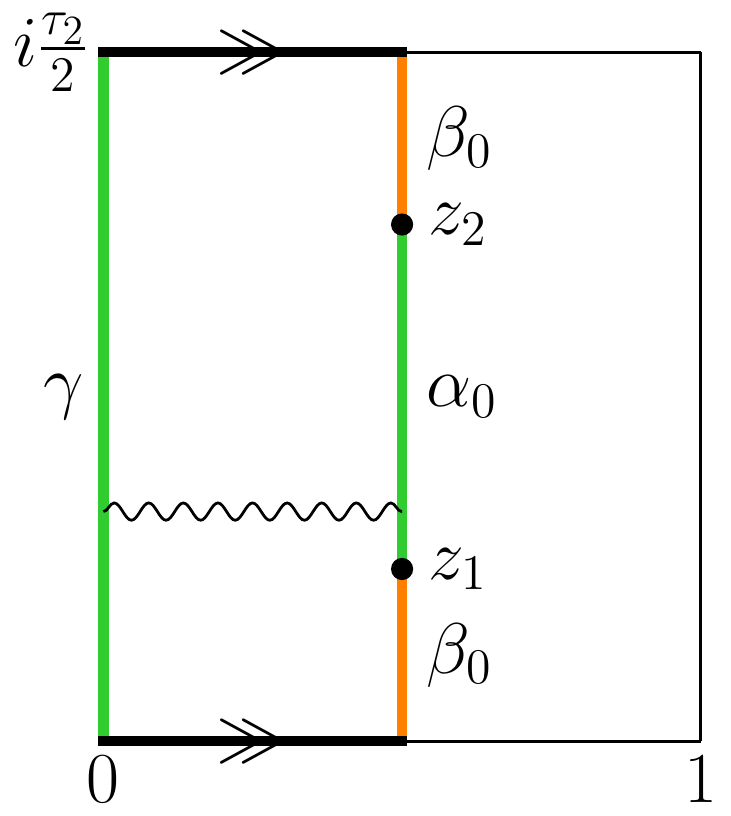}}\\[0.5cm]

\raisebox{-0.5\height}{\includegraphics[scale=0.60]{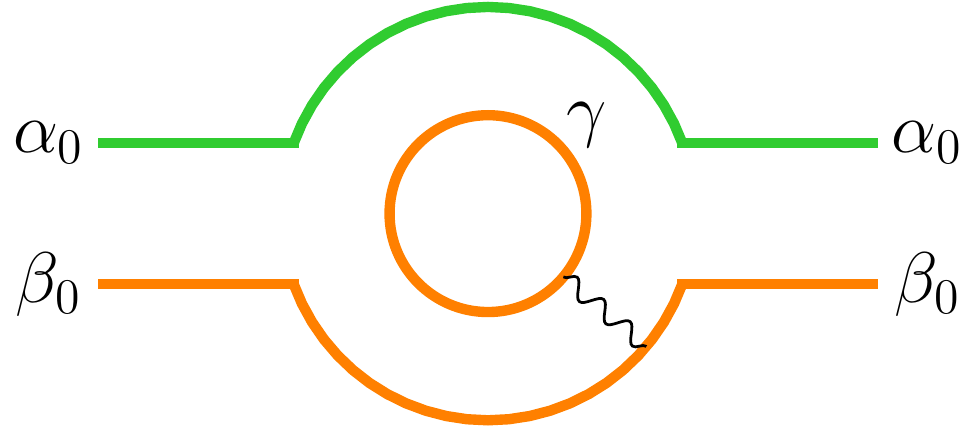}}
\quad~
\raisebox{-0.5\height}{\includegraphics[scale=0.60]{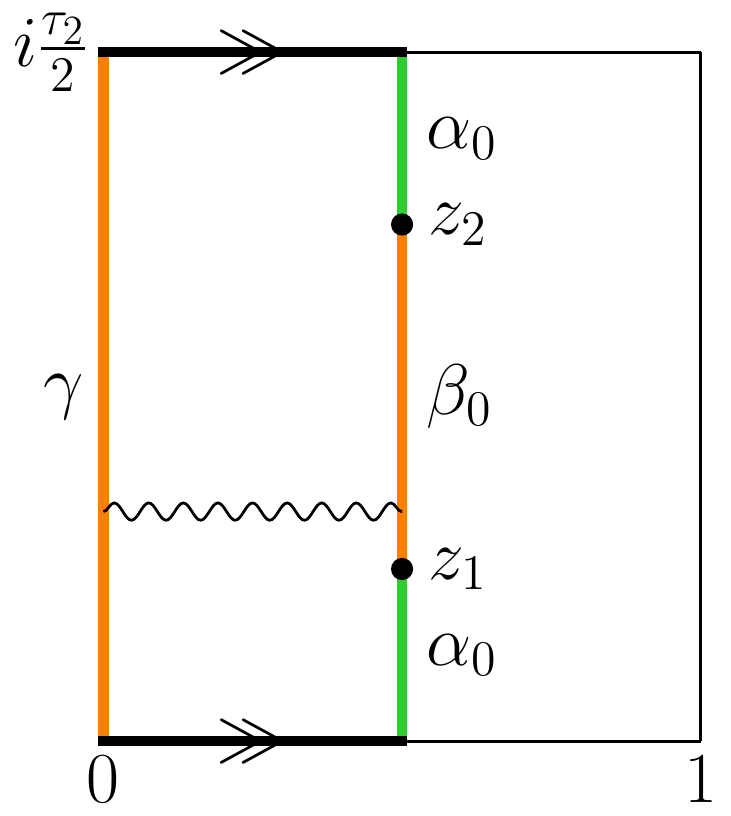}}
\quad~
\raisebox{-0.5\height}{\includegraphics[scale=0.60]{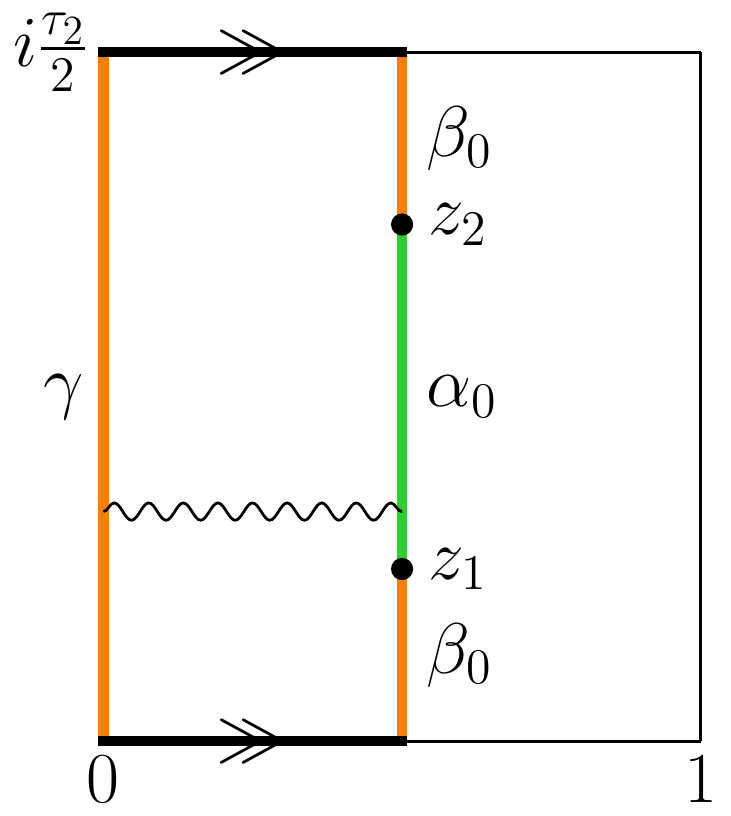}}\\[0.5cm]

\!\!\!\,\raisebox{-0.5\height}{\includegraphics[scale=0.60]{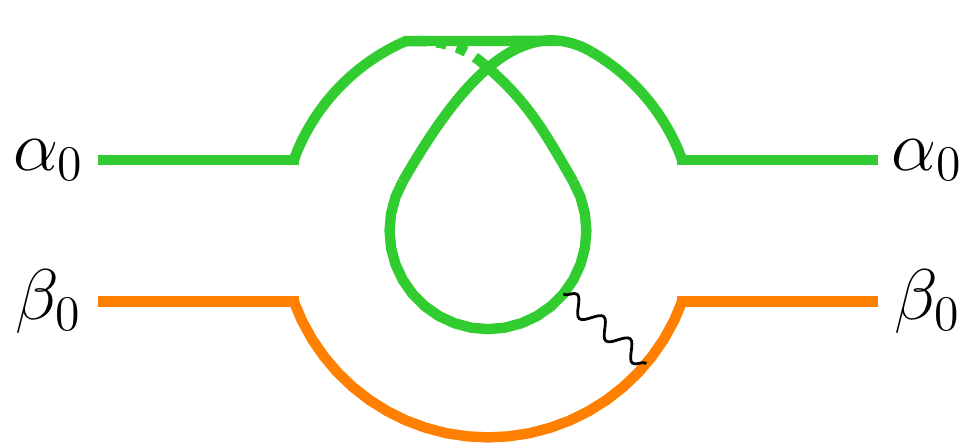}}
\!\!\!\!\!\!\!\!\!\!\!\!
\raisebox{-0.5\height}{\includegraphics[scale=0.60]{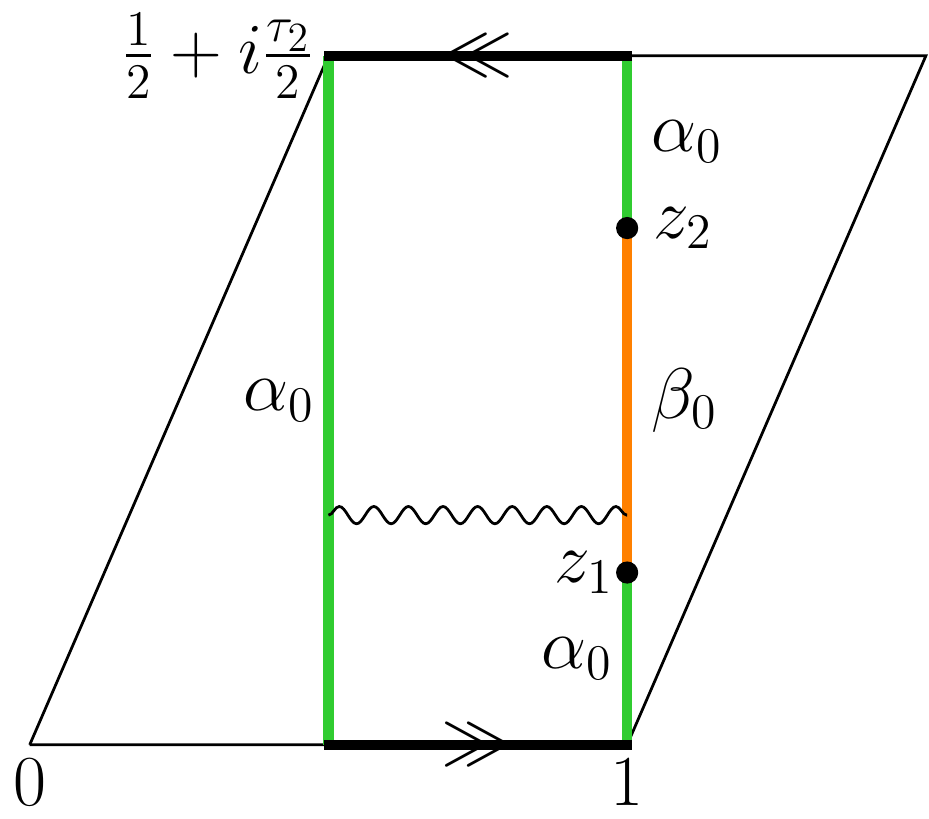}}
\!\!\!\!\!\!\!\!\!
\raisebox{-0.5\height}{\includegraphics[scale=0.60]{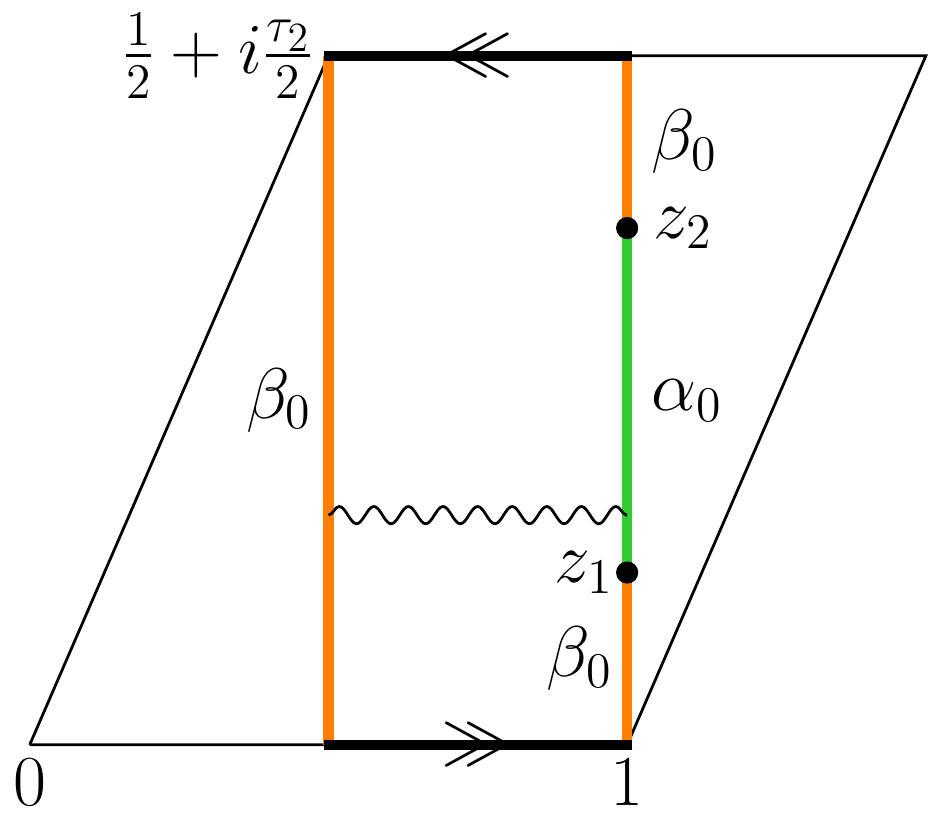}}
\end{center}
\caption{\footnotesize Open-string diagrams with two external legs in the ND and DN sectors (left panel). On the double-cover tori (center and right panels), the external legs are mapped to boundary-changing vertex operators at $z_1$, $z_2$. One switches from center to right panel by transporting $z_2$ along the entire edge it belongs to.}
\label{annulus+mobius_diagrams}
\end{figure}

To understand the meaning and necessity of  introducing operators $\sigma^u$, the open-string diagrams we want to compute are displayed in \Fig{annulus+mobius_diagrams} for some given $\alpha_0\in\{1,\dots,N_{i_0i_0'}\}$ and $\beta_0\in\{1,\dots,D_{j_0i_0'}\}$. The left panel shows two annuli  and one M\"obius strip amplitudes. Because the external legs bring quantum numbers  $(\lambda_{\alpha_0\beta_0},\epsilon)$ and $(\lambda^{\rm T}_{\beta_0\alpha_0},-\epsilon)$ of the ND and DN sectors, they must be attached to the \emph{same} boundary of the annulus. Therefore, the second boundary is sticked to another brane labelled $\gamma$, which can be any of the 32 D9-branes (in green) or 32 D5-branes (in orange). On the center and right panels, the same three diagrams are displayed, with the open-string worldsheets seen as fundamental domains of the involution
\be
 z\,\longrightarrow \,\I(z)\equiv1-\bar z
 \label{invo}
\ee
 acting on double-cover tori of Teichm\"uller parameters~\cite{Burgess1,Burgess2,ABFPT,review-1} 
\be
\taudc=i{\tau_2\over 2}~~\mbox{for the annulus} \quad ~~\and~~\quad  \taudc=\half+i{\tau_2\over 2}~~\mbox{for the M\"obius strip}\, .
\label{temo}
\ee
In this description, the external legs are conformally mapped to points $z_1$, $z_2$, where vertex  operators change the boundary conditions of the worldsheet fields $X^I(z)$ (\ie $Z^3(z)$, $Z^4(z)$)
\noindent
at one end of the intermediate open string running in the loop, from Neumann to Dirichlet or {\it vice versa}. The diagrams in the center and right panels are obtained from one another by transporting continuously $z_2$ along its entire boundary: $z_2\to z_2+i\,\Im \taudc$ for the annulus and  $z_2\to z_2+2i\,\Im \taudc$ for the M\"obius strip, modulo 1 and $\taudc$.

To conclude this subsection, notice that for consistency of the diagrams, the numbers of boundary-changing vertex operators must be even on each connected component of an open-string surface. Hence, all one-point functions \ie tadpoles of states in the ND or DN sectors vanish, which shows that the backgrounds we consider, \ie where no brane recombination is taking place~\cite{recomb1,recomb2,recomb3,recomb4}, imply the effective potential to be  extremal with respect to the scalars in the ND+DN sector.


\subsection{OPE's and ghost-picture changing}
\label{OPEpic}

In order to treat symmetrically both vertex operators when computing the correlation functions~(\ref{ampi}), we switch to the ghost picture $p=0$. This is done by  applying the formula~ 
\be
\begin{aligned}
V_{0}^{\alpha_0\beta_0}(z,k,\epsilon)&=\lim_{w\rightarrow z}e^{\phi}\,T_{\text{F}}(w)\,V^{\alpha_0\beta_0}_{-1}(z,k,\epsilon)\,,\\
V_{0}^{\beta_0\alpha_0}(z,-k,-\epsilon)&=\lim_{w\rightarrow z}e^{\phi}\,T_{\text{F}}(w)\,V^{\beta_0\alpha_0}_{-1}(z,-k,-\epsilon)\,,
\end{aligned}
\ee
where $T_{\rm F}$ is the  supercurrent given by 
\begin{align}
\begin{split}
T_{\text{F}}(z)=\frac{1}{\sqrt{\alpha'}}\,\partial X^{\mu}\psi_{\mu}(z)=\frac{1}{\sqrt{\alpha'}}\big(\partial \overbar{Z}^{u}\Psi^{u}(z)+\partial Z^{u}\overbar{\Psi}^{u}(z)\big)\,.
\end{split}
\end{align}
To this end, we display all necessary  operator product expansions (OPE's). First of all, for the ``ground-state boundary-changing fields'', we have 
\begin{align}
\begin{split}
\label{OPE_def1}
&\partial \X^u(z)\sigma^u(w)\underset{z\to w}{\sim} (z-w)^{-\half}\, \tau^u(w)+\mbox{finite}\, ,\\
&\partial\overbar{\X}^u(z)\sigma^u(w)\underset{z\to w}{\sim} (z-w)^{-\half}\,\tau^{\prime u}(w)+\mbox{finite}\,,
\end{split}
\end{align}
which introduces ``excited boundary-changing fields'' $\tau^u$, $\tau^{\prime u}$. Moreover, the other fields satisfy~\cite{Abel:2004ue,Schofield}\footnote{The definitions of $K^0$, $\overbar K^0$ apply to a Euclidean spacetime. In the Lorentzian case, replace $k^0\to ik^0$.}  
\begin{align}
e^{a\phi}(z)e^{b\phi}(w)&\underset{z\to w}{\sim} (z-w)^{-ab}\, e^{(a+b)\phi}(w)+\mbox{finite}\,,\nonumber \\
e^{iaH_u}(z)e^{ibH_u}(w)&\underset{z\to w}{\sim} (z-w)^{ab}\, e^{i(a+b)H_u}(w)+\mbox{finite}\,,\quad u\in\{3,4\}\, ,\\
\partial Z^u(z)\, e^{ik\cdot X}(w)&\underset{z\to w}{\sim}{iK^u\over z-w}\, e^{ik\cdot X}(w)+\mbox{finite}\,,\quad \where \quad K^u={k^{2u}+ik^{2u+1}\over \sqrt{2}}\, , \nonumber\\
\partial \overbar Z^u(z)\, e^{ik\cdot X}(w)&\underset{z\to w}{\sim}{i\overbar K^u\over z-w}\, e^{ik\cdot X}(w)+\mbox{finite}\, , \quad \where \quad \overbar K^u={k^{2u}-ik^{2u+1}\over \sqrt{2}}\, , \quad u\in\{0,1,2\}\, .\nonumber
\end{align}
Using these relations, we obtain for $\epsilon=+1$
\be
\begin{aligned}
V_0^{\alpha_0\beta_0}(z_1,k,+1)&=V_{0,\rm ext}^{\alpha_0\beta_0}(z_1,k,+1)+V_{0,\rm int}^{\alpha_0\beta_0}(z_1,k,+1)\,,\\
V_0^{\beta_0 \alpha_0}(z_2,-k,-1)&=V_{0,\rm ext}^{\beta_0\alpha_0}(z_1,-k,-1)+V_{0,\rm int}^{\beta_0\alpha_0}(z_1,-k,-1)\,,
\end{aligned}
\ee
where we have defined
\be
\begin{aligned}
V_{0,\rm ext}^{\alpha_0\beta_0}(z_1,k,+1)&=\sqrt{\alpha'}\, \lambda_{\alpha_0\beta_0}\, e^{i k\cdot X}\, i\sum_{u=0}^1(K^u\overbar\Psi^u+\bar K^u\Psi^u)\,e^{\frac{i}{2}(H_{3}-H_{4})}\, \sigma^{3}\sigma^{4}(\zu)\,,\\
V_{0,\rm int}^{\alpha_0\beta_0}(z_1,k,+1)&={\lambda_{\alpha_0\beta_0}\over \sqrt{\alpha'}}\, e^{i k\cdot X}\Big(e^{-\frac{i}{2}(H_{3}+H_{4})}\, \tau^{3}\sigma^{4}(\zu)+e^{\frac{i}{2}(H_{3}+H_{4})}\, \sigma^{3}\tau^{\prime4}(\zu)\Big)\,,\\
V_{0,\rm ext}^{\beta_0\alpha_0}(z_2,-k,-1)&=\sqrt{\alpha'}\, \lambda^{\rm T}_{\beta_0\alpha_0}\, e^{-i k\cdot X}\, (-i) \sum_{u=0}^1(K^u\overbar\Psi^u+\bar K^u\Psi^u)\,e^{-\frac{i}{2}(H_{3}-H_{4})}\, \sigma^{3}\sigma^{4}(z_2)\,,\\
V_{0,\rm int}^{\beta_0\alpha_0}(z_2,-k,-1)&={\lambda^{\rm T}_{\beta_0\alpha_0}\over \sqrt{\alpha'}}\, e^{-i k\cdot X}\Big(e^{-\frac{i}{2}(H_{3}+H_{4})}\, \sigma^{3}\tau^{4}(z_2)+e^{\frac{i}{2}(H_{3}+H_{4})}\, \tau^{\prime 3}\sigma^{4}(z_2)\Big)\,,
\end{aligned}
\ee
while the expressions for $\epsilon=-1$ are obtained by exchanging all subscripts and superscripts 3 and 4. Because we are interested in states massless at tree level, the \KK momentum in the $T^2$ complex direction $u=2$ is set to 0 in the ``external'' parts of the vertex operators.  In the ``internal'' parts, notice  the appearance of ``excited boundary-changing operators'' $\tau^3, \tau^{\prime 3},\tau^4,\tau^{\prime 4}$. 

Given the above definitions, the correlation functions~(\ref{ampi}) split accordingly into external and internal pieces. The former,
\be
\label{prelim_ext}
\begin{aligned}
\!\!\!\!A_{\text{ext}\Sigma}^{\alpha_0\beta_0}&\equiv \big\langle V_{0,\rm ext}^{\alpha_0\beta_0}(z_1,k,+1)V_{0,\rm ext}^{\beta_0\alpha_0}(z_2,-k,-1)\big\rangle^\Sigma\\
&=\alpha'\,\lambda_{\alpha_0\beta_0}\lambda^{\rm T}_{\beta_0\alpha_0}\,\langle e^{ik\cdot X}(\zu)e^{-ik\cdot X}(\zd)\rangle\,\langle e^{\frac{i}{2}H_3}(\zu)e^{-\frac{i}{2}H_3}(\zd)\rangle\,\langle e^{-\frac{i}{2}H_4}(\zu)e^{\frac{i}{2}H_4}(\zd)\rangle\times\esp\\
&\!\!\!\!\!\!\!\langle \sigma^{3}(\zu)\sigma^{3}(\zd)\rangle\,\langle \sigma^{4}(\zu)\sigma^{4}(\zd)\rangle\sum_{u=0}^1 K^u\overbar K^u\Big[\langle e^{iH_{u}}(\zu)e^{-iH_{u}}(\zd)\rangle+\langle e^{-iH_{u}}(\zu)e^{iH_{u}}(\zd)\rangle\Big]\,,
\end{aligned}
\ee
are useful to derive the one-loop corrections to the K\"ahler potential of the ND+DN sector massless scalars. Note that in order to bypass the issue that on shell $\sum_{u=0}^1 |K^u|^{2}\equiv k^2/2=0$, we may have kept the \KK momenta along $T^2$ arbitrary. On the contrary, the internal parts,~
\be
\label{prelim_int}
\begin{aligned}
A_{\text{int}\Sigma}^{\alpha_0\beta_0}&\equiv \big\langle V_{0,\rm int}^{\alpha_0\beta_0}(z_1,k,+1)V_{0,\rm int}^{\beta_0\alpha_0}(z_2,-k,-1)\big\rangle^\Sigma\\
&=\frac{1}{\alpha'}\,\lambda_{\alpha_0\beta_0}\lambda^{\rm T}_{\beta_0\alpha_0}\,\langle e^{ik\cdot X}(\zu)e^{-ik\cdot X}(\zd)\rangle\esp\\
&~~~\times\Big[\langle e^{-\frac{i}{2}H_3}(\zu)e^{\frac{i}{2}H_3}(\zd)\rangle\,\langle e^{-\frac{i}{2}H_4}(\zu)e^{\frac{i}{2}H_4}(\zd)\rangle\,\langle \tau^{3}(\zu)\tau'^{3}(\zd)\rangle\,\langle \sigma^{4}(\zu)\sigma^{4}(\zd)\rangle\\
&~~~~\,+\langle e^{\frac{i}{2}H_3}(\zu)e^{-\frac{i}{2}H_3}(\zd)\rangle\,\langle e^{\frac{i}{2}H_4}(\zu)e^{-\frac{i}{2}H_4}(\zd)\rangle\,\langle \sigma^{3}(\zu)\sigma^{3}(\zd)\rangle\,\langle \tau'^{4}(\zu)\tau^{4}(\zd)\rangle\Big]\,,
\end{aligned}
\ee
capture the mass corrections we are interested in. The amplitudes for $\epsilon=-1$ are obtained by exchanging all subscripts 3 and 4 in \Eq{prelim_ext} and all superscripts 3 and 4 in \Eq{prelim_int}. For $\Sigma=\A$, an implicit sum over a second boundary condition $\gamma$ is understood. Likewise, for $\Sigma=\A$, $\M$, sums over the spin structures of the fermions $\Psi^0$,  $\Psi^1$, $\Psi^2$ on the one hand, and $\Psi^3$,  $\Psi^4$ on the other hand are  implicit. 


\section{Genus-1 twist-field correlation functions}
\label{Atick}

 The main difficulty in computing the two-point functions in \Eqs{prelim_ext} and~(\ref{prelim_int}) is to evaluate the correlators of the boundary changing operators. However, it turns out that the OPE's of $\partial Z^u$, $\partial \overbar Z^u$ on these operators are identical to those found for the holomorphic part of ``$\Z_2$-twist fields'' inserted on a closed-string worldsheet, \ie for operators creating closed strings in the twisted sector of a $T^4/\Z_2$ orbifold.  
As a result, we may apply techniques relevant for the computation of correlation functions of twist fields in closed-string theory to our open-string case. In the present section, we review the relevant ingredients for computing correlators of twist fields at genus-1 in closed-string theory, or in the closed-string sector of an open-string theory, compactified on toroidal $\Z_N$ orbifolds, following the original works of  \Reffs{Atick,Dixon}. 


\subsection{Instanton decomposition of correlators}

In closed-string theory compactified on $T^2\times T^4/\Z_N$ where $N\in\natural^*$,  the complex fields defined in \Eq{Zu} depend on holomorphic and antiholomorphic worldsheet coordinates, $Z^u(z,\bar z)$. Moreover, upon parallel transport, the internal $Z^u$ undergo some $\Z_{N}$ rotations and translations,
\be
\begin{aligned}
Z^2&\,\longrightarrow \,Z^2+v^2\, , \\
\X^u&\,\longrightarrow \,e^{2i\pi {\kappa/ N}}\X^u+v^u\, ,\quad u\in\{3,4\}\, , ~~ \kappa\in\{0,\dots, N-1\}\,,
\end{aligned}
\ee
where the shifts $v^u$ and $v^2$ implement the $T^4$ and $T^2$ periodicities.

The twist fields create the states in the twisted sectors of the closed-string Hilbert space. For some given $\kappa\in\{1,\dots, N-1\}$ and $u\in\{3,4\}$, let us  denote by $\sigma^u(z,\bar z)$ the one that creates the ground state in the $\kappa$-th twisted sector. The requirement that positive frequency modes in the expansions of $\partial \X^u$ and $\partial \overbar{\X}^u$ annihilate the twisted ground state determines the OPE of $\partial \X^u(z)$ and $\partial \overbar{\X}^u(z)$ acting on  $\sigma^u(w,\bar{w})$ as $z$ approaches $w$,
\be
\label{OPE_def}
\begin{aligned}
&\partial \X^u(z)\sigma^u(w,\bar{w})\underset{z\to w}{\sim}(z-w)^{-(1-\kappa/N)}\, \tau^u(w,\bar{w})&&\!\!\!\!\!+\mbox{finite}\, ,\\
&\partial\overbar{\X}^u(z)\sigma^u(w,\bar{w})\underset{z\to w}{\sim}(z-w)^{-\kappa/N}\,\tau^{\prime u}(w,\bar{w})&&\!\!\!\!\!+\mbox{finite}\, ,\\
&\bar\partial \X^u(\bar z)\sigma^u(w,\bar{w})\underset{\bar z\to \bar w}{\sim}(\bar{z}-\bar{w})^{-\kappa/N}\,\tilde{\tau}^u(w,\bar{w})&&\!\!\!\!\!+\mbox{finite}\, ,\\
&\bar \partial \overbar{\X}^u(\bar z)\sigma^u(w,\bar{w})\underset{\bar z\to \bar w}{\sim}(\bar{z}-\bar{w})^{-(1-\kappa/N)}\,\tilde{\tau}^{\prime u}(w,\bar{w})&&\!\!\!\!\!+\mbox{finite}\, .
\end{aligned}
\ee
In the right-hand sides, $\tau^u$,  $\tau^{\prime u}$,  $\tilde \tau^u$,  $\tilde \tau^{\prime u}$ create excited states in the $\kappa$-th twisted sector.  The OPE's capture the local behavior corresponding to the rotations of the coordinates $\X^u$ but do not carry information about the global translations $v^u$. This data is recovered by imposing  \emph{global monodromy conditions} which describe how $\X^u(z,\bar z)$ and $\overbar{\X}^u(z,\bar{z})$ change when they are carried around a set of twist fields  with vanishing total twist. Splitting the coordinates of $T^2$ and $T^4$ into background values and quantum fluctuations, 
\be
\X^u(z,\bar{z})=\Xcl^u(z,\bar{z})+ \Xqu^u(z,\bar{z})\,,\quad u\in\{2,3,4\}\,,
\label{splitZ}
\ee
the whole global displacements arise from the classical parts $\Xcl^u(z,\bar{z})$. 

With this decomposition, the correlators of interest on a Riemann surface  $\Sigma$ of genus $g\ge 0$ involve, for each complex direction $u\in\{3,4\}$, $L\ge 2$ ground-state twist fields $\sigma_A^u$ of the $\kappa_A$-th twisted sector,
\be
\label{corsigma}
 \sum_{\Xcl}e^{-\Scl^{\Sigma }}\prod_{u=3}^4 \left\langle \prod_{A=1}^{L}\sigma_A^{u}(z_A,\bar z_A)\right\rangle_{\text{qu}},
\ee
where the total twist is trivial, $\sum_A\kappa_A=0$ modulo $N$, for the result not to vanish~\cite{Dixon}.  
In this expression, the sum is over instantons with worldsheet actions 
\be
\label{classical_action}
S_{\text{cl}}^{\Sigma}=\frac{i}{2\pi\alpha'}\int_{\Sigma}\dd z\wedge \dd\bar z\, \sum_{u=2}^4 \big(\partial\Xcl^u\, \bar \partial\Xclbar^u+\partial\Xclbar^u\,\bar \partial\Xcl^u\big)\,.
\ee
In the following, we first compute the ``quantum parts'' of the correlation functions and then derive the classical actions.   


\subsection{Stress-tensor method}
\label{Stress-tensor-method}

To determine for a given $u\in\{3,4\}$ the quantum part $\left\langle \prod_{A=1}^{L}\sigma_A^{u}(z_A,\bar z_A)\right\rangle_{\text{qu}}$ of the correlator~(\ref{corsigma}), \Reff{Atick} uses the \emph{stress-tensor method}. It consists in exploiting the OPE's between the stress tensor $T^u(z)$ and the primary fields $\sigma_A^{u}(z_A,\bar z_A)$ of conformal weights $h_A=\half(\kappa_A/N)(1-\kappa_A/N)$, namely
\be
\label{OPEtensor}
T^u(z)\sigma_A^{u}(z_A,\bar z_A)\underset{z\to z_A}{\sim}\frac{h_A}{(z-z_A)^2}\,\sigma_A^{u}(z_A,\bar z_A)+\frac{1}{z-z_A}\, \partial_{A}\sigma_A^{u}(z_A,\bar z_A)+\mbox{finite}\, .
\ee

To this  end, one considers the quantity
\be
\langle\langle T^u(z)\rangle\rangle\equiv\frac{\left\langle T^u(z)\prod_{A}\sigma^u_A(z_A,\bar{z}_A)\right\rangle_{\rm qu}}{\left\langle \prod_{A}\sigma^u_A(z_A,\bar{z}_A)\right\rangle_{\rm qu}}\,,
\ee
in terms of which we may write 
\be
\partial_B\ln\big\langle\prod_A\sigma^u_A(z_A,\bar{z}_A)\big\rangle_{\rm qu}=\lim_{z\rightarrow z_B}\left[(z-z_B)\,\langle\langle T^u(z)\rangle\rangle-\frac{h_B}{(z-z_B)}\right],
\label{dez}
\ee
upon using Eq.~(\ref{OPEtensor}). To evaluate $\langle\langle T^u(z)\rangle\rangle$, one considers the  \emph{Green's function in the presence of twist fields},\footnote{Because the integers $\kappa_A$ and insertion points $z_A$ are independent of $u\in\{3,4\}$, the Green's functions derived for $u=3$ and 4 are equal and do not need to be  distinguished by an index $u$.}
\be
\label{Green}
g(z,w)\equiv\frac{\langle -\partial \X^u_{\rm qu}(z)\,\partial \overbar{\X}^u_{\rm qu}(w)\prod_A\sigma^u_A(z_A,\bar{z}_A)\rangle_{\rm qu}}{\alpha'\, \langle\prod_A\sigma^u_A(z_A,\bar{z}_A)\rangle_{\rm qu}}\,,
\ee
and exploits the OPE
\be
\label{OPE_dX}
-{1\over \alpha'}\,\partial \X^u_{\rm qu}(z)\,\partial \overbar{\X}^u_{\rm qu}(w)\underset{z\to w}{\sim}\frac{1}{(z-w)^2}+T^u(w)+\O(z-w)
\ee
to obtain
\be
\langle\langle T^u(z)\rangle\rangle=\lim_{w\rightarrow z}\left[g(z,w)-\frac{1}{(z-w)^{2}}\right] .
\ee

To summarise, the stress-tensor method amounts to determining the Green's function $g(z,w)$, then deduce $\langle\langle T^u(z)\rangle\rangle$, and finally integrate the differential equations~(\ref{dez}).


\subsection{Ground-state twist field quantum correlators on the torus}
\label{gscf}

Let us specialize to the case where $\Sigma$ is a genus-1 surface. We will denote its \Teich parameter as $\tau^{\rm dc}$ for future use, when we see the genus-1 Riemann surface as the double cover of open-string surfaces.\footnote{Throughout \Sect{Atick}, the real part of $\tau^{\rm dc}$ is arbitrary.} 

In order to derive the quantum part of the correlator (\ref{corsigma}) for a given $u\in\{3,4\}$, $\left\langle \prod_{A=1}^{L}\sigma_A^{u}(z_A,\bar z_A)\right\rangle_{\text{qu}}$, the starting point is to write the most general ans\"atze  for $g(z,w)$ and the companion Green's function 
\be
\label{Green_aux}
h(\bar{z},w)\equiv\frac{\langle -\bar \partial \X^u_{\rm qu}(\bar z)\,\partial \overbar{\X}^u_{\rm qu}(w)\prod_A\sigma^u_A(z_A,\bar{z}_A)\rangle_{\rm qu}}{\alpha'\, \langle\prod_A\sigma^u_A(z_A,\bar{z}_A)\rangle_{\rm qu}}\, ,
\ee
satisfying the following properties: 
\begin{itemize}

\item Double periodicity $z\to z+1$, $z\to z+\taudc$ and $w\to w+1$, $w\to w+\taudc$ (and similarly for $\bar z$ in $h$). 

\item Local monodromies consistent with the OPE's given in \Eq{OPE_def}. For instance, when $z$ is transported along a tiny closed loop encircling some $z_A$, $g$ must transform as $e^{-2i\pi(1-\kappa_A/N)} g$.

\item A double pole for $g(z,w)$ as $z\to w$ dictated by \Eq{OPE_dX}, and finiteness of $h(\bar z,w)$ as $\bar z\to w$ thanks to the OPE $\bar\partial \X^u_{\rm qu}(\bar{z})\partial \overbar{\X}^u_{\rm qu}(w)\underset{\bar z\to w}{\sim} \mbox{finite}$.
\end{itemize}
This can be done by defining \emph{cut differentials} \cite{Atick} which form a basis of holomorphic one-forms on the torus that possess suitable monodromy behaviors as their arguments approach each of the insertion points  $z_A$. Denoting
\be
M=\sum_{A=1}^L {\kappa_A\over N}\, ,
\ee
which takes some value in the set $\{1,\dots, L-1\}$, and following the notations of \Reff{Atick}, such a basis is given by 
\be
\begin{aligned}
\omega_{N-\kappa}^{\alpha_A}(z)&=\gamma_{N-\kappa}(z)\,\vartheta_1(z-z_{\alpha_A}-y_{N-\kappa})\prod_{B\neq A}^{L-M}\vartheta_{1}(z-z_{\alpha_B})\,, &&\,A\in\{1,\dots,L-M\}\, ,\\
\omega_{\kappa}^{\beta_A}(z)&=\gamma_\kappa(z)\,\vartheta_1(z-z_{\beta_A}-y_\kappa)\prod_{B\neq A}^{M}\vartheta_{1}(z-z_{\beta_B})\, ,&&\,A\in\{1,\dots,M\}\, ,
\end{aligned}
\label{defcut}
\ee
where the second argument at $\taudc$  in the modular forms is implicit. In these formulas, we have defined the functions 
\be
\gamma_{N-\kappa}(z)=\prod_{A=1}^L\vartheta_1(z-z_{A})^{-(1-\kappa_{A}/N)}\, ,~~\quad\gamma_{\kappa}(z)=\prod_{A=1}^L\vartheta_1(z-z_{A})^{-\kappa_{A}/N}\, ,
\label{gfunc}
\ee
and denoted 
\be
y_{N-\kappa}=\sum_{A=1}^{L}\left(1-\frac{k_A}{N}\right)z_A-\sum_{B=1}^{L-M}z_{\alpha_B}\, ,~~\quad y_{\kappa}=\sum_{A=1}^{L}\frac{\kappa_A}{N}z_A-\sum_{B=1}^{M}z_{\beta_B}\, ,
\ee
while $\{z_{\alpha_1},\dots,z_{\alpha_{L-M}}\}$ and $\{z_{\beta_{1}},\dots,z_{\beta_M}\}$ are  subsets of twist insertion points chosen arbitrarily.\footnote{The subscripts ``$N-\kappa$'' and ``$\kappa$'' are  ``names''. They do not refer to varying indices. Moreover, the indices $\alpha_1,\dots,\alpha_{L-M}$ and $\beta_1,\dots,\beta_{M}$ here should not be confused with labels of branes also denoted by Greek letters elsewhere in our work.  } The functions $\gamma_{N-\kappa}$ and $\gamma_\kappa$ implement the monodromies around the $z_A$'s, while the extra $\vartheta_1$ modular forms in the definitions~(\ref{defcut}) lead to the double periodicity with respect to the variable $z$.\footnote{Note that no periodicity condition is imposed for the individual variables $z_A$ (which are kept implicit in the cut differentials). However, double periodicity in $z$ implies double periodicity of the whole set of points $z_A$, when they are moved together.} Given these notations, the Green's functions may be expressed as
\be
\label{Green2}
\begin{aligned}
g(z,w)&=g_{s}(z,w)+\sum_{A=1}^{L-M}\sum_{B=1}^M C_{AB}\,\omega_{N-\kappa}^{\alpha_A}(z)\,\omega_{\kappa}^{\beta_B}(w)\, ,\\
h(\bar{z},w)&=\sum_{A=1}^{M}\sum_{B=1}^M B_{AB}\,\bar{\omega}_{\kappa}^{\beta_A}(\bar{z})\,\omega_{\kappa}^{\beta_B}(w)\, ,
\end{aligned}
\ee
where $C_{AB}$ and $B_{AB}$ are ``constant coefficients.''\footnote{They depend only on the insertion points and $\taudc$.} The function $g_{s}(z,w)$ is doubly periodic in $z$ and $w$ and handles the double-pole structure of $g(z,w)$ as $z$ approaches $w$. It can be expressed as~\cite{Atick} 
\be
\label{gs}
g_{s}(z,w)=\gamma_{N-\kappa}(z)\,\gamma_{\kappa}(w)\left(\frac{\vartheta_1'(0)}{\vartheta_1(z-w)}\right)^2P(z,w)\, ,
\ee
where explicit knowledge of the function $P(z,w)$ is not required in the computation of correlation functions of ground-state twist fields. However, it does matter for correlators of excited twist fields, as will be seen in \Sect{corgen1}. We will come back to this issue at that stage.

The next step is to implement the global monodromy conditions, which by definition are ``trivial'' for the quantum fluctuations $Z_{\rm qu}^u(z,\bar z)$ (see below \Eq{splitZ}).  In practice, this implies that  
\be
0=\oint_{\gamma_a}\dd z\, g(z,w)+\oint_{\gamma_a}\dd\bar{z}\,h(\bar{z},w)\,,\quad a\in\{1,\dots,L\}\, ,
\ee
where $\{\gamma_a,\ a=1,\dots,L\}$ is a basis of the homology group of the genus-1 surface with $L$ punctures.\footnote{For a genus $g$ surface with $L$ punctures, the basis  has dimension $L+2g-2$ \cite{Atick}. } To solve these equations, it is convenient to define an $L\times L$  \emph{cut-period matrix} ${W_{a}}^A$ as follows,
\be
\label{cut-diff}
\begin{aligned}
{W_{a}}^{A}&=\oint_{\gamma_a}\dd z\, \omega_{N-\kappa}^{\alpha_A}(z)\, ,&&A\in\{1,\dots,L-M\}\,,\\
{W_{a}}^{L-M+A}&=\oint_{\gamma_a}\dd\bar{z}\,\bar{\omega}_{\kappa}^{\beta_A}(\bar{z})\, ,&&A\in\{1,\dots,M\}\,.
\end{aligned}
\ee
Indeed, it is easily checked that the expressions 
\be
\label{gtot}
\begin{aligned}
g(z,w)&=g_{s}(z,w)-\sum_{A=1}^{L-M}\omega_{N-\kappa}^{\alpha_A}(z)\sum_{a=1}^L{(W^{-1})_{A}}^{a}\oint_{\gamma_a}\dd \zeta \,g_s(\zeta,w)\,,\\
h(\bar{z},w)&=-\sum_{A=1}^{M}\omega_{\kappa}^{\beta_A}(\bar{z})\sum_{a=1}^L{(W^{-1})_{L-M+A}}^{a}\oint_{\gamma_a}\dd \zeta \,g_s(\zeta,w)\,,
\end{aligned}
\ee
satisfy the global monodromy conditions. 

Finally, the correlator can be found by applying  the stress-tensor method to find the holomorphic dependence on the $z_A$'s, and then a second time using the Green's functions $\bar{g}(\bar{z},\bar{w})$ and $\bar{h}(z,\bar{w})$ to determine the antiholomorphic part. The result is for $u\in\{3,4\}$ 
\begin{align}
\label{sigmasigma_full_qu}
\left\langle \prod_{A=1}^{L}\sigma_A^{u}(z_A,\bar z_A)\right\rangle_{\text{qu}}= &\;  f(\taudc;\kappa_1,\dots,\kappa_L)\, {1\over \det W }\;  \vartheta_1(y_{N-\kappa})^{L-M-1}\;  \overbar{\vartheta_1(y_{\kappa})}^{M-1}\nonumber \\
&\times\prod_{\substack{A,B=1\\ A<B}}^{L-M}\vartheta_1(z_{\alpha_A}-z_{\alpha_B})\,\prod_{\substack{A,B=1\\ A<B}}^M\overbar{\vartheta_1(z_{\beta_A}-z_{\beta_B})}\\
&\times\prod_{\substack{A,B=1\\ A<B}}^L\vartheta_1(z_A-z_B)^{-(1-\kappa_A/N)(1-\kappa_B/N)}\; \overbar{\vartheta_1(z_A-z_B)}^{-(\kappa_A/N)(\kappa_B/N)}\,,\nonumber 
\end{align}
where $f(\taudc,\kappa_1,\dots,\kappa_L)$ is a function arising as an ``integration constant''. The latter can be determined by coalescing all insertion points, since the left-hand side reduces in this case to $\langle 1\rangle$, which is the partition function. 


\subsection{Instanton actions}

In the OPE's~(\ref{OPE_def}), the actions of the background parts  $\partial\Xcl^u(z)$ and $\partial\overbar{\X}_{\rm cl}^u(z)$  on $\sigma^u(w,\bar w)$ for $u\in\{3,4\}$ are trivial multiplications. Hence, for the monodomy properties to be satisfied as $z$ is transported along a tiny closed loop encircling any $z_A$, the doubly-periodic $\partial\Xcl^u(z)$ and $\partial\overbar{\X}_{\rm cl}^u(z)$ must be linear sums of cut differentials.  To determine the coefficients, one imposes the global monodromy conditions 
\be
\oint_{\gamma_a}\dd z\, \partial\Xcl^u+\oint_{\gamma_a}\dd\bar{z}\, \bar \partial\Xcl^u=v^u_{a}\,,\quad a\in\{1,\dots,L\}\, ,
\label{displa}
\ee
where the $v^u_{a}$'s are displacement vectors. The solution of these equations can be expressed in terms of the inverse cut-period matrix,
\be
\partial\Xcl^u(z)=\omega_{A'}(z)\, {(W^{-1})_{A'}}^{a}\, v^u_a\, ,\qquad\bar\partial\Xcl^u(\bar{z})=\bar \omega_{A''}(\bar{z})\, {(W^{-1})_{A''}}^{a}\, v^u_a\, ,
\label{Zc}
\ee
where in the present context the index $A'$ is summed over $1,\dots,L-M$,  and $A''$ is summed over $L-M+1,\dots,L$. Moreover, we have redefined in the above formulas 
\be
\begin{aligned}
\omega_{A}(z)&\equiv\omega_{N-\kappa}^{\alpha_A}(z)\, ,&& \, A\in\{1,\dots,L-M\}\, ,\\
\omega_{L-M+A}(z)&\equiv\omega_{\kappa}^{\beta_A}(z)\, ,&& \, A\in\{1,\dots,M\}\, .\\
\end{aligned}
\ee

With the definition of the Hermitian product
\be
(\omega_i,\omega_j)\equiv i\int_{\Sigma}\dd z\wedge \dd\bar z\, \omega_i(z)\, \bar{\omega}_j(\bar{z})
\ee
of one-forms on the torus, the classical action for a single complex coordinate $u\in\{3,4\}$ reads 
\be
\label{Scl_torus}
\Scl^\Sigma\big|^u=\frac{v^u_a\bar v^u_b}{2\pi\alpha'}\Big[{(W^{-1})_{A'}}^{a}\, {(W^{-1})_{B'}}^{b}\, (\omega_{A'},\omega_{B'})+{(W^{-1})_{A''}}^{a}\, {(W^{-1})_{B''}}^{b}\, (\omega_{A''},\omega_{B''})\Big]\, .
\ee


\subsection{Useful correlators on the torus}
\label{corgen1}

In this subsection, we consider all correlators involved in the open-string amplitudes of \Eqs{prelim_ext} and~(\ref{prelim_int}), but display their values computed on a genus-1 surface. In this case, they have holomorphic and antiholomorphic dependencies. 


\paragraph{\em Correlator \bm $\langle \sigma^u(z_{1},\bar z_1)\sigma^u(z_{2},\bar z_2)\rangle_{\rm qu}$: } 

For the OPE's of the twist fields to match those of the boundary-changing fields we are interested in, we now consider the case where 
\be
N=2\, , ~~\quad L=2\, , ~~\quad {\kappa_1\over N}={\kappa_2\over N}=\half\, , ~~\quad M=1\, .
\ee 
Because $\kappa_1=\kappa_2$, we can omit from now on the subscripts $A$ of the twist fields.   Using \Eq{sigmasigma_full_qu}, we obtain for  $u=3,4$ 
\be
\langle \sigma^u(z_{1},\bar z_1)\sigma^u(z_{2},\bar z_2)\rangle_{\text{qu}}=f(\taudc;\mbox{$1\over2$},\mbox{$1\over2$})\, (\det W)^{-1}\, \thetau(z_{1}-z_{2})^{-\frac{1}{4}}\; \overbar{\thetau(z_{1}-z_{2})}^{-\frac{1}{4}}\, .
\ee

The $2\times 2$ cut-period matrix ${W_ a}^i$ defined in Eq~(\ref{cut-diff}) involves only one cut differential,
\be
\omega(z)=\vartheta_{1}(z-z_{1})^{-\half}\, \vartheta_{1}(z-z_{2})^{-\half}\,\vartheta_{1}\big(z-\frac{z_{1}+z_{2}}{2}\big)\,,
\ee 
to be integrated on the cycles of the genus-1 surface $\Sigma$,  $\gamma_{1}:~ z\rightarrow z+1$ and $\gamma_{2}:~ z\rightarrow z+\taudc$, which yields 
\be
W=
\begin{pmatrix}
W_1 & \overbar{W}_1\\
W_2 & \overbar{W}_2
\end{pmatrix},\quad \where \quad W_{a}=\oint_{\gamma_a}\dd z\, \omega\,,\quad a\in\{1,2\}\,.
\ee
In these notations, the background action written in Eq.~(\ref{Scl_torus}) reads for $u=\in\{3,4\}$ 
\be
\Scl^\Sigma\big|^u={1\over 4\pi\alpha'\,{\text{Im}}(\overbar W_{1}W_{2})}\left(\big|\overbar{W}_{2}v_{1}^u-\overbar{W}_{1}v_{2}^u\big|^{2}+\big|W_{2}v_{1}^u-W_{1}v_{2}^u\big|^{2}\right) ,
\label{scla}
\ee
where $v^u_a$, $a\in\{1,2\}$,  are the displacements introduced in \Eq{displa}. In our case of interest, given an instanton solution, the real-coordinate background $X^I_{\rm cl}(z,\bar z)$, $I\in\{6,\dots,9\}$, winds $n_I$ times and $l_I$ times the circle $S^1(R_I)$ as  $z$ is transported along $\gamma_1$ and $\gamma_2$, so that 
\be
\label{vs}
\begin{aligned}
v_{1}^{u}=\frac{2\pi R_{2u}n_{2u}+2i\pi R_{2u+1}n_{2u+1}}{\sqrt{2}}\, ,~~\quad v_{2}^{u}=\frac{2\pi R_{2u}l_{2u}+2i\pi R_{2u+1}l_{2u+1}}{\sqrt{2}}\, .
\end{aligned}
\ee
For the $T^2$ coordinate $u=2$, which is not twisted, the above formula apply with cut differentials that induce trivial local monodromies. In other words, replacing $\omega(z)$ by 1, the relevant  cut-period matrix becomes 
\be
\begin{pmatrix}
1 & 1\\
\taudc & \bar\tau^{\rm dc}
\end{pmatrix}.
\ee
Defining displacements $v^u_a$ for $u=2$ exactly as those given in \Eq{vs}, 
one obtains
\be
\Scl^\Sigma\big|^2={\pi\over \alpha' \,\Im \taudc}\big(R_4^2|n_4\taudc-l_4|^2+R_5^2|n_5\taudc-l_5|^2\big)\,,
\label{scla2}
\ee
which is the well know result for the instanton action on a two-torus \cite{Kiritsis_book}. 
The sum over instantons appearing in Eq.~(\ref{corsigma}) translates therefore into a sum over winding numbers $n_{I'}$, $n_I$, and wrapping numbers $l_{I'}$, $l_I$, where $I'\in\{4,5\}$ and $I\in\{6,\dots,9\}$.


\paragraph{\em Correlator \bm$\langle \tau^u(z_{1},\bar z_1)\tau^{\prime u}(z_{2},\bar z_2)\rangle_{\rm qu}$: } 

To derive the correlator of excited twist fields, we will follow the technique described in \Reffs{Abel:2004ue,markG1,markG2,markG3}. Thanks to the OPE's between $\partial Z^u, \partial \overbar{Z}^u$ and the ground-state twist fields given in \Eq{OPE_def}, and using the splitting defined Eq.~(\ref{splitZ}), we may divide accordingly this correlation function for $u\in\{3,4\}$ into two pieces,~ 
\be
\langle \tau^u(z_{1},\bar z_1)\tau^{\prime u}(z_{2},\bar z_2)\rangle_{\text{qu}}=\langle \tau^u(z_{1},\bar z_1)\tau^{\prime u}(z_{2},\bar z_2)\rangle_{\text{qu}}^{(1)}+\langle \tau^u(z_{1},\bar z_1)\tau^{\prime u}(z_{2},\bar z_2)\rangle_{\text{qu}}^{(2)}\, ,
\ee
where we have defined 
\begin{align}
\begin{split}
\langle \tau^u(z_{1},\bar z_1)\tau^{\prime u}(z_{2},\bar z_2)\rangle_{\text{qu}}^{(1)}&=\langle \sigma^u(z_{1},\bar z_1)\sigma^u(z_{2},\bar z_2)\rangle_{\rm qu}\lim_{\substack{z\rightarrow z_{1}\\ w\rightarrow z_{2}}}\!\left[(z-\zu)^{\half}(w-\zd)^{\half}\,\partial \Xcl^u(z)\partial\Xclbar^u(w)\right] ,\\
\langle \tau^u(z_{1},\bar z_1)\tau^{\prime u}(z_{2},\bar z_2)\, \rangle_{\text{qu}}^{(2)}&=\lim_{\substack{z\rightarrow z_{1}\\ w\rightarrow z_{2}}}\!\left[(z-\zu)^{\half}(w-\zd)^{\half}\langle \partial \Xqu^u(z)\partial \Xqubar^u(w)\sigma^u(z_{1},\bar z_1)\sigma^u(z_{2},\bar z_2)\rangle_{\rm qu}\right] .
\end{split}
\label{limites12}
\end{align}

To derive part $(1)$ of the correlator, we use Eq.~(\ref{Zc}) which becomes 
\be
\partial Z_{\rm cl}^u(z)=\omega(z)\,c^u_1\, ,\quad\bar \partial Z_{\rm cl}^u(\bar z)=\bar \omega(\bar z)\,c^u_2\, ,\quad \where\quad c_A^u  ={(W^{-1})_A}^a \,v^u_a\, .
\ee
Remember that due to their local monodromy behaviors, these expressions diverge at the insertion points. Hence, the limits defined in Eq.~(\ref{limites12}) contribute a finite result which is~
\be
\langle \tau^u(z_{1},\bar z_1)\tau^{\prime u}(z_{2},\bar z_2)\rangle_{\text{qu}}^{(1)}=s\,i\,c^u_{1}\bar{c}^u_{2}\,\frac{\thetau(\frac{\zu-\zd}{2})^{2}}{\thetau'(0)\, \thetau(\zu-\zd)}\, \langle \sigma^u(z_{1},\bar z_1)\sigma^u(z_{2},\bar z_2)\rangle_{\rm qu}\, ,
\label{part1}
\ee
where we denote 
\be
s\,i~\equiv~\left({z_2-z_1\over z_1-z_2}\right)^\half .
\ee

Using the Green's function $g(z,w)$ defined in Eq.~(\ref{Green}), part $(2)$ of the correlator can be expressed as
\be
\label{tautauprimelim}
\langle \tau^u(z_{1},\bar z_1)\tau^{\prime u}(z_{2},\bar z_2)\, \rangle_{\text{qu}}^{(2)}=-\alpha'\,\langle \sigma^u(z_{1},\bar z_1)\sigma^u(z_{2},\bar z_2)\rangle_{\rm qu} \lim_{\substack{z\rightarrow z_{1}\\ w\rightarrow z_{2}}}\!\big[(z-\zu)^{\half}(w-\zd)^{\half}\,g(z,w)\big]\,.
\ee
In the present case, \Eqs{Green2} and~(\ref{gtot}) become 
\be
\label{2de}
\begin{aligned}
g(z,w)&=g_s (z,w)+C\,\omega(z)\,\omega(w)\\
&=g_{s}(z,w)-\omega(z){(W^{-1})_{1}}^{a}\oint_{\gamma_a}\dd \zeta \,g_s(\zeta,w)\,,
\end{aligned}
\ee
where $g_s (z,w)$ is defined in Eq.~(\ref{gs}). The latter involves a function $P(z,w)$ derived in Ref.~\cite{Atick}, and whose expression is given by
\be
\label{gs2}
\begin{aligned}
g_s (z,w)=\gamma(z)\gamma(w)\left(\frac{\thetau'(0)}{\thetau(z-w)}\right)^{2}\half\,\Big[&F_{1}(z,w)\thetau(w-\zu)\thetau(z-\zd)\\
+&F_{2}(z,w)\thetau(w-\zd)\thetau(z-\zu)\Big]\,.
\end{aligned}
\ee
The right-hand side  is written in terms of  a unique function $\gamma$ (see Eq.~(\ref{gfunc}))
\be
\gamma(z)=\thetau(z-\zu)^{-\half}\,\thetau(z-\zd)^{-\half}\, ,
\ee
as well as 
\be
\label{FA}
\begin{aligned}
&F_{A}(z,w)=\frac{\thetau(z-w+U_{A})}{\thetau(U_{A})}\, \frac{\thetau(z-w+Y_{A}-U_{A})}{\thetau(Y_{A}-U_{A})}\, ,\quad A\in\{1,2\}\, ,\\
\where\quad& Y_{A}=\frac{\zu+\zd}{2}-z_{A}\quad\and\quad  U_{A}\text{ is such that } \partial_{z}F_{A}(z,w)\big|_{z=w}=0\, .
\end{aligned}
\ee
Notice that in the above formula, we adopt the notations of Ref.~\cite{Atick} but it turns out that $F_1(z,w)\equiv F_2(w,z)$.
Computing the limits in \Eq{tautauprimelim}, we find
\be
\begin{aligned}
\langle \tau^u(z_{1},\bar z_1)\tau^{\prime u}(z_{2},\bar z_2)\, \rangle_{\text{qu}}^{(2)}=&-\alpha' s\,i\,\langle \sigma^u(z_{1},\bar z_1)\sigma^u(z_{2},\bar z_2)\rangle_{\rm qu}\\
&\times \Bigg[\frac{\thetau'(0)\,F_{1}(\zu,\zd)}{2\,\thetau(\zu-\zd)}+C\,\frac{\thetau(\frac{\zu-\zd}{2})^{2}}{\thetau'(0)\,\thetau(\zu-\zd)}\Bigg]\,.
\end{aligned}
\ee
Moreover, using in the derivation the second expression in \Eq{2de}, an explicit expression for the term linear in $C$ is obtained,
\be
C\,\frac{\thetau(\frac{\zu-\zd}{2})^{2}}{\thetau'(0)\,\thetau(\zu-\zd)} = -\half \, \vartheta_1'(0)\, \vartheta_1\big({z_1-z_2\over 2}\big) {(W^{-1})_{1}}^{a}\oint_{\gamma_a}\dd z \,{F_1(z,z_2)\over \vartheta_1(z-z_1)^\half \, \vartheta_1(z-z_2)^{3\over 2}}\,.
\label{Cexp}
\ee

Adding the pieces (1) and (2) of the correlator, we obtain for $u\in\{2,3\}$ 
\be
\begin{aligned}
\langle \tau^u(z_{1},\bar z_1)\tau^{\prime u}(z_{2},\bar z_2)\, \rangle_{\text{qu}}=&-s\,i\, \langle \sigma^u(z_{1},\bar z_1)\sigma^u(z_{2},\bar z_2)\rangle_{\rm qu}\\
&\times\! \Bigg[\left(\alpha'C-c^u_{1}\bar{c}^u_{2}\right)\frac{\thetau(\frac{z_1-z_2}{2})^{2}}{\thetau'(0)\,\thetau(z_1-z_2)}+\alpha'\frac{\thetau'(0)\, F_{1}(\zu,\zd)}{2\, \thetau(z_1-z_2)}\Bigg]\, .
\end{aligned}
\ee


\paragraph{\em Correlator \bm$\langle \tau^{\prime u}(z_{1},\bar z_1)\tau^{u}(z_{2},\bar z_2)\rangle_{\rm qu}$: } 

Proceeding the same way, and using the fact that $F_2(z_2,z_1)=F_1(z_1,z_2)$, we obtain the identity
\be
\langle \tau^{\prime u}(z_{1},\bar z_1)\tau^{u}(z_{2},\bar z_2)\, \rangle_{\text{qu}}=\langle \tau^u(z_{1},\bar z_1)\tau^{\prime u}(z_{2},\bar z_2)\, \rangle_{\text{qu}}\, .
\ee


\paragraph{\em Bosonic correlator: }

The propagator of the spacetime coordinates $X^\mu$ is given by
\be
\langle X^\mu(\zu,\bar z_1)X_\nu(\zd,\bar z_2)\rangle=\delta^{\mu}_{\nu}\bigg[-\frac{\alpha'}{2}\ln\left|\frac{\thetau(\zu-\zd)}{\thetau'(0)}\right|^{2}+\frac{\alpha'\pi\,[\Im\!(\zu-\zd)]^{2}}{\Im\tau}\bigg]\,,
\ee
which leads to 
\be
\langle e^{ik\cdot X}(\zu,\bar z_1)e^{-ik\cdot X}(\zd,\bar z_2)\rangle=\left(\left|\frac{\vartheta_1(z_1-z_2)}{\thetau'(0)}\right|e^{-\frac{\pi[\text{Im}(z_1-z_2)]^{2}}{\Im \tau}}\right)^{-\alpha'k^{2}} \,.
\ee


\paragraph{\em Bosonized-fermion correlators: } 

Each complex fermion $\Psi^u$, $u\in\{0,\dots,4\}$, has one out of four pairs of periodic/antiperiodic boundary conditions on the genus-1 surface $\Sigma$, which corresponds to a spin structure $\nu\in\{1,2,3,4\}$. In bosonized picture, for any $H_u$-charge $q$, one finds by applying the stress-tensor method that the following correlator depends accordingly on~$\nu$, 
\be
\langle e^{iqH_{u}}(\zu)e^{-iqH_{u}}(\zd)\rangle_{\nu}=K_{\nu,|q|}\, \thetan(q(z_1-z_2))\, \vartheta_1(z_1-z_2)^{-q^{2}}\, ,
\label{corfer}
\ee
where  $K_{\nu,|q|}$ is a $\taudc$-dependent normalization factor~\cite{Abel:2004ue}. 


\section{Full amplitudes of massless ND and DN states}
\label{NDmasses}

We are now ready to use all ingredients introduced in \Sects{ndmass} and~\ref{Atick} to compute the two-point functions of massless bosonic states in the ND and DN sectors. As depicted in \Fig{annulus+mobius_diagrams}, the annulus and M\"obius strip can be described as tori with \Teich parameters given in \Eq{temo} and modded by the involution~(\ref{invo}). The boundaries of the open-string surfaces being the fixed points, we may choose the insertion points of the boundary-changing vertex operators to be 
\be
\label{points}
z_A\equiv x_A+iy_A\, ,\quad \where\quad \left\{\begin{array}{ll}
0\leq y_A\leq \Im \taudc\, ,&A\in\{1,2\}\, ,\\
x_1=x_2\in\dis\left\{0,\half\right\} &\for ~~\Sigma=\A\, , \esp\\
x_1, x_2\in\dis\left\{0,\half\right\} &\for~~\Sigma=\M \, .\esp
\end{array}
\right.
\ee


\subsection{Useful correlators on the annulus and M\"obius strip }

Let us first collect the correlators presented in the previous section now evaluated on the open-string surfaces $\Sigma=\A$ and $\M$, and to be used to express the amplitudes $A_{\text{ext}\Sigma}^{\alpha_0\beta_0}$ and $A_{\text{int}\Sigma}^{\alpha_0\beta_0}$. 


\paragraph{\em Correlator \bm$\langle \sigma^u(z_{1})\sigma^{ u}(z_{2})\rangle_{\rm qu}$: } 

In \Reff{Abel:2004ue}, the method of images was applied on the Green's functions of \Sects{Stress-tensor-method} and~\ref{gscf} to define their open-string counterparts. The latter were used to derive the correlator between two ground-state boundary-changing fields by using the stress-tensor method.  The result amounts essentially to take the ``square root''  of the closed-string result, \ie for $u\in\{3,4\}$, 
\be
\langle \sigma^u(z_{1})\sigma^u(z_{2})\rangle_{\text{qu}}=f_{\rm op}(\taudc;\mbox{$1\over2$},\mbox{$1\over2$})\,(\det W)^{-\half}\,  \thetau(z_{1}-z_{2})^{-\frac{1}{4}}\, ,
\ee
where $f_{\rm op}$ is a normalization function. 
Notice that the product $\langle \sigma^3\sigma^3\rangle_{\text{qu}}\, \langle \sigma^4\sigma^4\rangle_{\text{qu}}$ involves $\vartheta_1(z_1-z_2)^{-\half}$, which is well defined up to a sign.  We will see  in the next subsection how such ambiguities can be lifted.

The instanton actions can also be derived from the closed-string result given in \Eqs{scla} and~(\ref{scla2}). These expressions must be divided by 2, the order of the involution $\I$,  to account for the fact that the open-string worldsheets are halves of their genus-1 double-covers. Moreover, we have to consider instantonic worldsheets with NN,  DD, ND or DN boundary conditions for $\A$, and  N or D boundary conditions for $\M$. In the NN and N case, all winding numbers along $T^2\times T^4$ must vanish, $n_{I'}=n_I=0$. The DD and D case is similar, up to the T-duality transformation $R_I\to \alpha'/R_I$. Denoting the T-dual wrapping numbers with ``tildes'', we  have $\tilde n_{I'}=\tilde n_I=0$. Finally, for worldsheets with ND or DN boundary conditions in the annulus case, non-trivial instantons wrap $T^2$ only, \ie satisfy $n_{I'}=n_I=l_I=0$ or $\tilde n_{I'}=\tilde n_I=\tilde l_I=0$. In total, we thus have for $\Sigma=\A$ or $\M$
\be
S^\Sigma_{\text{cl}}={\pi[(R_4l_4)^2+(R_5l_5)^2]\over \alpha' \tau_2}+{|W_1|^2\over 4\pi\alpha'\,{\text{Im}}(\overbar W_{1}W_{2})}\times \left\{ \begin{array}{ll}{\dis\sum_{u=3}^4 |v_{2}^u|^{2}} & \mbox{for NN and N ,}\\
{\dis\sum_{u=3}^4 |\tilde v_{2}^u|^{2}}& \mbox{for DD and D ,}\\
0 & \mbox{for ND and DN ,}\esps\end{array}
\right.
\ee
where the displacements and their T-dual counterparts are given by 
\be
 v_{2}^{u}=\frac{2\pi R_{2u}l_{2u}+2i\pi R_{2u+1}l_{2u+1}}{\sqrt{2}}\, ,\quad  \tilde v_{2}^{u}=\frac{2\pi { \alpha'\over R_{2u}}\, \tilde l_{2u}+2i\pi {\alpha'\over R_{2u+1}}\,\tilde l_{2u+1}}{\sqrt{2}}\, ,\quad u\in\{3,4\}\, .
 \ee
 

\paragraph{\em Correlators \bm$\langle \tau^u(z_{1})\tau^{\prime u}(z_{2})\rangle_{\rm qu}$ and \bm$\langle \tau^{\prime u}(z_{1})\tau^{ u}(z_{2})\rangle_{\rm qu}$: }  

On the genus-1 surfaces, the twist fields $\tau^u(z,\bar z)$ and $\tau^{\prime u}(z,\bar z)$ are excited only on their holomorphic sides (see \Eq{OPE_def}). Therefore, their correlation functions take formally the same forms as those of the excited boundary-changing fields $\tau^u(z)$ and $\tau^{\prime u}(z)$ evaluated on $\A$ and $\M$.  There is however a subtlety concerning part (1) of the correlator, \Eq{part1}.  

When considering the \emph{full} amplitudes \ie with the instanton dressings $e^{-S_{\rm cl}^\Sigma}$, the open-string actions are divided by 2 compared to the closed-string case. Hence, for the full open-string correlators to preserve their interpretations on double-cover tori, one should rescale the displacements  in parts (1) as follows, $|v_2^u|^2\to |v_2^u|^2/2$ and  $|\tilde v_2^u|^2\to |\tilde v_2^u|^2/2$. As a result we have for $u\in\{3,4\}$
\be
\begin{aligned}
\langle \tau^u(z_{1})&\tau^{\prime u}(z_{2}) \rangle_{\text{qu}}=\langle \tau^{\prime u}(z_{1})\tau^{u}(z_{2}) \rangle_{\text{qu}}=-s\,i\, \langle \sigma^u(z_{1})\sigma^u(z_{2})\rangle_{\rm qu}~\times \\
&~~~~~~~~~~~~~~~\Bigg[\bigg(\alpha'C+ {\overbar{W}_1^2\, |v_2^u|^2\over 8[\Im (\overbar{W}_1 W_2)]^2}\bigg)\frac{\thetau(\frac{z_1-z_2}{2})^{2}}{\thetau'(0)\,\thetau(z_1-z_2)}+\alpha'\frac{\thetau'(0)\, F_{1}(\zu,\zd)}{2\, \thetau(z_1-z_2)}\Bigg]\, ,\esp
\end{aligned}
\ee
when the worldsheet has NN or N boundary conditions on the annulus or M\"obius strip. For DD or D boundary conditions, the correlators take identical forms up to the change $v_2^u\to \tilde v_2^u$. Finally, for boundary conditions ND or DN on the annulus, the classical displacements vanish and only the pure quantum contributions proportional to $\alpha'$ survive. 

Note that $\langle \tau^3\tau^{\prime 3}\rangle_{\text{qu}}\,  \langle \sigma^4\sigma^4\rangle_{\text{qu}}$ and $\langle \sigma^3\sigma^3\rangle_{\text{qu}}\,  \langle \tau^{\prime 4}\tau^{4}\rangle_{\text{qu}}$ contain factors $\vartheta_1(z_1-z_2)^{-\half}$, which yield signs ambiguities.


\paragraph{\em Bosonic correlator: }  

The spacetime-coordinate propagators on the annulus and M\"obius strip can be expressed in terms of those on the double-cover tori by symmetrizing with respect to the involution. The result is  
\be
\langle X^\mu(\zu)X_\nu(\zd)\rangle=\delta^{\mu}_{\nu}\bigg[-\alpha'\,\ln\left|\frac{\thetau\left(\zu-\zd\right)}{\thetau'(0)}\right|^{2}+\frac{\alpha'4\pi\,[\text{Im}(\zu-\zd)]^2}{\tau_{2}}\bigg]\, ,
\ee
which can be used to derive 
\be
\langle e^{ik\cdot X}(\zu)e^{-ik\cdot X}(\zd)\rangle=\left(\left|\frac{\thetau(\zu-\zd)}{\thetau'(0)}\right|e^{-\frac{2\pi[\text{Im}(z_1-z_2)]^{2}}{\tau_{2}}}\right)^{-2\alpha'k^{2}}\, .
\ee


\paragraph{\em Bosonized-fermion correlators: } 

In \Reff{Abel:2004ue}, it is shown by applying the method of images on the Green's functions used in the stress-tensor method that the correlators of bosonized-fermion  on the open-string worldsheets are identical to those on the double-cover tori. They are given  in \Eq{corfer}. For $q=\pm \half$, products of two such correlators are well defined up to signs. 


\subsection{Full expressions of the amplitudes}

Putting everything together and defining $\zud=\zu-\zd$ to lighten notations, the full expression~(\ref{prelim_ext})  of the external part of the one-loop two-point function of massless bosonic states in the ND and DN sectors is, for $\Sigma\in\{\A,\M\}$, 
\be
\label{Aext}
\begin{aligned}
A_{\text{ext}\Sigma}^{\alpha_0\beta_0}=&\; \alpha'k^2\, \lambda_{\alpha_0\beta_0}\lambda^{\rm T}_{\beta_0\alpha_0}\left[\left|\frac{\thetauzud}{\thetau'(0)}\right|e^{-\frac{2\pi}{\tau_2}[\text{Im}(\zud)]^{2}}    \right]^{-2\alpha'k^{2}}\, {1\over \det W\, \thetauzud^{2}} \\
& \times\sum_{\nue\neq 1}  K_{\nue,1}\, \vartheta_{\nue}(z_{12})\,\sum_{\nui}(-1)^{\delta_{\nui,1}}\, \vartheta_{\nui}\big({\zud\over 2}\big)^2\\
&\times\sum_{\vec l'}e^{-{\pi\over \alpha'\tau_2}\sum_{I'}(R_{I'}l_{I'})^2}\Bigg[\sum_{\vec{l}}e^{-\frac{|W_{1}|^{2}(|{v}_{2}^{3}|^{2}+|{v}_{2}^{4}|^{2})}{4\pi\alpha'\Im(\overbar{W}_1W_2)}}\C_{\nui}^{\Sigma\vec l'\vec l}+\sum_{\vec \tl}e^{-\frac{|W_{1}|^{2}(|\tilde{v}_{2}^{3}|^{2}+|\tilde{v}_{2}^{4}|^{2})}{4\pi\alpha'\Im(\overbar{W}_1W_2)}}\tilde \C_{\nui}^{\Sigma\vec l'\vec \tl}\Bigg]\,,
\end{aligned}
\ee
which is independent of the choice of $\epsilon\in\{-1,+1\}$. 
In this expression, we use the following notations:
\begin{itemize}
\item $\vec l'$ stands for $(l_4,l_5)$, and $\vec l$, $\vec \tl$ are the four-vectors whose  components are $l_I$ and $\tl_I$. 

\item $\nue, \nui\in\{1,2,3,4\}$ denote the spin structures of the worldsheet complex fermions: The former for  $\Psi^0,\Psi^1$ (and $\Psi^2$), and the latter for $\Psi^3,\Psi^4$. 

\item The normalization factor of the external fermion correlators is given by~\cite{Kiritsis_book} 
\be
K_{\nue,1}={\vartheta_1'(0)\over \vartheta_{\nue}(0)}\, ,\quad  \nue\in\{2,3,4\}\, ,\quad \where \quad \vartheta_1'(0)=- 2\pi \eta^3\, .
\ee

\item $\C_{\nui}^{\Sigma\vec l'\vec l}$ and $\tilde \C_{\nui}^{\Sigma\vec l'\vec \tl}$ are normalization functions to be determined. They stand for products of the form  
\be
K_{\nui,\half}(\taudc)^2 \, f_{\rm op}(\taudc;\mbox{$1\over2$},\mbox{$1\over2$})\, f_{\rm op}(\taudc;\mbox{$1\over2$},\mbox{$1\over2$})\, ,
\ee
possibly dressed by signs that may depend on the instanton numbers $\vec l,\vec l'$ or $\vec \tl,\vec l'$. Indeed, as stressed before, pairs of correlators of twist fields as well as pairs of correlators of spin fields yield signs ambiguities. Moreover, for the amplitude computed on the annulus, the normalization functions should contain sums over the free boundary condition denoted $\gamma$. Furthermore,  $\C_{\nui}^{\A\vec l'\vec 0}$ takes into account two contributions associated with the NN and ND worldsheet boundary conditions, while  $\tilde \C_{\nui}^{\A\vec l'\vec 0}$ describes those arising from DD and DN boundary conditions. 
\end{itemize}

Similarly, the internal piece~(\ref{prelim_int}) of the amplitude is independent of $\epsilon$ and can be expressed in terms of the same normalization functions, 
\be
\label{Aint}
\begin{aligned}
A_{\text{int}\Sigma}^{\alpha_0\beta_0}=&-\frac{s\,i}{\alpha'}\, \lambda_{\alpha_0\beta_0}\lambda^{\rm T}_{\beta_0\alpha_0}\left[\left|\frac{\thetauzud}{\thetau'(0)}\right|e^{-\frac{2\pi}{\tau_2}[\text{Im}(\zud)]^{2}}    \right]^{-2\alpha'k^{2}}\, {\vartheta_1({z_{12}\over 2})^2\over \det W\, \thetauzud^{2}\,  \vartheta_1'(0)} \\
& \times 4\,\sum_{\nui} \vartheta_{\nui}\big({\zud\over 2}\big)^2\, \sum_{\vec l'}e^{-{\pi\over \alpha'\tau_2}\sum_{I'}(R_{I'}l_{I'})^2}\esps \\
&\times\Bigg\{\sum_{\vec l}e^{-\frac{|W_{1}|^{2}(|{v}_{2}^{3}|^{2}+|{v}_{2}^{4}|^{2})}{4\pi\alpha'\Im(\overbar{W}_1W_2)}}\C_{\nui}^{\Sigma\vec l'\vec l}\Bigg[ {\overbar{W}_1^2\, (|v_2^3|^2+|v_2^4|^2)\over 8[\Im (\overbar{W}_1 W_2)]^2}+2\alpha'(C+\hat C)\Bigg]\\
&\,\;\;\;+\sum_{\vec \tl}e^{-\frac{|W_{1}|^{2}(|{\tilde v}_{2}^{3}|^{2}+|{\tilde v}_{2}^{4}|^{2})}{4\pi\alpha'\Im(\overbar{W}_1W_2)}}\tilde \C_{\nui}^{\Sigma\vec l'\vec \tl}\Bigg[{\overbar{W}_1^2\, (|\tilde v_2^3|^2+|\tilde v_2^4|^2)\over 8[\Im (\overbar{W}_1 W_2)]^2}+2\alpha'(C+\hat C)\Bigg]\Bigg\}\,,
\end{aligned}
\ee
where the factor 4 accounts for the trivial sum over the external-fermion spin structure $\nue$, $C$ is given in \Eq{Cexp},  and we have defined
\be
\hat C\equiv {\thetau'(0)^2\over 2\,\vartheta_1({z_{12}\over 2})^2}\, F_{1}(\zu,\zd)\, .
\label{Ch}
\ee
In the following, we will not consider anymore in \Eqs{Aext} and~(\ref{Aint}) the irrelevant contributions of the external bosonic correlators, $[\,\cdots]^{-2\alpha' k^2}$, which are equal to 1 on shell. We stress again that we could have introduced non-trivial \KK momenta along $T^2$ to avoid ambiguities in extracting information from the amplitude $A_{{\rm ext}\Sigma}^{\alpha_0\beta_0}$. 

\begin{figure}[H]
\begin{center}
\raisebox{-0.5\height}{\includegraphics[scale=0.60]{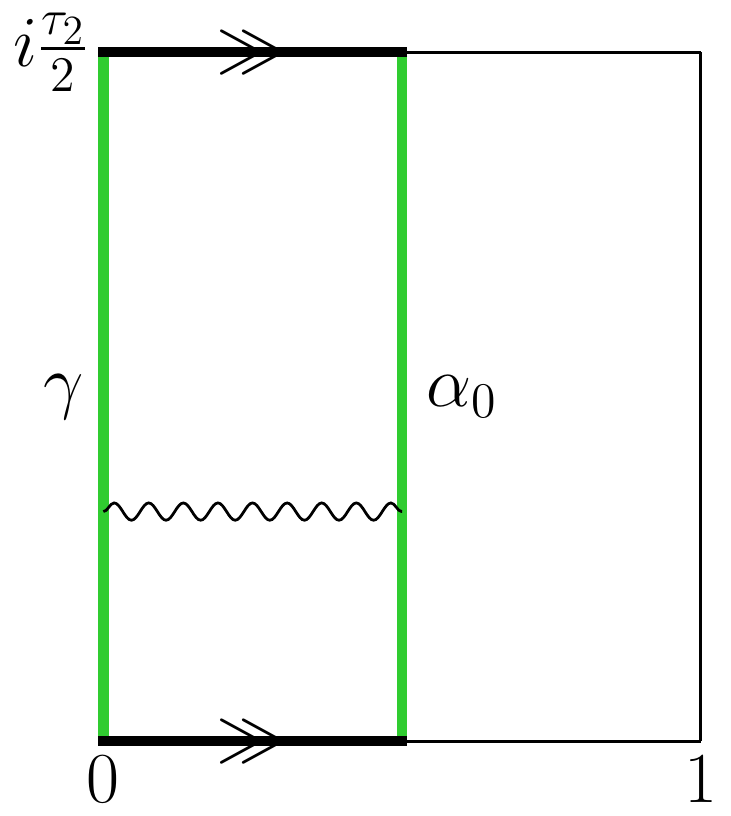}}
\qquad\qquad~
\raisebox{-0.5\height}{\includegraphics[scale=0.60]{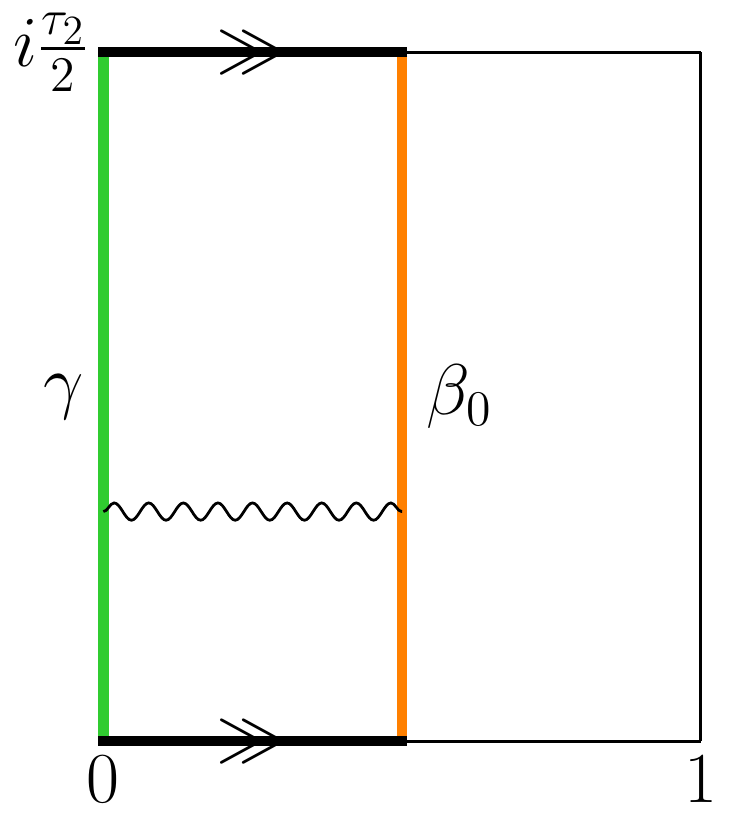}}\\[0.5cm]

\raisebox{-0.5\height}{\includegraphics[scale=0.60]{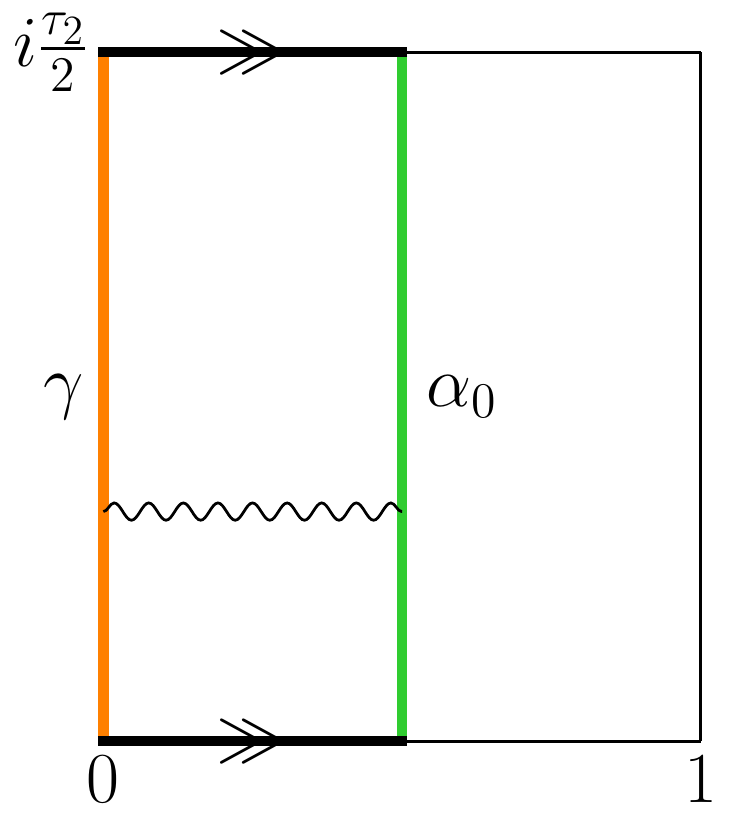}}
\qquad\qquad~
\raisebox{-0.5\height}{\includegraphics[scale=0.60]{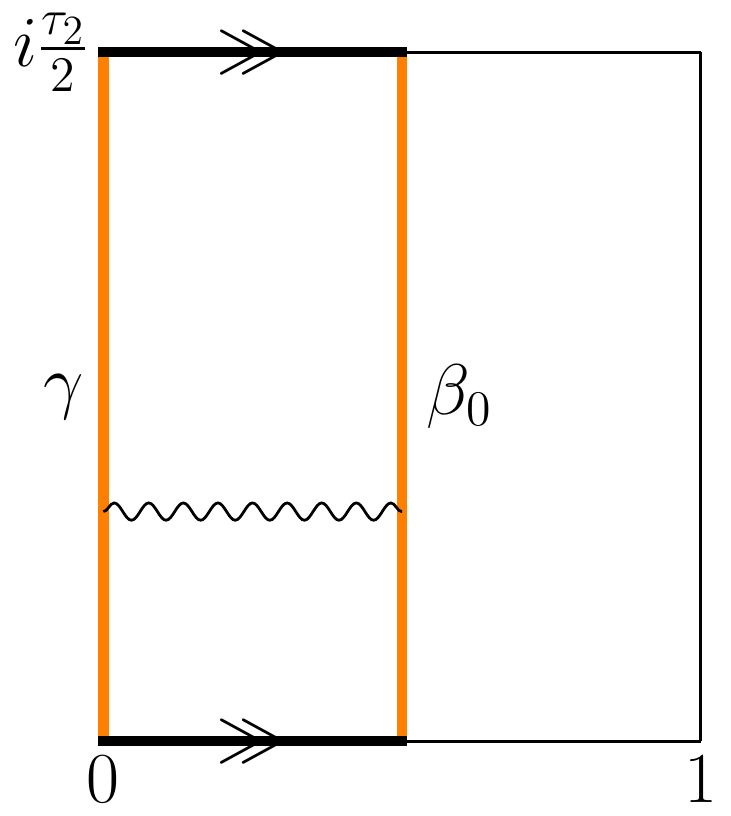}}\\[0.5cm]

\!\!\!\!\!\!\!\!\!\!\!\!\!\!\!\!\!\!\!\!\!\!\raisebox{-0.5\height}{\includegraphics[scale=0.60]{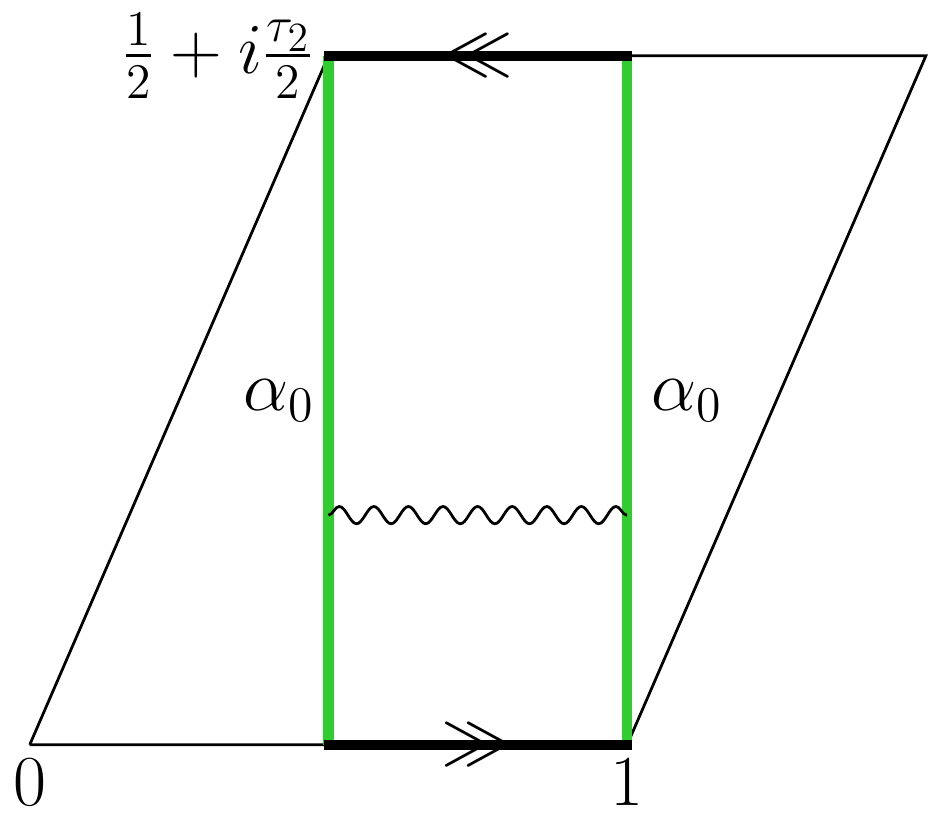}}
\quad\;\,
\raisebox{-0.5\height}{\includegraphics[scale=0.60]{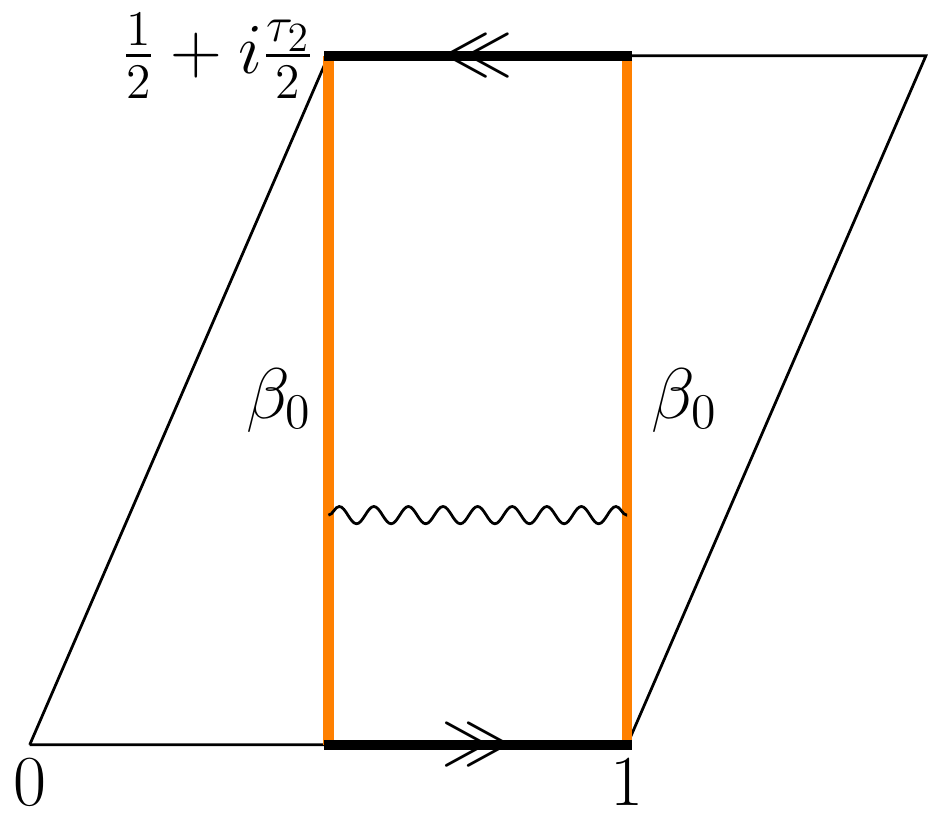}}
\end{center}
\caption{\footnotesize Open-string diagrams of \Fig{annulus+mobius_diagrams} in the limit $z_{12}\to 0$. }
\label{annulus+mobius_diagrams_limite}
\end{figure}

\paragraph{\em \bm Normalization functions $\C_{\nui}^{\Sigma\vec l'\vec l}$ and $\tilde \C_{\nui}^{\Sigma\vec l'\vec \tl}$: } 
As said in the remark below \Eq{sigmasigma_full_qu}, $\C^{\Sigma\vec l'\vec l}_{\nui}$ and $\tilde \C^{\Sigma\vec l'\vec \tl}_{\nui}$ may be determined by using the fact that when $z_1$ and $z_2$ coalesce, the effects of the ground-state boundary-changing operators compensate each other. Hence, the external part of the amplitude reduces, up to a multiplicative factors, to selected pieces of the open-string contributions to the partition function.  To identify precisely which pieces are relevant, \Fig{annulus+mobius_diagrams_limite} shows what the diagrams in \Fig{annulus+mobius_diagrams} become when $z_{12}\to 0$. In this limit, the cut differential associated with either of the complex directions $u\in\{3,4\}$ becomes trivial, $\omega(z)\to 1$, so that 
\be
W_1\underset{z_{12}\to 0}{\longrightarrow} 1\, ,~~\quad W_2\underset{z_{12}\to 0}{\longrightarrow}  \taudc\, .
\ee
This leads to 
\be 
\begin{aligned}
A_{\text{ext}\Sigma}^{\alpha_0\beta_0}\underset{z_{12}\to 0}{\sim} & \, {3\over 2i\pi}\, \alpha'k^2\, \lambda_{\alpha_0\beta_0}\lambda^{\rm T}_{\beta_0\alpha_0}\,  {1\over z_{12}^2}\,  \frac{1}{\tau_2\eta^3}\,  \sum_{\nui}(-1)^{\delta_{\nui,1}}\, \vartheta_{\nui}^2\, \times\\
&\sum_{\vec l'}e^{-{\pi\over \alpha'\tau_2}\sum_{I'}(R_{I'}l_{I'})^2}\Bigg[\sum_{\vec{l}}e^{-{\pi\over \alpha'\tau_2}\sum_I(R_Il_I)^2}\C_{\nui}^{\Sigma\vec l'\vec l}+\sum_{\vec \tl}e^{-{\pi\alpha'\over \tau_2}\sum_I(\tl_I/R_I)^2}\tilde \C_{\nui}^{\Sigma\vec l'\vec \tl}\Bigg]\,,
\end{aligned}
\label{aext}
\ee
which has to be identified with 
\be
\begin{aligned}
& {3\over 2i\pi}\,\alpha'k^2\, \lambda_{\alpha_0\beta_0}\lambda^{\rm T}_{\beta_0\alpha_0}\,{8\C\over z_{12}^2}\times {1\over\tau_2^2}\, \sum_{\gamma=1}^{32+32} \underset{\alpha_0\gamma+\beta_0\gamma}{\Str}  \half\, \frac{1+g}{2}\, q^{\half(L_0-1)}&&\; \mbox{for $\A$}\, ,\\
\and \;\;\; \;&{3\over 2i\pi}\,\alpha'k^2\, \lambda_{\alpha_0\beta_0}\lambda^{\rm T}_{\beta_0\alpha_0}\,{8\C\over z_{12}^2}\times {1\over\tau_2^2}\, \underset{\alpha_0\alpha_0+\beta_0\beta_0}{\Str} {\Omega\over 2}\, \frac{1+g}{2}\, q^{\half(L_0-1)}&&\; \mbox{for $\M$}\, .
\end{aligned}
\label{Zz120}
\ee

\noindent
In these expressions, $\C$ is a constant number,\footnote{It can be determined by replacing in \Eq{prelim_ext} all correlators by their dominant poles, which can be found by the OPE's.}  while the supertraces are restricted to the open-string modes with ends attached to branes as shown in \Fig{annulus+mobius_diagrams_limite}. For the identification to be  possible, one has to switch the $T^2$, $T^4$ and $\tilde T^4$ lattices  in the partition functions $Z_\A$ and $Z_\M$ from Hamiltonian to instantonic forms, which is done  by Poisson summations, 
\be
\begin{aligned}
\sum_{\vec m} P^{(4)}_{\vec m+\vec a_i-\vec a_j}&= {\rmv_4\over \alpha'^2\, \tau_2^2}\sum_{\vec l}e^{-{\pi\over \alpha'\tau_2}\sum_I(R_Il_I)^2}e^{2i\pi \vec l\cdot (\vec a_i-\vec a_j)}\, , \\
\sum_{\vec m} W^{(4)}_{\vec n+\vec a_i-\vec a_j}&= {\alpha'^2\over \rmv_4\,  \tau_2^2}\sum_{\vec \tl}e^{-{\pi\alpha'\over \tau_2}\sum_I(\tl_I/R_I)^2}e^{2i\pi \vec \tl\cdot (\vec a_i-\vec a_j)}\, , \\
\sum_{\vec m'} P^{(2)}_{\vec m'+F\vec a_{\rm S}'+\vec a_{i'}-\vec a_{j'}}&= {\rmv_2\over \alpha'\, \tau_2}\sum_{\vec l'}e^{-{\pi\over \alpha'\tau_2}\sum_I(R_{I'}l_{I'})^2} e^{2i\pi \vec l'\cdot (\vec a_{i'}-\vec a_{j'})}e^{2i\pi F\vec l'\cdot \vec a_{\rm S}'}\, ,
\end{aligned}
\ee
where we have defined 
\be
\rmv_4=R_6R_7R_8R_9\, , ~~\quad \rmv_2=R_4R_5\, .
\ee

For the case of the annulus, using the  definitions of the characters given  in Eq.~(\ref{eq:CharacterDef}), we identify 
\begin{align}
\label{Cnu}
&\C_{1}^{\A\vec l' \vec l}=\rho\, {\C\over \tau_2^2\, \eta^3}\, f_{\alpha_0\text{D}}^{\A\vec l' \vec l}\, ,&&\qquad\tilde{\C}_{1}^{\A\vec l'\vec \tl}=\rho\, {\C\over \tau_2^2\, \eta^3}\, f_{\beta_0\text{N}}^{\A\vec l'\vec \tl}\, ,\nonumber\\
&\C_{2}^{\A\vec l'\vec l}={\C\over \tau_2^2\, \eta^3}\,{\vartheta_3^2\over \vartheta_4^2}\,  f_{\alpha_0\text{D}}^{\A\vec l' \vec l}-{\C\,\vartheta_2^2\over \tau_2^4\, \eta^9}\, f_{\alpha_0\text{N}}^{\A\vec l'\vec l}\, e^{2i\pi \vec l'\cdot \vec a_{\rm S}'}\, ,&&\qquad\tilde{\C}_{2}^{\A\vec l'\vec \tl}={\C\over \tau_2^2\, \eta^3}\, {\vartheta_3^2\over \vartheta_4^2}\,f_{\beta_0\text{N}}^{\A\vec l'\vec \tl}-{\C\, \vartheta_2^2\over \tau_2^4\, \eta^9}\, f_{\beta_0\text{D}}^{\A\vec l'\vec \tl}\, e^{2i\pi \vec l'\cdot \vec a_{\rm S}'}\, ,\nonumber\\
&\C_{3}^{\A\vec l'\vec l}={\C\, \vartheta_3^2\over \tau_2^4\, \eta^9}\, f_{\alpha_0\text{N}}^{\A\vec l'\vec l}-{\C\over \tau_2^2\, \eta^3}\, {\vartheta_2^2\over \vartheta_4^2}\, f_{\alpha_0\text{D}}^{\A\vec l'\vec l}\, e^{2i\pi \vec l'\cdot \vec a_{\rm S}'}\, ,&&\qquad \tilde{\C}_{3}^{\A\vec l'\vec \tl}={\C\, \vartheta_3^2\over \tau_2^4\, \eta^9}\, f_{\beta_0\text{D}}^{\A\vec l'\vec \tl}-{\C\over \tau_2^2\, \eta^3}\, {\vartheta_2^2\over \vartheta_4^2}\, f_{\beta_0\text{N}}^{\A\vec l'\vec \tl}\, e^{2i\pi \vec l'\cdot \vec a_{\rm S}'}\, ,\nonumber \\
&\C_{4}^{\A\vec l'\vec l}=-{\C\, \vartheta_4^2\over \tau_2^4\, \eta^9}\, f_{\alpha_0\text{N}}^{\A\vec l'\vec l}\,,&&\qquad \tilde{\C}_{4}^{\A\vec l'\vec \tl}=-{\C\, \vartheta_4^2\over \tau_2^4\, \eta^9}\, f_{\beta_0\text{D}}^{\A\vec l'\vec \tl}\, ,
\end{align}
where we have defined 
\begin{align}
&f_{\alpha_0\text{N}}^{\A\vec l'\vec l}={\rmv_2\rmv_4\over \alpha'^3}\, \sum_{i,i'}N_{ii'}\,e^{2i\pi\vec{l}\cdot(\vec{a}_{i_0}-\vec{a}_i)} e^{2i\pi\vec l'\cdot(\vec{a}_{i'_0}-\vec a_{i'})}\,,&&f_{\beta_0\text{\text{D}}}^{\A\vec l'\vec \tl}={\rmv_2\,\alpha'^2\over \alpha'\,\rmv_4}\, \sum_{i,i'}D_{ii'}\, e^{2i\pi\vec \tl\cdot(\vec{a}_{j_0}-\vec{a}_i)} e^{2i\pi\vec l'\cdot(\vec{a}_{i'_0}-\vec a_{i'})}\,,\nonumber\\
&f_{\alpha_0\text{D}}^{\A\vec l'\vec{l}}=\delta_{\vec{l},\vec{0}}\, {\rmv_2\over \alpha'}\, \sum_{i,i'}D_{ii'}\, e^{2i\pi \vec l'\cdot (\vec a_{i_0'}-\vec a_{i'})}\, ,&&f_{\beta_0\text{N}}^{\A\vec l'\vec \tl}=\delta_{\vec \tl,\vec{0}}\, {\rmv_2\over \alpha'}\, \sum_{i,i'}N_{ii'}\, e^{2i\pi \vec l'\cdot (\vec a_{i_0'}-\vec a_{i'})}\, .
\end{align}
To better understand how the discrete sums and coefficients $N_{ii'}$ and $D_{ii'}$ arise, let us display as an example what the first product of traces in \Eq{annulus} becomes, when restricting to open strings attached (in the T-dual picture) to the D3-brane $\alpha_0$, and to a D3-brane $\gamma$ in the Neumann sector,
 \be
\sum_{\substack{i,i'\\ j,j'}} \tr(\gamma_{{\rm N},1}^{ii'})\tr(\gamma_{{\rm N},1}^{jj'\,-1})\quad \longrightarrow\quad (\gamma_{{\rm N},1}^{i_0i_0'})_{\alpha_0\alpha_0} \sum_{j,j'}\sum_{\gamma=1}^{N_{jj'}}(\gamma_{{\rm N},1}^{jj'\,-1})_{\gamma\gamma}=\sum_{j,j'}N_{jj'}\, .
 \ee
On the contrary, all terms associated with the group generator $g$ vanish, due to the fact that the diagonal components of the matrices $J_{k}$ are zero.  In the expressions of $\C_{1}^{\A\vec l' \vec l}$ and $\tilde \C_{1}^{\A\vec l' \vec \tl}$, we have introduced a coefficient $\rho$ that accounts for the ambiguity arising in their  determination, since  $\vartheta_{\nu_{\rm int}}^2=0$ for $\nu_{\rm int}=1$.  This coefficient will be determined in the sequel. Finally, notice that the identification has lifted all sign ambiguities associated with the twist- and spin-field correlators. These signs depend on the instanton numbers $\vec l'$, $\vec l$, $\vec \tl$ and the positions of the D3-branes $\alpha_0$, $\beta_0$, $\gamma$. 

To perform the similar computation in the M\"obius strip case, note that $Z_\M$ is expressed in terms of ``hatted characters'' defined in \Eq{hatc}. However, in light-cone gauge, the characters associated with the worldsheet fermions multiplied by $1/\hat\eta^8$ arising from the bosonic coordinates yield low-lying states at the massless level. Hence, all phases $e^{-i\pi(h-c/24)}$ appearing in the definitions of the hatted characters cancel each other and we may simply remove all ``hats'' on the $\vartheta$  and $\eta$ functions when identifying \Eq{Zz120} with the amplitude~(\ref{aext}).  In that case, the normalization functions are found to be 
\be
\begin{aligned}
\label{Cnu_mobius}
& \C_{1}^{\M\vec l'\vec l}=0\, ,\qquad&&{\tilde{\C}}_{1}^{\M\vec l'\vec \tl}=0\, ,\\
& \C_{2}^{\M\vec l'\vec l}={\C\, \vartheta_2^2\over \tau_2^4\, \eta^9}\, {\rmv_2\rmv_4\over \alpha'^3}\, e^{2i\pi\vec l'\cdot \vec a_{\rm S}'}\, , \qquad&&{\tilde{\C}}_{2}^{\M\vec l'\vec \tl}={\C\, \vartheta_2^2\over \tau_2^4\, \eta^9}\, {\rmv_2\,\alpha'^2\over \alpha'\, \rmv_4}\, e^{2i\pi\vec l'\cdot \vec a_{\rm S}'}\,,\\
& \C_{3}^{\M\vec l'\vec l}=-{\C\, \vartheta_3^2\over \tau_2^4\, \eta^9}\, {\rmv_2\rmv_4\over \alpha'^3}\, ,\qquad&&{\tilde{\C}}_{3}^{\M\vec l'\vec \tl}=-{\C\, \vartheta_3^2\over \tau_2^4\, \eta^9}\, {\rmv_2\,\alpha'^2\over \alpha'\, \rmv_4}\, ,\\
& \C_{4}^{\M\vec l'\vec l}={\C\, \vartheta_4^2\over \tau_2^4\, \eta^9}\, {\rmv_2\rmv_4\over \alpha'^3}\, ,\qquad&&{\tilde{\C}}_{4}^{\M\vec l'\vec \tl}={\C\,\vartheta_4^2\over \tau_2^4\, \eta^9}\, {\rmv_2\,\alpha'^2\over \alpha'\, \rmv_4}\, .
\end{aligned}
\ee
For instance,  when one restricts to the boundary conditions shown in \Fig{annulus+mobius_diagrams_limite}, the first trace appearing in the expression of $Z_\M$ yields,
 \be
\tr(\gamma_{{\rm N},\Omega}^{ii'\, \rm T}\gamma_{{\rm N},\Omega}^{jj'\,-1})\quad \longrightarrow\quad (\gamma_{{\rm N},\Omega}^{i_0i_0'\, \rm T})_{\alpha_0\alpha_0}(\gamma_{{\rm N},\Omega}^{jj'\,-1})_{\alpha_0\alpha_0}=1\, .
 \ee
On the contrary, all traces in the second line of \Eq{mobius} yield vanishing contributions since the selected diagonal matrix elements are zero.  


\paragraph{\em \bm Consistency when supersymmetry is restored: } 
All normalization functions can be injected back in  \Eqs{Aext} and~(\ref{Aint}) to obtain the full expressions of the amplitudes  $A_{\rm ext\Sigma}^{\alpha_0\beta_0}$ and $A_{\rm int\Sigma}^{\alpha_0\beta_0}$ for arbitrary $z_1$, $z_2$. To analyze their structures in more details, let us focus on the instantonic sums. For the internal part of the amplitude computed on the annulus, we obtain for each given $\vec l'$, $\vec l$ a contribution of the form~ 
\begin{align}
\sum_{\nui}\vartheta_{\nui}\big({\zud\over 2}\big)^2\, \C_{\nui}^{\A\vec l' \vec l}&=\C\, {f_{\alpha_0\rm N}^{\A\vec l'\vec l}\over \tau_2^4\, \eta^9}\,\bigg\{\vartheta_3^2\vartheta_3\big({z_{12}\over 2}\big)^2-\vartheta_4^2\vartheta_4\big({z_{12}\over 2}\big)^2-e^{2i\pi\vec l'\cdot \vec a_{\rm S}'}\,\vartheta_2^2\vartheta_2\big({z_{12}\over 2}\big)^2\bigg\}\nonumber \\
&\;\;\;\;+\C\, {f_{\alpha_0\rm D}^{\A\vec l'\vec l}\over \tau_2^2\, \eta^3\, \vartheta_4^2}\,\bigg\{\vartheta_3^2\vartheta_2\big({z_{12}\over 2}\big)^2+\rho\, \vartheta_4^2\vartheta_1\big({z_{12}\over 2}\big)^2-e^{2i\pi\vec l'\cdot \vec a_{\rm S}'}\, \vartheta_2^2\vartheta_3\big({z_{12}\over 2}\big)^2\bigg\}\nonumber \\
&=\C\,\big(1-(-1)^{l_5}\big) \bigg\{{f_{\alpha_0\rm N}^{\A\vec l'\vec l}\over \tau_2^4\, \eta^9}\, \vartheta_2^2\vartheta_2\big({z_{12}\over 2}\big)^2+{f_{\alpha_0\rm D}^{\A\vec l'\vec l}\over \tau_2^2\, \eta^3\, \vartheta_4^2}\, \vartheta_2^2\vartheta_3\big({z_{12}\over 2}\big)^2\bigg\}  \label{s1}\\
&\;\;\;\;+\C\, (\rho-1)\, {f_{\alpha_0\rm D}^{\A\vec l'\vec l}\over \tau_2^2\, \eta^3}\, \vartheta_1\big({z_{12}\over 2}\big)^2\, ,\nonumber 
\end{align}
where the second equality is obtained by applying a generalized Jacobi identities~\cite{Kiritsis_book} with non-zero first arguments, as well as the specific form of the vector $\vec a_{\rm S}'$. For given $\vec l'$, $\vec \tl$, the similar sum for the coefficients $\tilde \C_{\nui}^{\A\vec l' \vec \tl}$ is obtained by changing $\alpha_0\to \beta_0$ and ${\rm N}\leftrightarrow {\rm D}$. We are now ready to determine the constant $\rho$ by taking the  limit $R_5\to +\infty$ in \Eq{Aint}. Indeed, $l_5=0$ is the only  contribution  in the sum over $l_5$ that survives in this limit. Hence, $A_{\text{int}\A}^{\alpha_0\beta_0}$ vanishes when supersymmetry is restored if and only if $\rho=1$, since in that case only,  \Eq{s1} projects out all even values of the wrapping number $l_5$.  Indeed, in the supersymmetric case, the effective potential cannot be corrected perturbatively, which implies $\rho$ to be such that the one-loop corrections to the masses we are computing  vanish.

We can proceed the same way for the internal part of the amplitude computed on the M\"obius strip. For fixed $\vec l'$, $\vec l$, we have 
\be
\begin{aligned}
\sum_{\nui}\vartheta_{\nui}\big({\zud\over 2}\big)^2\, \C_{\nui}^{\M\vec l' \vec l}&={\C\, \rmv_2\rmv_4\over \alpha'^3\, \tau_2^4\, \eta^9}\, \bigg\{e^{2i\pi \vec l'\cdot \vec a_{\rm S}'}\, \vartheta_2^2\vartheta_2\big({z_{12}\over 2}\big)^2 - \vartheta_3^2\vartheta_3\big({z_{12}\over 2}\big)^2+\vartheta_4^2\vartheta_4\big({z_{12}\over 2}\big)^2\bigg\}\\
&=-{\C\, \rmv_2\rmv_4\over \alpha'^3\, \tau_2^4\, \eta^9}\,\big(1-(-1)^{l_5}\big) \, \vartheta_2^2\vartheta_2\big({z_{12}\over 2}\big)^2\, , 
\end{aligned}
\label{s3}
\ee
while for given $\vec l'$, $\vec \tl$, the analogous sum for $\tilde \C_{\nui}^{\M\vec l' \vec l}$ is obtained by changing $\rmv_4/\alpha'^2\to \alpha'^2/\rmv_4$. In the limit $R_5\to +\infty$ where supersymmetry is restored, the amplitude $A_{\text{int}\M}^{\alpha_0\beta_0}$  vanishes consistently. 

As can be seen from \Eqs{Aext} and~(\ref{Aint}), the sums over the spin structure $\nu_{\rm int}$ in the external and internal parts of the amplitudes  are identical, up to the insertion of the sign $(-1)^{\delta_{\nui,1}}$ for $A_{\text{ext}\Sigma}^{\alpha_0\beta_0}$. Of course, this does not make any difference  in the case of the M\"obius strip since the normalization functions for $\nu_{\rm int}=1$ vanish. On the contrary, for $\Sigma=\A$, the extra sign amounts to changing $\rho\to -\rho$ in \Eq{s1}. As a result, the external part of the amplitude, $A_{\text{ext}\A}^{\alpha_0\beta_0}$, does not vanish in the decompactification limit, and yields a one-loop correction to the K\"ahler potential of the massless scalars in the ND+DN sector, even in the supersymmetric case.


\paragraph{\em \bm Integration over the moduli and vertex positions: } What remains to be done is to integrate the amplitudes over the moduli of the open-string surfaces and vertex operator positions modulo the conformal Killing group~\cite{PolV1}. 
The moduli of $\A$ and $\M$ are the imaginary parts of the \Teich parameters of the double-cover tori,  $\Im \taudc$. Moreover, instead of integrating over the locations of both insertion points and dividing by the volume of the conformal Killing group, we may simply fix to an arbitrary value the position of one vertex operator, say $z_2\equiv \half$,  and integrate over the location of the other. 

In the case of the annulus, both vertices must be located on the same boundary, so that $z_1\equiv \half+iy_1$. As a result, denoting the integrated amplitudes by calligraphic letters, the internal part reads
\be
\EuScript A_{{\rm int}\A}^{\alpha_0\beta_0}= \int_0^{+\infty} \dd\;\,\!\! \Im \taudc\int_0^{\Im \taudc}\!\!\!\dd y_1 \left.A_{{\rm int}\A}^{\alpha_0\beta_0}\right|_{\half+iy_1,\half} ,
\label{AtotA}
\ee
and likewise for the external amplitude.
Similarly, for the two-point function computed on the M\"obius strip, $z_1$ must follow the entire boundary. However, the latter being twice longer than the one considered on the annulus, $z_1$ can actually be parametrized as $z_1=x_1+iy_1$, where $x_1\in\{0,\half\}$. As a result, the internal part of the  integrated amplitude is 
 \be
\EuScript A_{{\rm int}\M}^{\alpha_0\beta_0}= \int_0^{+\infty} \dd\;\,\!\! \Im \taudc\int_0^{\Im \taudc}\!\!\!\dd y_1\, \big(\!\!\left.A_{{\rm int}\M}^{\alpha_0\beta_0}\right|_{\half+iy_1,\half}+ \left.A_{{\rm int}\M}^{\alpha_0\beta_0}\right|_{iy_1,\half}\big)\, ,
\label{AtotM}
\ee
and similarly for the external part. 

In these forms, the full two-point functions are not particularly illuminating, while performing explicitly the integrals is certainly a hard task. Hence, our goal in the next section is  to extract simpler answers valid in the case where the scale of supersymmetry breaking is low.


\section{Limit of low supersymmetry breaking scale}
\label{alpha'0}

The analysis of \Reff{ACP} is valid in regions of moduli space where the supersymmetry breaking scale $M_{3/2}$ is lower than all other non-vanishing scales present in the model. The reason of this restriction is that extrema of the one-loop effective potential are then easily found, and correspond in the open-string sector to distributing all D3-branes (in the T-dual pictures) on O3-planes. In such a case, the squared masses acquired at one loop by the moduli fields arising from the NN and DD sectors take particularly simple forms, up to exponentially suppressed corrections of order $e^{-\pi \frac{c\Ms }{M_{3/ 2}}}$, in the notations of \Eq{vd}. In practice, the fact that $M_{3/2}$ is lower than the string scale as well as all other scales generated by  compactification means that, effectively, the dominant contributions of the effective potential and masses derived in \Reff{ACP} match those found in a  \KK field theory in $4+1$ dimensions. 
 
 In the present section, we would like to find similar results for the masses of the moduli fields present in the ND+DN sector of the theory. This will be done by imposing all mass scales other than $M_{3/2}$ to be proportional to $\Ms=1/\sqrt{\alpha'}$ and then taking the small   $\alpha'$ limit. 
 

\subsection{Limit of super heavy oscillator states}
\label{osci}

In order to treat all massive string-oscillator states as super heavy in the Hamiltonian forms of the partition functions, let us rescale the \Teich parameters of the open-string surfaces as follows\footnote{This rescaling also implies that the imaginary parts of $\tau$ and $2i\tau_2$, the \Teich parameters of the torus and Klein bottle, are large. Hence, the massive oscillator states are super heavy also in the closed-string sector.}
\be
\Im \taudc\equiv {\tau_2\over 2}\equiv {t\over 2\pi \alpha'}\gg 1\, , \quad \where \quad t\in(0,+\infty)\, .
\label{lim1}
\ee
Physically, this amounts to stretching the surfaces along their proper times in order to look like field-theory worldlines with topology of a circle.  The main practical consequence of the rescaling is the approximation
\be
\begin{aligned}
\vartheta_1(z)&\equiv -2\,q_{\rm dc}^{1\over 8}\sin (\pi z) \prod_{n\ge 1}\big[(1-\qdc^n)(1-\qdc^nz^{-2i\pi z})(1-\qdc^nz^{2i\pi z})\big]\, , \quad q_{\rm dc}\equiv e^{2i\pi\taudc}\, , \\
&=-2\,q_{\rm dc}^{1\over 8}\sin (\pi z) (1+\cdots)\, ,\quad \when \quad |\Im z|<\Im \taudc\, , 
\end{aligned}
\label{condi}
\ee
where from now on, ellipses stand for terms exponentially suppressed when $\alpha'\to 0$, \ie of order $e^{-L^2/\alpha'}$ for lengths $L>0$. In particular, the cut differential associated with the complex directions $u\in\{3,4\}$ becomes
\be
\omega(z)={\sin\big(\pi(z-{z_1+z_2\over 2})\big)\over \sin\!\big(\pi(z-z_1)\big)^\half\, \sin\!\big(\pi(z-z_2)\big)^\half}+\cdots\, ,\quad \when\quad |\Im z|,\, y_1,\,y_2 <\Im \taudc\, .
\label{win}
\ee 
Periodicity $z\to z+1$ remains explicit, while periodicity $z\to z+\taudc$ is hidden in the ellipsis. 
Notice that compared to \Eq{points}, we impose $y_1$, $y_2$ to be strictly lower than $\Im \taudc$ for the above formula to always be valid. More generally, throughout the derivations to come, we will write formulas in their generic forms. Indeed, because in the end all quantities will have to be integrated, taking into account extra contributions arising only at special values of the integration variables results in subdominant corrections for small $\alpha'$.  We will come back to this issue at the end of this section.

Keeping this in mind, we redefine  
\be
y_A\equiv u_A\, \Im\taudc={t\,u_A\over 2\pi\alpha'}\gg 1\,, \quad u_A\in(0,1)\, ,~~A\in\{1,2\}\,, ~~\quad \and ~~\quad u\equiv |u_1-u_2|\, , 
\label{lim2}
\ee
in terms of which the components of the cut-period matrix can be expressed like
\be
\label{cuts}
\begin{aligned}
W_1&=1+\cdots\, ,\\
W_2&=\taudc-\xi(z_1-z_2) +{i\over\pi} \ln 4+\cdots\, , \quad \where \quad \xi\equiv \sign(y_1-y_2)\, .
\end{aligned}
\ee
The first expression is easily found by integrating over $z$ finite along  $\gamma_1$ and replacing all sines in \Eq{win} by their dominant exponentials when $\Im \taudc$ is large. By contrast, $W_2$ can be derived by integrating $z$ between $x_0$ and $x_0+\taudc$,  $x_0\in \R$, using a primitive of $\omega$ in its form given in \Eq{win}. As a result, we  obtain that
\be
\Im(\overbar W_1W_2)={t(1-u)\over 2\pi \alpha'}+{\ln 4\over \pi}+\cdots\, .
\ee 

When taking the limit of small $\alpha'$ in \Eq{s1}, it turns out that the terms proportional to $f_{\alpha_0\rm D}^{\A \vec l'\vec l}$ (and $f_{\beta_0\rm N}^{\A \vec l'\vec \tl}$ for the formula involving $\tilde \C^{\A\vec l'\vec \tl}_{\nu_{\rm int}}$) are exponentially suppressed. Notice that they arise from the ND and DN sectors of the partition function $Z_\A$, which therefore cease to contribute to the amplitudes in this limit. In the case of the annulus, we then arrive at the expression 
\begin{align}
A_{{\rm int}\A}^{\alpha_0\beta_0}=&-4\C s\, \lambda_{\alpha_0\beta_0}\lambda^{\rm T}_{\beta_0\alpha_0} \,\sum_{\vec l'}e^{-{\pi^2\over t}\sum_{I'}(R_{I'}l_{I'})^2}\, \sum_{\vec l}e^{-{\pi^2\over t(1-u)+\alpha'(2\ln 4+\cdots\mbox{\tiny$\!\,$})}\sum_{I}(R_{I}l_{I})^2}\, \times\nonumber \\
&\,\big(1-(-1)^{l_5}\big)\, 2\pi^3\, {\alpha'^4\over t^4}\, {\rmv_2\rmv_4\over \alpha'^3}\, \sum_{ii'}N_{ii'}\,e^{2i\pi\vec{l}\cdot(\vec{a}_{i_0}-\vec{a}_i)} e^{2i\pi\vec l'\cdot(\vec{a}_{i'_0}-\vec a_{i'})}\, \times\label{imw12}
\\
&\, {\pi\over t(1-u)+\alpha'(2\ln 4+\cdots)}\, \bigg[{\pi^4\, \alpha'^2\over [t(1-u)+\alpha'(2\ln 4+\cdots)]^2}\sum_{J}(R_Jl_J)^2+2\alpha'(C+\hat C)\bigg]\nonumber\\
& + (\vec l, i_0,N_{ii'},R_I)\to (\vec \tl , j_0,D_{ii'},\alpha'/R_I)+\cdots\, ,\esps\nonumber 
\end{align}
while for the M\"obius strip we obtain 
\begin{align}
A_{{\rm int}\M}^{\alpha_0\beta_0}=&\;4\C s\, \lambda_{\alpha_0\beta_0}\lambda^{\rm T}_{\beta_0\alpha_0} \,\sum_{\vec l'}e^{-{\pi^2\over t}\sum_{I'}(R_{I'}l_{I'})^2}\, \sum_{\vec l}e^{-{\pi^2\over t(1-u)+\alpha'(2\ln 4+\cdots\mbox{\tiny$\!\,$})}\sum_{I}(R_{I}l_{I})^2}\, \times\nonumber\\
&\,\big(1-(-1)^{l_5}\big)\, 2\pi^3\, {\alpha'^4\over t^4}\, {\rmv_2\rmv_4\over \alpha'^3}\, \times\\
&\, {\pi\over t(1-u)+\alpha'(2\ln 4+\cdots)}\, \bigg[{\pi^4\, \alpha'^2\over [t(1-u)+\alpha'(2\ln 4+\cdots)]^2}\sum_{J}(R_Jl_J)^2+2\alpha'(C+\hat C)\bigg]\nonumber\\   
& + (\vec l, R_I)\to (\vec \tl ,\alpha'/R_I)+\cdots\, .\esps\nonumber
\end{align}
In the above formulas, the limit of small $\alpha'$ in $C$ and $\hat C$ will be derived  in \Sects{F1} and~\ref{termA}. 


\subsection{Limits of large compactification scales}
\label{radii}

We would like now to have all compactification mass scales other than $M_{3/2}$ very large. In practice, this amounts to taking small radii $R_4$, $R_I$ and dual radii $\alpha'/R_I$ limits. In order to avoid having to consider very large instanton numbers in such a regime, it is convenient to apply Poisson summations over $l_4$, $\vec l$ and $\vec \tl$~\cite{Kiritsis_book}, which lead for $\Sigma=\A$ to
\be
\label{Ann_all_states}
\begin{aligned}
A_{{\rm int}\A}^{\alpha_0\beta_0}=&-16\C s\sqrt{\pi}\, \lambda_{\alpha_0\beta_0}\lambda^{\rm T}_{\beta_0\alpha_0}\, {\alpha'^3 R_5\over t^{7\over 2}} \,\sum_{l_5}e^{-{\pi^2\over t}R_5^2(2l_5+1)^2}\, \sum_{ii'}N_{ii'}\,e^{2i\pi (a^5_{i_0'}-a^5_{i'})}\\
&\times \sum_{m_4}e^{-t\big({m_4+a^4_{i_0'}-a^4_{i'}\over R_4}\big)^2}\,  \sum_{\vec m}e^{-[t(1-u)+\alpha'(2\ln 4+\cdots\mbox{\tiny$\!\,$})]\sum_I\big({m_I+a^I_{i_0}-a^I_{i}\over R_I}\big)^2} \\
&\times\bigg\{\pi^3\bigg[2-[t(1-u)+\alpha'(2\ln 4+\cdots)] \sum_J\Big({m_J+a^J_{i_0}-a^J_i\over R_J}\Big)^2\bigg]\\
&\;\;\;\;\;+2\pi \bigg[{t(1-u)\over \alpha'}+2\ln 4+\cdots\bigg] (C+\hat C)\bigg\}\\
& \;+ (\vec m, i_0,N_{ii'},R_I)\to (\vec n , j_0,D_{ii'},\alpha'/R_I)+\cdots\, , \esps
\end{aligned}
\ee
and  for $\Sigma=\M$
\be
\label{Mob_all_states}
\begin{aligned}
A_{{\rm int}\M}^{\alpha_0\beta_0}=&\,16\C s\sqrt{\pi}\, \lambda_{\alpha_0\beta_0}\lambda^{\rm T}_{\beta_0\alpha_0}\, {\alpha'^3 R_5\over t^{7\over 2}} \,\sum_{l_5}e^{-{\pi^2\over t}R_5^2(2l_5+1)^2}\\
&\times \sum_{m_4}e^{-t({m_4\over R_4})^2}\,  \sum_{\vec m}e^{-[t(1-u)+\alpha'(2\ln 4+\cdots\mbox{\tiny$\!\,$})]\sum_I({m_I\over R_I})^2} \\
&\times\bigg\{\pi^3\bigg[2-[t(1-u)+\alpha'(2\ln 4+\cdots)]  \sum_J\Big({m_J\over R_J}\Big)^2\bigg]\\
&\;\;\;\;\;+2\pi \bigg[{t(1-u)\over \alpha'}+2\ln 4+\cdots\bigg](C+\hat C)\bigg\}\\
& \;+ (\vec m, R_I)\to (\vec n ,\alpha'/R_I)+\cdots\, . \esps
\end{aligned}
\ee

One may think that considering $T^4$ to be small would imply having the T-dual torus $\tilde T^4$  large. This is not true, as can be seen by redefining the radii as follows,
\be
 R_4=r_4\sqrt{\alpha'}\, ,~~\quad R_I=r_I\sqrt{\alpha'}\, , ~~\quad {\alpha'\over R_I}={\sqrt{\alpha'}\over r_I}\, ,
\ee
where $r_4$, $r_I$ are fixed and dimensionless. Indeed, all radii and dual radii  vanish as $\alpha'\to 0$. As a consequence, the limit of small $R_4$ implies that we may restrict the dominant term in $A_{{\rm int}\A}^{\alpha_0\beta_0}$ to $m_4=0$ and $i'\in\{i_0', \hat \imath_0'\}$, where 
\be
\mbox{$\hat \imath_0'$  is the fixed point in $\tilde T^2$ that faces $i_0'$ along the direction $\tilde X^5$}
\ee
in the T-dual pictures. Similarly, the limits of small $R_I$ and $\alpha'/R_I$ force $\vec m=0$, $i=i_0$ on the one hand, and $\vec n=0$, $i=j_0$ on the other hand. All other contributions can be absorbed in the ellipsis. In total, we obtain for the amplitude computed on the annulus 
\be
\begin{aligned}
A_{{\rm int}\A}^{\alpha_0\beta_0}=&-16\C s\sqrt{\pi}\, \lambda_{\alpha_0\beta_0}\lambda^{\rm T}_{\beta_0\alpha_0}\, {\alpha'^3 R_5\over t^{7\over 2}} \,\sum_{l_5}e^{-{\pi^2\over t}R_5^2(2l_5+1)^2}\, \big(N_{i_0i_0'}-N_{i_0\hat \imath_0'}+D_{j_0i_0'}-D_{j_0\hat \imath_0'}\big)\\
&\times\bigg\{2\pi^3+2\pi \bigg[{t(1-u)\over \alpha'}+2\ln 4+\cdots\bigg] (C+\hat C)\bigg\}+\cdots\, ,
\end{aligned}
\label{a1}
\ee
while on the M\"obius strip we have similarly
\be
\begin{aligned}
A_{{\rm int}\M}^{\alpha_0\beta_0}=&\;16\C s\sqrt{\pi}\, \lambda_{\alpha_0\beta_0}\lambda^{\rm T}_{\beta_0\alpha_0}\, {\alpha'^3 R_5\over t^{7\over 2}} \,\sum_{l_5}e^{-{\pi^2\over t}R_5^2(2l_5+1)^2}\, (1+1)\\
&\times\bigg\{2\pi^3+2\pi \bigg[{t(1-u)\over \alpha'}+2\ln 4+\cdots\bigg](C+\hat C)\bigg\}+\cdots\,.
\end{aligned}
\label{a2}
\ee


\subsection{\bm Limit $\alpha'\to 0$ of $U_1$ and $F_1(z,z_2)$}
\label{F1}

In this subsection and the following, our aim is to derive the limits of $C$ and $\hat C$ for small $\alpha'$, \ie the contributions arising from parts~$(2)$ of the correlators $\langle \tau^u\tau^{\prime u}\rangle_{\text{qu}}=\langle \tau^{\prime u}\tau^{u} \rangle_{\text{qu}}$. Because the results can be obtained with no more effort for any  \Teich parameter, we will keep the real part of $\taudc$ arbitrary, and $z_1$, $z_2$  will be chosen anywhere  in the torus represented by $\P$ in the complex plane, where 
\be
\mbox{$\P$ is the parallelogram with corners at 0, 1 and $\taudc$}\, .
\ee
The important thing, though, is that \Eqs{lim1} and~(\ref{lim2}) hold. Hence, our computations of parts~$(2)$ are valid for excited twist fields (for closed strings) and excited boundary-changing fields (for open strings). Let us start by deriving the limits of $U_1$ and $F_1(z,z_2)$, which will be used in the next subsection to derive those of $C$ and $\hat C$. 


\paragraph{\em \bm $U_1$ when $\alpha'\to 0$: } 

The function $F_1(z,w)$ defined in \Eq{FA} involves $U_1$ which is a root of 
\be
\Omega(U)\equiv \partial_zF_1(z,w)|_{z=w}={\vartheta_1'\over \vartheta_1}(U)+{\vartheta_1'\over \vartheta_1}(Y_1-U)\, ,\quad \where \quad Y_1=-{z_{12}\over 2}\, . 
\ee
To see that this definition makes sense, notice that the meromorphic function $\Omega(U)$ is doubly periodic on the genus-1 Riemann surface $\Sigma$, with two simple poles at $U=0$ and $U=Y_1$. Therefore, it has two simple zeros. Denoting one of them $U_1$, the second is $Y_1-U_1$.\footnote{It is understood that poles and zeros are defined modulo $1$ and $\taudc$.}  

When considering the limit $\alpha'\to 0$ in the equation $\Omega(U_1)=0$, a difficulty we have to face is the following: If we look for $U_1$ such that $0<\Im U_1<\Im \taudc$, using the fact that $|\Im Y_1|<\half \,\Im \taudc$ we obtain that 
\be
-{3\over 2}\, \Im \taudc<\Im(Y_1-U_1)<\half\,  \Im\taudc\,.
\ee
Hence, it is not clear when we can apply \Eq{condi} or not. For this reason, let us consider two cases:
\begin{itemize}
\item When $0<\Im Y_1<\half\, \Im \taudc$, we obtain that $|\Im (Y_1-U_1)|<\half\, \Im\taudc$ which allows to write
\be
\begin{aligned}
0=\Omega(U_1)&=  \pi\big(\!\cot(\pi U_1)+\cot[\pi(Y_1-U_1)]\big)+\cdots\\
&= \pi\,{\sin(\pi Y_1)\over \sin(\pi U_1)\sin[\pi(Y_1-U_1)]}+\cdots\, . 
\end{aligned}
\label{essai}
\ee
For the right-hand side to vanish when $\alpha'\to 0$, we see that $U_1$ must satisfy  \mbox{$\Im U_1\to +\infty$} and $\Im (Y_1-U_1)<0$. To determine $U_1$ more precisely, let us keep the first subdominant term in the ellipsis of \Eq{essai}, which is $2i\pi \,\qdc \,e^{-2i\pi U_1}$.\footnote{It can be found from \Eq{Heq} where the function $H$ is defined in \Eq{Hdef}.\label{footn}}  In that case, the equation becomes
\be
i\sin(\pi Y_1)=2 \sin(\pi U_1)\sin[\pi(Y_1-U_1)]\, \qdc \,e^{-2i\pi U_1}(1+\cdots)\, , 
\ee
which implies the asymptotic equivalence 
\be
-e^{-i\pi Y_1}\underset{\alpha'\to 0} \sim e^{-i\pi U_1}\, e^{i\pi(Y_1-U_1)}\, \qdc\, e^{-2i\pi U_1}\, .
\label{equiva}
\ee
Redefining
\be
U_1\equiv {\taudc+Y_1\over 2}+{1\over 4}+{m\over 2}+\varepsilon  \quad \mbox{for some }m\in\Z\, ,\quad \with\quad  |\Re \varepsilon|<\half\, ,
\label{change}
\ee
\Eq{equiva} shows that  $\varepsilon\to 0$ when $\alpha'\to 0$. 
\item When $-\half\, \Im \taudc<\Im Y_1<0$, we can apply the change of variable~(\ref{change}) which yields
\be
-{3\over 4}\, \Im \taudc -\Im \varepsilon<\Im(Y_1-U_1)<-\half\, \Im \taudc -\Im \varepsilon\, .
\ee
Hence, assuming that $\varepsilon$ is bounded when $\alpha'\to 0$, \Eq{essai} is legitimate for small enough $\alpha'$. However, the first dominant term in the ellipsis is now $-2i\pi\, \qdc\, e^{2i\pi(Y_1-U_1)}$,$^{\ref{footn}}$ and the equation becomes
\be
i\sin(\pi Y_1)=- 2 \sin(\pi U_1)\sin[\pi(Y_1-U_1)]\, \qdc \,e^{2i\pi (Y_1- U_1)}(1+\cdots)\, .
\ee
Hence, we obtain that 
\be
e^{i\pi Y_1}\underset{\alpha'\to 0} \sim -e^{-i\pi U_1}\, e^{i\pi(Y_1-U_1)}\, \qdc\, e^{2i\pi (Y_1-U_1)}\, ,
\label{equiva2}
\ee
which is equivalent to \Eq{equiva} and leads to $\varepsilon\to 0$ when $\alpha'\to 0$. The assumption on the boundedness of $\varepsilon$ being consistent, we have also found solutions in the present case. 
\end{itemize}
In both instances, $\varepsilon$ can be expressed in terms of exponentially suppressed contributions subdominant to those we have taken into account explicitly. Its leading behavior is derived in Appendix~\ref{lvar}. 

By imposing $U_1$ to be located in $\P$, we find the two roots of $\Omega(U)$,
\be
U_1= {\taudc+Y_1\over 2}+{1\over 4}+\cdots\quad \mbox{or}\quad  {\taudc+Y_1\over 2}+{3\over 4}+\cdots\, .
\ee
Since we know that there cannot be other solutions modulo 1 and $\taudc$, a cross-check of this  result is to observe that $Y_1-U_1$ satisfies consistently
\be
Y_1-U_1+1+\taudc = {\taudc+Y_1\over 2}+{3\over 4}+\cdots\quad \mbox{or}\quad {\taudc+Y_1\over 2}+{1\over 4}+\cdots\, .
\ee


\paragraph{\em \bm $F_1(z,z_2)$ when $\alpha'\to 0$: } 

Both possible choices of $U_1$ yield the same function $F_1(z,w)$. What we need to analyze in order to derive the limits of $C$ and $\hat C$ is  its expression for $w=z_2$,~
\be
F_1(z,z_2)={\vartheta_1\big(z+{\taudc\over 2}-{1\over 4}\, z_1-{3\over 4}\, z_2+{1\over 4}+\cdots\big)\,\vartheta_1\big(z-{ \taudc\over 2}-{1\over 4}\, z_1-{3\over 4}\, z_2-{1\over 4}+\cdots\big)\over\vartheta_1\big({\taudc\over 2}-{z_{12}\over 4}+{1\over 4}+\cdots \big)\,\vartheta_1\big(-{\taudc\over 2}-{z_{12}\over 4}-{1\over 4}+\cdots \big)}\, ,
\label{F12}
\ee
where $z\equiv x+iy\in \P$, and $x$, $y\in\R$. Notice that \Eq{condi} can be applied to both  $\vartheta_1$ functions appearing in the denominator (for small enough $\alpha'$). 

For $0<{1\over 2}\, y_1+{3\over 2}\, y_2<\Im \taudc$, the second $\vartheta_1$ function in the numerator fulfils the hypothesis of \Eq{condi}, while the first one requires more scrutiny:
\begin{itemize}
\item When $0<y<\half\,\Im \taudc+{1\over 4}\, y_1+{3\over 4}\, y_2$, \Eq{condi} applies to the first $\vartheta_1$.
\item When $\half\,\Im \taudc+{1\over 4}\, y_1+{3\over 4}\, y_2<y<\Im \taudc$, we have (for small enough $\alpha'$)
\be
\Im \taudc<\Im \Big(z+{\taudc\over 2}-{1\over 4}\, z_1-{3\over 4}\, z_2+{1\over 4}+\cdots\Big)<{3\over 2} \,\Im \taudc\, ,
\ee
which shows that we have to multiply the second line of \Eq{condi} by an extra factor $-\qdc e^{-2i\pi z}$ to apply it to the first $\vartheta_1$ function in the numerator of \Eq{F12}. 
\end{itemize}
In the end, we obtain that 
\begin{align}
&\when\quad 0<\mbox{$\half$}\, y_1+\mbox{${3\over 2}$}\, y_2<\Im \taudc\, , \quad \mbox{then}\label{exp1} \\
&F_1(z,z_2)=\left\{\begin{array}{ll}1+\cdots&\mbox{if~~$0<y<\half\,\Im \taudc+{1\over 4}\, y_1+{3\over 4}\, y_2$}\, , \\ e^{-4i\pi(z-{\taudc\over 2}-{1\over 4} z_1-{3\over 4} z_2)}\,(1+\cdots)&\mbox{if~~$\half\,\Im \taudc+{1\over 4}\, y_1+{3\over 4}\, y_2<y<\Im\taudc$\,.}\end{array}\right. \nonumber  
\end{align}

Conversely, for $\Im \taudc<\half\, y_1+{3\over 2}\, y_2<2\, \Im \taudc$, it is the first $\vartheta_1$ function in the numerator that satisfies the condition of validity of \Eq{condi}, while for the second one we have to consider two possibilities:
\begin{itemize}
\item When ${1\over 4}\, y_1+{3\over 4}\, y_2-\half \, \Im \taudc<y<\Im \taudc$, \Eq{condi} applies to this  $\vartheta_1$ function.
\item When $0<y<{1\over 4}\, y_1+{3\over 4}\, y_2-\half \, \Im \taudc$, we have (for small enough $\alpha'$)
\be
-{3\over 2}\, \Im \taudc<\Im \Big(z-{ \taudc\over 2}-{1\over 4}\, z_1-{3\over 4}\, z_2-{1\over 4}+\cdots\Big)<-{1\over 2} \,\Im \taudc\, .
\ee
Hence, the second line of \Eq{condi} must contain an extra factor $-\qdc e^{2i\pi z}$ to apply it to the second $\vartheta_1$ function in the numerator of $F_1(z,z_2)$.
\end{itemize}
All in all,
\begin{align}
&\when\quad \Im \taudc<\mbox{$\half$}\, y_1+\mbox{${3\over 2}$}\, y_2<2\, \Im \taudc \, , \quad \mbox{then}\label{exp2} \\
&F_1(z,z_2)=\left\{\begin{array}{ll}  e^{4i\pi(z-{1\over 4}z _1-{3\over 4}z_2+{\taudc\over 2})}\, (1+\cdots)&\mbox{if~~$0<y<{1\over 4}\, y_1+{3\over 4}\, y_2-\half\,\Im \taudc$\,,}\\
1+\cdots&\mbox{if~~${1\over 4}\, y_1+{3\over 4}\, y_2-\half\,\Im \taudc<y<\Im \taudc$}\, .\esps\end{array}\right. \nonumber 
\end{align}


\subsection{\bm Limit $\alpha'\to 0$ of $C$ and $\hat C$}
\label{termA}

We are now ready to evaluate the limits of $C$ and $\hat C$ for small $\alpha'$. 


\paragraph{\em \bm $\hat C$ when $\alpha'\to 0$: } The expression of $\hat C$ given  in \Eq{Ch} involves $F_1(z_1,z_2)$. If this quantity can certainly be obtained by taking $z=z_1$ in the results we have just derived, it can also be computed from scratch by reasoning in the same way, which turns out to be easier. The result is 
\be
F_1(z_1,z_2)= \left\{ \begin{array}{llrll}e^{2i\pi(\taudc-{3\over 2}z_{12})} (1+\cdots)& \mbox{if} \!\!&{2\over 3}\, \Im \taudc\!\!\!\!&<\Im z_{12}\!\!\!\!&<\Im \taudc\, , \\
(1+\cdots)& \mbox{if} \!\! &-{2\over 3}\, \Im \taudc\!\!\!\!& <\Im z_{12}\!\!\!\!&<{2\over 3}\, \Im \taudc\, , \esps\\
e^{2i\pi(\taudc+{3\over 2}z_{12})} (1+\cdots) &\mbox{if} \!\!\! &- \Im \taudc\!\!\!&<\Im z_{12}\!\!\!\!&<-{2\over 3}\, \Im \taudc\, .\esps
\end{array}
\right.
\ee
As a result, the contributions proportional to $\hat C$ in \Eqs{a1} and~(\ref{a2})  are of the form 
\begin{align}
2\pi\bigg[{t(1-u)\over \alpha'}+2\ln 4+\cdots\bigg] \hat C&=-4\pi^3 \bigg[{t(1-u)\over \alpha'}+2\ln 4+\cdots\bigg]\times\nonumber \\
&~~~\left\{ \begin{array}{llrll}e^{2i\pi(\taudc-z_{12})} (1+\cdots)& \mbox{if} \!\!&{2\over 3}\, \Im \taudc\!\!\!\!&<\Im z_{12}\!\!\!\!&<\Im \taudc\, , \\
e^{i\pi z_{12}}(1+\cdots) & \mbox{if}  \!\! &0\!\!\!\!& <\Im z_{12}\!\!\!\!&<{2\over 3}\, \Im \taudc\, , \esps  \\
e^{-i\pi z_{12}}(1+\cdots)& \mbox{if}  \!\! &-{2\over 3}\, \Im \taudc\!\!\!\!& <\Im z_{12}\!\!\!\!&<0\, ,\esps  \\
e^{2i\pi(\taudc+z_{12})} (1+\cdots) &\mbox{if}  \!\!\! &- \Im \taudc\!\!\!&<\Im z_{12}\!\!\!\!&<-{2\over 3}\, \Im \taudc\, ,\esps
\end{array}
\right.\nonumber\\
&=\cdots\, ,
\label{C_hat_lim}
\end{align}
which is exponentially suppressed. Note however that this statement is valid when the intervals of $\Im z_{12}$ are open. 


\paragraph{\em \bm $C$ when $\alpha'\to 0$: } Using the relation given in \Eq{Cexp}, the term linear in $C$ in \Eqs{a1} and~(\ref{a2}) can be written as 
\be
\begin{aligned}
&2\pi\bigg[{t(1-u)\over \alpha'}+2\ln 4+\cdots\bigg] C=-2i\pi^4 \cos\!\Big({\pi\over 2}\,z_{12}\Big)\big[\overbar W _2 \,\F_1-\F_2\big]\, ,\\
\where\quad &\F_a=\oint_{\gamma_a}\dd z\, {F_1(z,z_2)\, (1+\cdots)\over \sin[\pi(z-z_1)]^\half\sin[\pi(z-z_2)]^{3\over 2}}\, ,\quad a\in\{1,2\}\, , \esp
\end{aligned}
\label{Cequa}
\ee
and $W_2$ is given in \Eq{cuts}. We are going to show that $\F_1$ contributes exponentially suppressed terms while $\F_2$ yields a finite result. 

In order to evaluate $\F_1$, we impose the points $z$ of the representative path of the cycle $\gamma_1$ to satisfy $\Im z\equiv \half \,\Im \taudc$. The advantage of this choice is that $F_1(z,z_2)$ can be replaced by $1+\cdots$ all along the path,
\be
\F_1=\int_0^1 \dd x\, {1+\cdots\over  \sin[\pi(x+{\taudc\over 2}-z_1)]^\half\sin[\pi(x+{\taudc\over 2}-z_2)]^{3\over 2}}\, .
\ee
Omitting the ellipsis, an explicit integration using a primitive of the integrand yields an exactly  vanishing result. However, the exponentially suppressed terms in the numerator may be large once multiplied by $\cos(\pi z_{12}/2)$. To take them into account, one can find upper bounds valid for given signs of $\half\, \Im \taudc-y_1$, $\half \, \Im\taudc-y_2$ and $y_1-y_2$. As an example, when $(\half\, \Im \taudc-y_1)(\half \, \Im\taudc-y_2)<0$ we obtain 
\be
\Big|\cos\!\Big({\pi\over 2}\,z_{12}\Big)\,\F_1\Big|\leq \int_0^1 \dd x \, 4\, e^{-\pi |\half \Im \taudc-y_2|} \, K=\cdots\, , 
\ee
where the constant $K$ is any majorant of $|1+\cdots|$ for small enough $\alpha'$. It turns out that in all instances the contributions proportional to $\F_1$ are suppressed.

To compute $\F_2$, we have to consider two cases. When $0<\mbox{$\half$}\, y_1+\mbox{${3\over 2}$}\, y_2<\Im \taudc$, \Eq{exp1} allows us to decompose the integral into two pieces, 
\be
\begin{aligned}
&\F_2=\F_2^{(1)}+\F_2^{(2)}\, , \\
\where\quad &\F_2^{(1)}=\int_{0}^{{\taudc\over2}+{1\over 4}z_1+{3\over 4}z_2}\dd z\, {1+\cdots\over \sin[\pi(z-z_1)]^\half\sin[\pi(z-z_2)]^{3\over 2}}\, ,\\
&\F_2^{(2)}=\int^{\taudc}_{{\taudc\over2}+{1\over 4}z_1+{3\over 4}z_2}\dd z\, {e^{-4i\pi(z-{\taudc\over 2}-{1\over 4} z_1-{3\over 4} z_2)}\, (1+\cdots)\over \sin[\pi(z-z_1)]^\half\sin[\pi(z-z_2)]^{3\over 2}}\, .
\end{aligned}
\ee
A primitive of the leading term of the integrand of $\F_2^{(1)}$ can be found and the limit of small $\alpha'$ taken after integration. This second step requires considering two cases, namely 
\be
(a):~y_1-y_2<{2\over 3}\, \Im \taudc \quad ~~ \and ~~\quad (b):~ y_1-y_2>{2\over 3}\, \Im \taudc\, , 
\ee 
which turn out to yield identical finite results, 
\be
\cos\!\Big({\pi\over 2}\,z_{12}\Big)\int_{0}^{{\taudc\over2}+{1\over 4}z_1+{3\over 4}z_2}\dd z\, {1\over \sin[\pi(z-z_1)]^\half\sin[\pi(z-z_2)]^{3\over 2}}={2i\over \pi}+\cdots\, .
\ee
The extra contribution arising from the ellipsis in the integrand of $\F_2^{(1)}$ turns out to be exponentially suppressed. However, this is not totally obvious  since the dominant implicit term in the numerator is of the form $e^{4i\pi({\taudc\over 2}+{1\over 4}z_1+{3\over 4}z_2-z)}$, which is 1 at the upper bound of the integral. Hence, keeping only the ellipsis in the numerator, we divide the domain of integration from 0 to some $z_0$ and from  $z_0$ to ${\taudc\over2}+{1\over 4}z_1+{3\over 4}z_2$. The value of $z_0$ is chosen such that the first domain yields an integral multiplied by $\cos(\pi z_{12}/2)$ admitting a trivial exponentially suppressed majorant, and also such that in the second domain it is legitimate to replace the two sines in the denominator by a single large exponential allowing an easy integration. In both cases~$(a)$ and~$(b)$, we obtain that 
\be
\cos\!\Big({\pi\over 2}\,z_{12}\Big)\F_2^{(1)}={2i\over \pi}+\cdots\, .
\label{f11}
\ee

In the integrand of $\F_2^{(2)}$, $\sin[\pi(z-z_2)]$ can always be replaced by a large exponential thanks to the fact that $\Im(z-z_2)\neq 0$ throughout the integration domain. While the same is true for $\sin[\pi(z-z_1)]$ in case~$(a)$, it turns out that  $\Im(z-z_1)$ vanishes for some $z$ in case~$(b)$.  In the first instance~$(a)$, it is therefore valid to write 
\be
\begin{aligned}
|\F_2^{(2)}|&=\left|\int^{\taudc}_{{\taudc\over2}+{1\over 4}z_1+{3\over 4}z_2}\dd z\, 4\, e^{-i\pi(2z-2\taudc-{1\over 2} z_1-{3\over 2} z_2)}\, (1+\cdots)\right|\\
&<\int^{\Im\taudc}_{{1\over2}\Im \taudc+{1\over 4}y_1+{3\over 4}y_2}\dd y\, {4K\over \sin\phi}\, e^{-\pi(\half z_1-{3\over 2} z_2)}\, e^{2\pi(y-\Im\taudc)}\, ,
\end{aligned}
\ee 
where we have chosen the path of integration for $z$ to be the straight segment in the complex plane, which forms an angle $\phi\in(0,\pi)$ with the horizontal axis. 
Integrating the majorant, one obtains
\be
\begin{aligned}
\Big|\cos\!\Big({\pi\over 2}\,z_{12}\Big)\, \F_2^{(2)}\Big|&<{K\over \pi \sin\phi}\times \left\{ \begin{array}{ll}e^{-2\pi y_2} & \mbox{if }~y_1-y_2>0 \, , \\
e^{-\pi(y_1+ y_2)} & \mbox{if }~y_1-y_2<0 \, , \esps\\
\end{array}\right.\\
&=\cdots\, , 
\end{aligned}
\label{f22}
\ee
where in the last line we use the fact that $\sin\phi\to 1$ when $\alpha'\to 0$. On the contrary, in case~$(b)$, only $\sin[\pi(z-z_2)]$ can be replaced by a single large  exponential. However, it is possible to integrate the dominant term of the integrand, and show as we did for $\F_2^{(1)}$ that the result dominates the integral arising from the ellipsis. 
Combining both pieces, we find that the conclusion of \Eq{f22}  holds again.

Let us move on to the second case, namely  $\Im \taudc<\mbox{$\half$}\, y_1+\mbox{${3\over 2}$}\, y_2<2\Im \taudc$, which can be treated by following the same steps as before. The starting point is \Eq{exp2} which leads to the decomposition  
\be
\begin{aligned}
&\F_2=\F_2^{(1)}+\F_2^{(2)}\, , \\
\where\quad &\F_2^{(1)}=\int^{\taudc}_{{1\over 4}z_1+{3\over 4}z_2-{\taudc\over2}}\dd z\, {1+\cdots\over \sin[\pi(z-z_1)]^\half\sin[\pi(z-z_2)]^{3\over 2}}\, ,\\
&\F_2^{(2)}=\int_{0}^{{1\over 4}z_1+{3\over 4}z_2-{\taudc\over2}}\dd z\, {e^{4i\pi(z-{1\over 4} z_1-{3\over 4} z_2+{\taudc\over 2})}\, (1+\cdots)\over \sin[\pi(z-z_1)]^\half\sin[\pi(z-z_2)]^{3\over 2}}\, .
\end{aligned}
\ee
Omitting the ellipsis in the integrand of $\F_1^{(1)}$, a direct integration yields for 
\be
(c):~y_1-y_2>-{2\over 3}\, \Im \taudc \quad ~~ \and ~~\quad (d):~ y_1-y_2<-{2\over 3}\, \Im \taudc\, , 
\ee 
the same finite result we found  in the previous case
\be
\cos\!\Big({\pi\over 2}\,z_{12}\Big)\int^{\taudc}_{{1\over 4}z_1+{3\over 4}z_2-{\taudc\over2}}\dd z\, {1\over \sin[\pi(z-z_1)]^\half\sin[\pi(z-z_2)]^{3\over 2}}={2i\over \pi}+\cdots\, .
\ee
Moreover, even if the ellipsis in the integrand of $\F_2^{(1)}$ equals 1 at the lower bound of the integral, it can be shown as before that it yields an extra exponentially suppressed contribution after integration.  Therefore, \Eq{f11} remains valid. 

In the integrand of $\F_2^{(2)}$, it is always safe to replace $\sin[\pi(z-z_2)]$ by a large exponential. This is also the case for $\sin[\pi(z-z_1)]$ in case~$(c)$, for which we  can write
\be
\begin{aligned}
|\F_2^{(2)}|&=\left|\int_{0}^{{1\over 4}z_1+{3\over 4}z_2-{\taudc\over2}}\dd z\, 4\, e^{i\pi(2z+2\taudc-{1\over 2} z_1-{3\over 2} z_2)}\, (1+\cdots)\right|\\
&<\int_{0}^{{1\over 4}y_1+{3\over 4}y_2-\half \Im \taudc}\dd y\, {4K\over \sin\phi}\, e^{-\pi(2\Im \taudc-\half z_1-{3\over 2} z_2)}\, e^{-2\pi y}\, .
\end{aligned}
\ee 
In the first line, the path of integration for $z$ is the segment that forms an angle $\phi\in(0,\pi)$ with the horizontal axis. Integrating the upper bound, we conclude that 
\be
\begin{aligned}
\Big|\cos\!\Big({\pi\over 2}\,z_{12}\Big)\, \F_2^{(2)}\Big|&<{K\over \pi \sin\phi}\times \left\{ \begin{array}{ll}e^{-\pi(2\Im \taudc-y_1- y_2)} & \mbox{if }~y_1-y_2>0 \, , \\
e^{-2\pi(\Im \taudc - y_2)} & \mbox{if }~y_1-y_2<0 \, , \esps\\
\end{array}\right.\\
&=\cdots\, .
\end{aligned}
\label{f22'}
\ee
In case~$(d)$, only $\sin[\pi(z-z_2)]$ can be replaced by a large exponential. The integration using a primitive as before shows that the conclusion of \Eq{f22'} is again true.  

Taking into account all of the above results for the integrals $\F_a$, and using the fact that $W_2$ does not grow exponentially fast as $\alpha'\to 0$, \Eq{Cequa} leads to the contribution
\be
2\pi\bigg[{t(1-u)\over \alpha'}+2\ln 4+\cdots\bigg] C=-4\pi^3+\cdots\, . 
\ee


\subsection{\bm Integration over $\tau_2$, $z_1$, $z_2$ and final result}
\label{integra}

Collecting the contributions of parts~$(1)$ and~$(2)$ (involving $C$ and $\hat C$) of the correlators $\langle \tau^u\tau^{\prime u}\rangle_{\text{qu}}=\langle \tau^{\prime u}\tau^{u} \rangle_{\text{qu}}$, the braces in the amplitudes~(\ref{a1}) and~(\ref{a2}) reduce to 
\be
\{2\pi^3-4\pi^3+\cdots\}=-2\pi^3+\cdots\, .
\ee
Because all the dependence  in $z_1$ and $z_2$ of the amplitudes is now hidden in ellipses, the integrations in \Eqs{AtotA} and~(\ref{AtotM}) can be  performed easily.  
Using the identity
\be
\int_0^{+\infty} \dd\;\,\!\! \Im \taudc\int_0^{\Im \taudc}\!\!\!\dd y_1\equiv {1\over (2\pi \alpha')^2}\int_0^{+\infty} t\, \dd t \int_0^1 d u_1\, , 
\label{integ}
\ee
and the fact that the integration over $u_1$ of the dominant contributions of the two-point functions are trivial, we obtain
\be
\begin{aligned}
\EuScript A_{{\rm int}\A}^{\alpha_0\beta_0}+\EuScript A_{{\rm int}\M}^{\alpha_0\beta_0}=&\;{4\over \pi}\, \C s \, \lambda_{\alpha_0\beta_0}\lambda^{\rm T}_{\beta_0\alpha_0}\,\sum_{l_5}{1\over |2l_5+1|^3}\\
&\times{\alpha'\over R_5^2}\big(N_{i_0i_0'}-N_{i_0\hat \imath_0'}-2+D_{j_0i_0'}-D_{j_0\hat \imath_0'}-2\big)\!+ \O\left(\frac{\alpha'^2}{R_5^4}\right) .
\end{aligned}
\label{amplif}
\ee
The origin of the terms of order $\alpha'^2/R_5^4$ will be explained at the end of this section. 

We are now ready to display the main result of our work. Implementing the correct dimension, the  mass squared acquired at one loop by the classically massless state $(\lambda,\epsilon)$ belonging to the ND+DN bosonic sector realized by strings ``stretched'' between the stack of $N_{i_0i_0'}\ge 2$ D3-branes (T-dual to D9-branes) and  the stack of $D_{j_0i_0'}\ge 2$ D3-branes (T-dual to D5-branes) is given by
\newpage 
\begin{align}
M^2&={1\over \alpha'}\sum_{\alpha_0=1}^{N_{i_0i_0'}}\sum_{\beta_0=1}^{D_{j_0i_0'}}\!\big(\EuScript A_{{\rm int}\A}^{\alpha_0\beta_0}+\EuScript A_{{\rm int}\M}^{\alpha_0\beta_0}\big)\label{resul}\\
&={32\over \pi}\, \C s \sum_{l_5}{1\over  |2l_5+1|^3} \,\tr(\lambda\lambda^{\rm T})\,M_{3/2}^2\,(n_{i_0i_0'}-n_{i_0\hat \imath_0'}-1+d_{j_0i_0'}-d_{j_0\hat \imath_0'}-1)+ \O\left(\frac{\alpha'}{R_5^4}\right) ,\nonumber 
\end{align}
where the supersymmetry breaking scale~(\ref{breakingscale}) reduces to 
\be
M_{3/2}={1\over 2R_5}\, .
\ee
From a field theory point of view, it can be seen that the bosonic (fermionic) degrees of freedom charged under $U(n_{i_0i_0'})$ or $U(d_{j_0i_0'})$ which are running in the loop contribute positively (negatively) to the  mass-squared term. Because $\tr(\lambda\lambda^{\rm T})>0$,\footnote{Since $\lambda\lambda^{\rm T}$ is a  real symmetric $n_{i_0i_0'}\times n_{i_0i_0'}$ matrix, it is diagonalizable and its eigenvalues are real. Moreover, since for any real $n_{i_0i_0'}$-vector $V$ we have $V^{\rm T}(\lambda^T\lambda) V=(\lambda V)^{\rm T}(\lambda V)=||\lambda V||^2\ge 0$, we conclude that all eigenvalues are nonnegative.} this implies  that $\C s>0$. 

To conclude this section, notice that \Eq{resul} guaranties or rules out the stability of moduli fields in the ND+DN sector only when the coefficient in parenthesis is not zero. When the latter vanishes, one has to compute four-point functions to conclude. 


\paragraph{\em \bm Subdominant contributions: } As announced below \Eq{win}, all our derivations have been presented at generic insertion points $z_1$, $z_2$. However, for special values of these variables, contributions we included in ellipses  are actually no more  exponentially suppressed when $\alpha'\to 0$. 

When the limit is taken up to $\alpha'=0$, all ellipses are identically zero, except at these particular values of $z_1$, $z_2$ which are loci of zero measure in the final integrals. Hence, the existence of such points does not affect the end result, which in this case is \Eq{resul} with no subdominant term at all. 
One may expect this expression to  match exactly the outcome of the computation of the masses in a pure Kaluza--Klein field theory in $4+1$ dimensions, with the field-theory \SS mechanism implemented along the circle of radius $R_5$. The field content should include the Kaluza--Klein towers of modes (propagating along $S^1(R_5)$) present in the string model and associated with the massless states or their superpartners charged under $U(n_{i_0i_0'})\times U(d_{j_0i_0'})$. However, this is not quite the case. Indeed, we already noticed above \Eq{imw12} that the ND and DN sectors do not run in the loop when $\alpha'=0$. Indeed, this has to be the case since otherwise they would contribute  extra terms proportional to $\delta_{\vec l,\vec 0}$ or $\delta_{\vec \tl,\vec 0}$ in \Eq{imw12}. Because of the first factor $\pi/[t(1-u_1)]$ (for $\alpha'=0$ and $y_2=0$) in the third line, such contributions would yield a divergence when integrating over~$u_1$. On the contrary, the Poisson summations over the $\vec l$- or $\vec \tl$-dependent contributions cancel the factor $\pi/[t(1-u_1)]$ and yield a finite result interpreted as the contributions of states in the NN and DD sectors running in the loop. Hence, from the point of view of a Kaluza--Klein field theory in $4+1$ dimensions,  the states in the ND+DN sector are treated semi-classically \ie as classical backgrounds (with vanishing vev's) in interaction with quantum matter in the NN and DD sectors. 
Moreover, notice  that the presence of infinite towers of Kaluza--Klein states associated with the \SS circle and running in the Feynman diagrams prevents all ultraviolet divergences from occurring, in exactly the same way as it happens in the string computation or in a supersymmetric quantum field theory at finite temperature.

By contrast, when $\alpha'$ is not strictly zero, the neighborhoods of the special points in which some ellipses are not exponentially small are no longer of measure zero. For instance, the term given in \Eq{C_hat_lim} is finite for all $\alpha'$ at the particular values $y_1=\Im \taudc$, $y_2=0$, \ie $u=1$. However, integrating it over $y_1\in[{2\over 3}\,\Im\taudc,\Im\taudc]$, one obtains a contribution $\O(\alpha'/t)$ which after insertion in the full amplitude and integration over $t$ leads to a contribution $\O(\alpha'/R_5^2)$ smaller than the dominant one shown in \Eq{amplif}. Another example is given by the contributions to the amplitudes arising from the {\em massive} Kaluza--Klein modes propagating along $T^4/\Z_2$ or $\tilde T^4/\Z_2$. As seen in \Eqs{Ann_all_states} and~(\ref{Mob_all_states}), they  involve in the former case factors
\be
e^{-\left[{t(1-u)\over \alpha'}+2\ln 4+\cdots\mbox{\tiny$\!\,$}\right]\sum_I\big({m_I+a^I_{i_0}-a^I_{i}\over r_I}\big)^2}\, , 
\ee
where the discrete sum is non-zero. However, at the particular values $y_1=\Im \taudc$, $y_2=0$, \ie $u=1$, this factor is finite for all $\alpha'$. Implementing the integrals shown in \Eq{integ}, one obtains again corrections $\O(\alpha'^2/R_5^4)$ in \Eq{amplif}. However, throughout the computation of the amplitudes, terms similar to the above examples are numerous and we have not dealt with them in full detail.  


\section{Stability analysis  at one loop}
\label{stanaly}

As seen in \Reff{ACP}, most of the brane configurations implying the one-loop effective potential to be extremal and tachyon free\footnote{Up to exponentially suppressed terms as shown in \Eq{vd}.\label{f1}}  yield a run away behavior of $M_{3/2}$ with $\nF-\nB<0$.  However, setups that lead to exponentially suppressed  or positive values of $\Vone$ may be of particular interest.  Indeed, for $\nF-\nB=0$, it is conceivable that the suppressed terms at one loop combine with higher loop corrections to stabilize the dilaton and $M_{3/2}$ in a perturbative regime. In that case, the resulting cosmological constant should be small, and the issue raised in~\Reff{DineSei} avoided. Moreover, cases where $\nF-\nB>0$ may  shed light on the existence or non-existence of de Sitter vacua after stabilization of the string coupling and supersymmetry breaking scale.

To be specific, the existence of 2 brane configurations without tachyons at one loop$^{\ref{f1}}$ and satisfying classically a  Bose/Fermi degeneracy at the massless level were shown to exist in \Reff{ACP}. Moreover,  4 more tachyon free setups with $\nF-\nB>0$ were also found. In all these instances, reaching these conclusions was possible thanks to the absence of moduli in the ND+DN sectors, and thanks to anomaly-induced supersymmetric masses for all blowing-up modes of $T^4/\Z_2$. Furthermore, 2 extra brane configurations were  presented~\cite{ACP}, one with  $\nF-\nB=0$  and the  other with  $\nF-\nB>0$, which we analyze further in the present section. Indeed, it was established that they both yield nonnegative squared masses at one loop for all moduli in the NN, DD and untwisted closed-string sectors, and that they possess moduli fields in the ND+DN sectors.  Given the result of the previous section, we are going to see that the latter are non-tachyonic. 


\subsection{NN, DD and closed-string sector moduli masses at one loop}

Before describing the two brane configurations of interest, let us review the stability conditions established in \Reff{ACP} for all moduli fields that are not in the ND+DN sector. 


\paragraph{\em Moduli in the {\rm \bf NN} and {\rm \bf DD} open-string sectors: } 

The number of these scalars and their masses can be determined in two steps. To start with, we can count the number of positions in $\tilde T^2\times T^4/\Z_2$ of the D3-branes T-dual to the D5-branes that are allowed at genus-0 to vary consistently with the orbifold and orientifold symmetries. In $\tilde T^2$, we have explained in \Sect{BSGP} that there are 16 independent locations,\footnote{We will see in a second step  that  some of the $16+16$ positions in $\tilde T^2$ of the D3-branes T-dual to the D5- or D9-branes have a tree-level mass proportional to  the open-string coupling.} which are associated with the pairs of brane/mirror brane under $\Omega$. Moreover, we have seen that when 2 modulo 4 D3-branes sit on one of the 64 corners of the six-dimensional box in \Fig{D5D9D5D9}, 2 D3-branes have rigid coordinates in $T^4/\Z_2$, which reduces the maximum number of 8 independent dynamical positions in this orbifold space.  Hence, for the D3-branes T-dual to the D5-branes and similarly for those T-dual to the D9-branes, the numbers of moduli fields describing the positions in $T^4/\Z_2$ and  $\tilde T^4/\Z_2$ are given by 
\be
\begin{aligned}
\sum_{i,i'}\left\lfloor\frac{d_{ii'}}{2}\right\rfloor \quad \and \quad \sum_{i,i'}\left\lfloor\frac{n_{ii'}}{2}\right\rfloor .
\end{aligned} 
\ee

If perturbatively all choices of coefficients $d_{ii'}$ and $n_{ii'}$ are allowed, it turns out that they must satisfy constraints, whose origin is six-dimensional, in order to guaranty  the consistency of the model at the non-perturbative level~\cite{GimonPolchinski2}. Before stating these conditions, we need to define few quantities. Let us denote by  $D_i\equiv \sum_{i'}D_{ii'}$  the number of D5-branes (in the initial picture) sitting at the fixed point $i$ of $T^4/\Z_2$. Defining $\cR$ the number of coefficients $D_i$, $i\in\{1,\dots,16\}$, that are equal to 2 modulo 4, a little thought allows to conclude that $\cR$ is even, \ie $\cR\in\{0,2,\dots, 16\}$. Moreover, there are at most $8-\cR/2$ independent dynamical locations in $T^4/\Z_2$.\footnote{This number is not reached when the Wilson-line backgrounds  on the worldvolumes of the D5-branes  lead to some $D_i=0$ modulo 4 with some $D_{ii'}=2$ modulo 4. Physically, this corresponds to ``eating'' moduli fields when the gauge group is Higgsed in its Coulomb branch. } Hence, denoting $\tilde \cR$ the counterpart of $\cR$ for the D9-branes, the following inequalities hold,  
\be
\sum_{i,i'}\left\lfloor\frac{d_{ii'}}{2}\right\rfloor\le \sum_{i}\left\lfloor\frac{\sum_{i'} d_{ii'}}{2}\right\rfloor\equiv 8-{\cR\over 2}\,,\quad~~
\sum_{i,i'}\left\lfloor\frac{n_{ii'}}{2}\right\rfloor\le \sum_{i}\left\lfloor\frac{\sum_{i'}n_{ii'}}{2}\right\rfloor\equiv8-{\tilde \cR\over 2}\, .
\label{ine}
\ee
Notice that $(\cR,\tilde \cR)$ characterizes disconnected components of the moduli space. Indeed, when the model is decompactified to six dimensions, $\cR$ is the number of rigid pairs of D5-branes  in $T^4/\Z_2$, while $\tilde \cR$ counts the pairs of D5-branes T-dual to the D9-branes that are rigid in $\tilde T^4/\Z_2$. There is no gauge-theory phase transition in six dimensions that can describe a variation of $(\cR,\tilde \cR)$. The components that are fully consistent non-perturbatively have $\cR$ and $\tilde \cR$ equal to 0, 8 or 16~\cite{GimonPolchinski2}.\footnote{In four dimensions, this results in constraints on $\sum_{i'}d_{ii'}$ and $\sum_{i'}n_{ii'}$.} Moreover, when $\cR$ (or~$\tilde \cR$) is 8, the rigid pairs of D5-branes (D5-branes T-dual to the D9-branes) must be located on the 8 fixed points of one of the hyperplanes $X^I=0$ or $\pi$, $I\in\{6,\dots,9\}$, ($\tilde X^I=0$ or $\pi$). Up to T-duality, there are therefore six inequivalent consistent classes of brane configurations in six dimensions, which are  characterized by 
\be
(\cR,\tilde \cR)\in\big\{(0,0), (0,8), (0,16),(8,8), (8,16),(16,16)\big\}\, . 
\ee

In the language of D3-branes, the mass-squared  terms of the positions around the fixed points $ii'$ have been derived in~\cite{ACP} by computing the effective potential at one loop and extracting the quadratic contributions in moduli fields. Up to irrelevant positive numerical factors, they are equal to $M_{3/2}^2$ multiplied by coefficients listed below. For the locations in $\tilde T^2$, they are given by 
\be
\begin{aligned}
&d_{ii'}-d_{i\hat \imath'}-1+{1\over 4}\sum_{j=1}^{16}(n_{ji'}-n_{j\hat \imath'})&\mbox{for the D3's T-dual to the D5's, ~~when $d_{ii'}\ge 1$}\, , \\
&n_{ii'}-n_{i\hat \imath'}-1+{1\over 4}\sum_{j=1}^{16}(d_{ji'}-d_{j\hat \imath'})&\mbox{for the D3's T-dual to the D9's, ~~when $n_{ii'}\ge 1$}\, ,
\end{aligned}
\ee
while for the positions in  $T^4/\Z_2$ and $\tilde T^4/\Z_2$, they are 
\be
\begin{aligned}
&d_{ii'}-d_{i\hat \imath'}-1&\mbox{for the D3's T-dual to the D5's, ~~when $d_{ii'}\ge 2$}\, , \\
&n_{ii'}-n_{i\hat \imath'}-1&\mbox{for the D3's T-dual to the D9's, ~~when $n_{ii'}\ge 2$}\, .
\end{aligned}
\label{cot4}
\ee

In order to avoid tachyons, all mass-term coefficients of the locations in $T^4/\Z_2$ and $\tilde T^4/\Z_2$ should be nonnegative. However, for the positions in $\tilde T^2$, this is not  necessarily the case. As announced before, this follows from the fact that the true number of  positions free to move classically in $\tilde T^2$ is less than $16 +16$ for the  D3-branes T-dual to D5-branes or D9-branes. In fact, the product of unitary open-string gauge-group factors present in six dimensions (when all D5-branes and D5-branes T-dual to D9-branes sit on fixed points) contains anomalous $U(1)$'s. To cancel the anomalies, a generalized Green--Schwarz mechanism takes place, which implies the existence of large tree-level masses proportional to the open-string coupling  for the associated vector bosons~\cite{ACP, GimonPolchinski2}. The net result is that if there are 16 or fewer unitary factors in six dimensions, they  are all broken to $SU$ groups, while if there are more than 16 unitary factors, exactly 16 of them are broken to $SU$ groups. Compatifying down to four dimensions, no Wilson-line backgrounds on $T^2$ can be switched on for the massive vector bosons. Hence, at least 2 and at most 16 linear combinations of the $16+16$ D3-brane positions in $\tilde T^2$ are identically vanishing. Imposing these relations in the mass terms derived from the effective potential, one obtains the true mass matrix for the remaining degrees of freedom associated with the Wilson lines along~$T^2$. For the configuration to be potentially tachyon free at one loop, it is only this matrix that should have non-negative eigenvalues.


\paragraph{\em Moduli in the twisted closed-string sector: } 

The anomalous $U(1)$'s in six dimensions become massive by ``eating'' Stueckelberg fields, which turn out to be blowing-up modes of  $T^4/\Z_2$. Hence, there are classically between 0 and 14 surviving moduli fields in the twisted closed-string sector. When such modes exist, one may derive their squared masses at one loop by computing two-point functions of massless twisted scalars, which can be done using the results of \Reff{Atick} and/or our \Sects{Atick} and~\ref{alpha'0}.   


\paragraph{\em Moduli in the untwisted closed-string sector: } 

The expression of the one-loop effective potential in the orientifold model with background~(\ref{back}) was computed in \Reff{ACP}, for vanishing vev's of the RR two-form moduli. As can be seen in \Eq{vd}, when all D3-branes sit on O3-planes (in the T-dual languages), all dependency on the metric components  $G_{I'J'}$ and $G_{IJ}$ disappears up to exponentially small contributions, except for $G^{55}\equiv 4M_{3/2}^2/\Ms^2$. Hence, up to neglected corrections, these moduli remain flat directions at one loop, along with the dilaton at this order in string coupling. Moreover, in configurations showing a Bose-Fermi degeneracy at the massless level, the tadpole of  $M_{3/2}$ vanishes and the latter is an extra flat direction. In supergravity language, the spontaneous supersymmetry breaking  scale is known as  the  ``no-scale modulus,'' which parametrizes a flat direction of the classical potential in Minkowski space~\cite{noscale}.  Hence, in the particular string setups satisfying $\nF-\nB=0$, the no-scale structures valid in the classical backgrounds are preserved at one loop, up to exponentially suppressed terms. For this reason, they are designated as ``super no-scale models''~\cite{SNS1,SNS2,FR}. 

As explained in \Reff{ACP}, the heterotic/type~I duality can be used to show that the dependency of the one-loop effective potential on the RR two-form components $C_{I'J'}$ and $C_{IJ}$ appears only in the exponentially suppressed terms (even when the D3-branes are located in the bulk of the internal space). Hence, the expression~(\ref{vd})  of critical points of $\Vone$ remains valid when $C_{I'J'}$ and $C_{IJ}$ are switched on. In other words, these moduli  parametrize flat directions (up to the suppressed terms). 


\subsection{Configurations with non-tachyonic ND+DN-sector moduli}

A complete computer scan of the $32+32$ D3-brane distributions on O3-planes allowed in the non-perturbatively consistent components of the moduli space was performed in \Reff{ACP}. Six tachyon-free configurations with $\nF-\nB\ge 0$ were found, while two more deserved further investigation due to the existence of moduli fields in their ND+DN sectors. Let us analyze them. 

\begin{figure}[H]
\captionsetup[subfigure]{position=t}
\begin{center}
\begin{subfigure}[t]{0.48\textwidth}
\begin{center}
\includegraphics [scale=0.70]{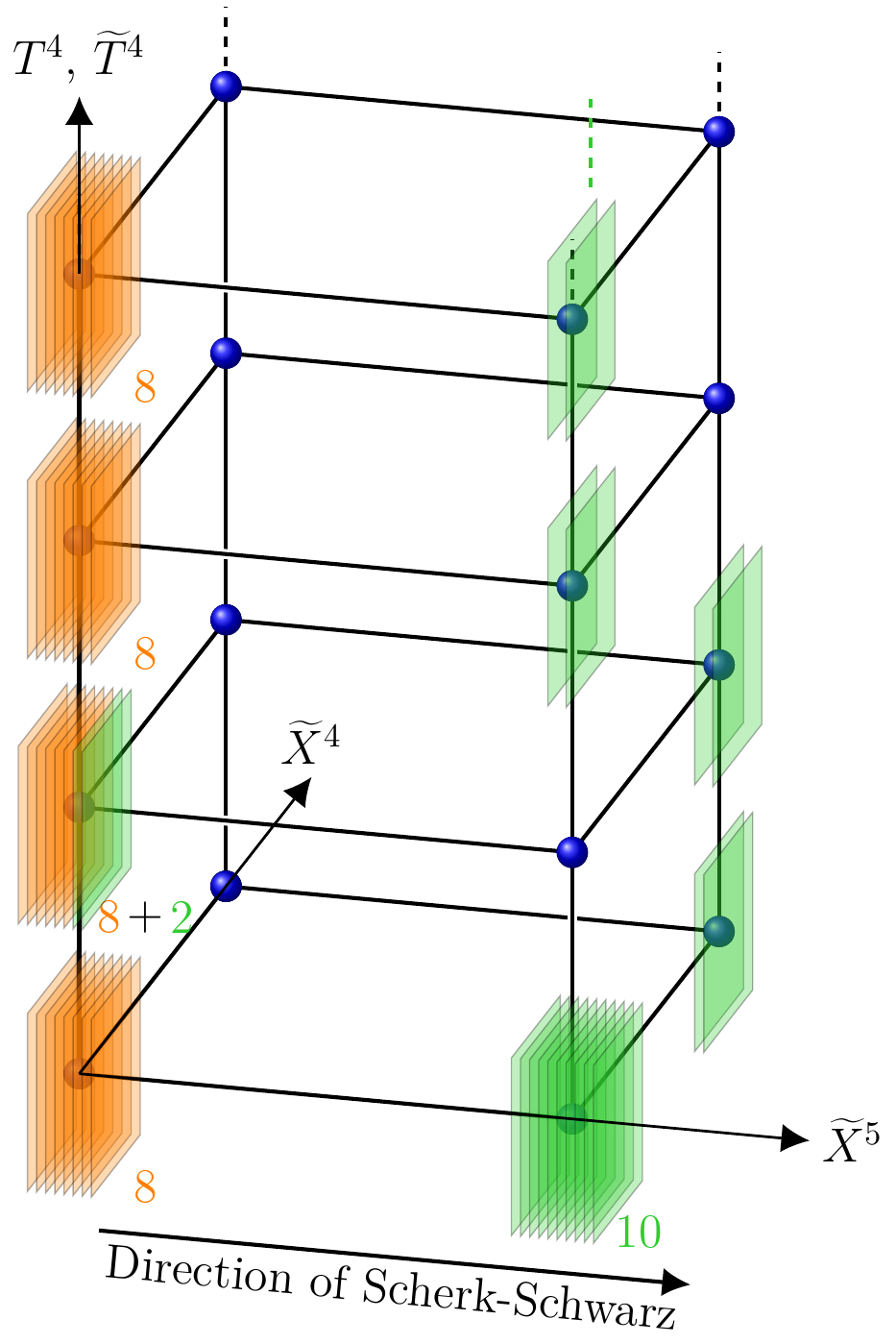}
\end{center}
\caption{\footnotesize Configuration  in the component $(\cR,\tilde \cR)=(0,8)$ of the moduli space, with $\nF-\nB=0$. The masses at one loop of two blowing-up moduli in the twisted closed-string  sector of $T^4/\Z_2$ deserve further study.  }
\label{nfnb0_c}
\end{subfigure}
\quad
\begin{subfigure}[t]{0.48\textwidth}
\begin{center}
\includegraphics [scale=0.70]{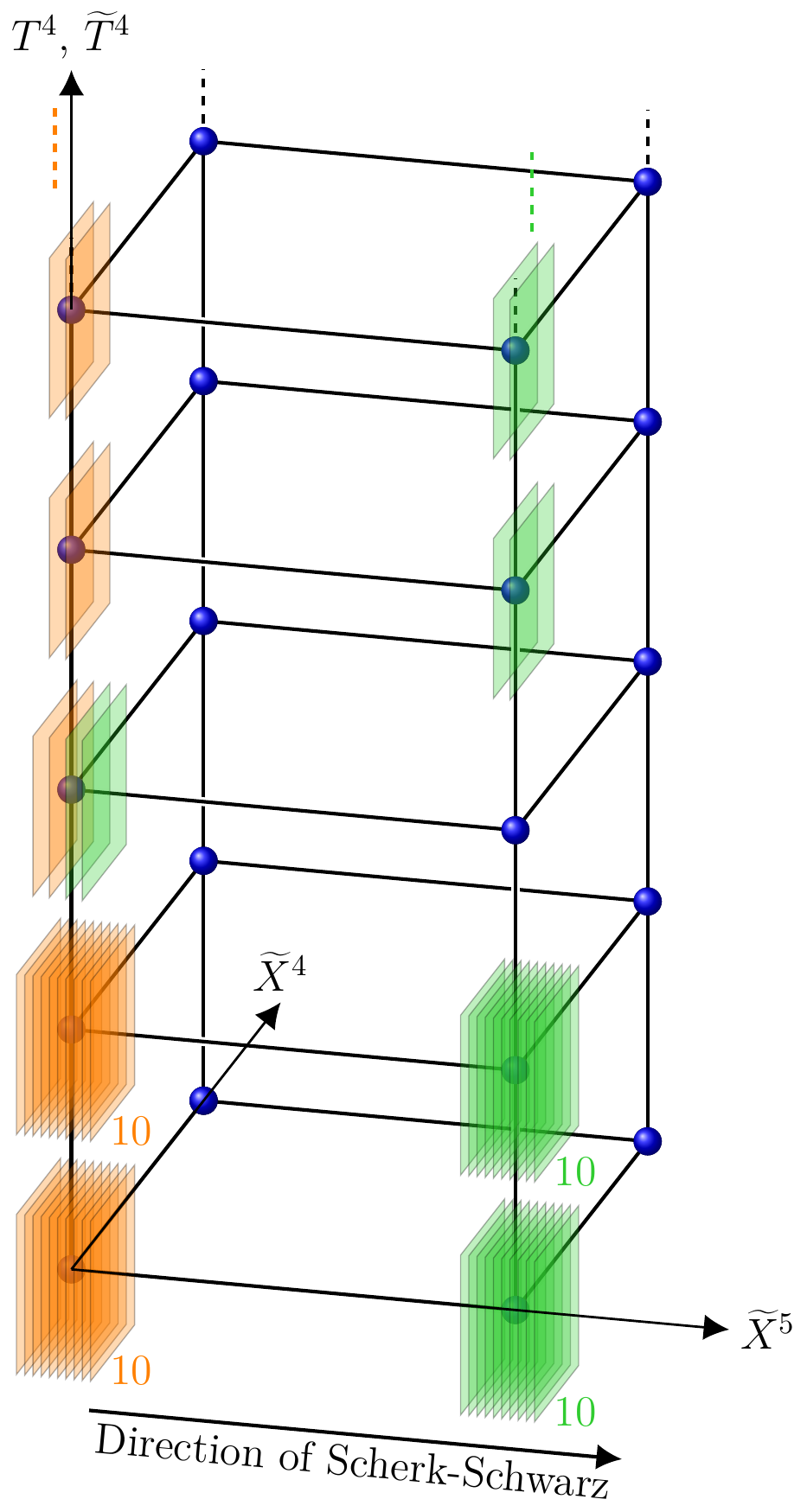}
\end{center}
\caption{\footnotesize Configuration  in the component $(\cR,\tilde \cR)=(8,8)$ of the moduli space, with \mbox{$\nF-\nB=32$}. All twisted-sector moduli are massive.  }
\label{nfnbpos_e}
\end{subfigure}
\end{center}
\caption{\footnotesize D3-brane configurations without tachyons at one loop in the NN, DD and ND+DN open-string sectors, as well as in the untwisted closed string sector. }
\label{nfnb0}
\end{figure}

\paragraph{\em Exponentially suppressed one-loop potential:} The first of these D3-brane distributions satisfies  Bose/Fermi degeneracy at the massless level and genus-0, $\nF-\nB=0$. It lies in the moduli-space component $(\cR,\tilde \cR)=(0,8)$ and is shown in \Fig{nfnb0_c}.\footnote{It corresponds to Fig.~5c of \Reff{ACP}.}  

The D3-branes T-dual to the D5-branes are distributed in $T^4/\Z_2$ as 4 stacks of 8. The D3-branes T-dual to the D9-branes are distributed as 8 pairs (with rigid positions in $\tilde T^4/\Z_2$),  one stack of 4 divided  into $2+2$ in $\tilde T^2$, and one stack of 12 divided into $10+2$  in $\tilde T^2$. All precise locations in $\tilde T^2$ can be read from \Fig{nfnb0_c}. 

The open-string gauge group including anomalous $U(1)$'s is
\begin{equation}
\label{u16c}
\left[U(4)^4\right]_{\text{DD}}\times\left[U(1)^8\times U(1)^2\times \big(U(1)\times U(5)\big)\right]_{\text{NN}}\, .
\end{equation}
It descends from the gauge symmetry $[U(4)^4]_{\text{DD}} \times [U(1)^{8}\times U(2)\times U(6)]_{\text{NN}}$ present in six dimensions, which contains 14 unitary factors. As a result, there are 14 anomalous $U(1)$'s becoming massive by ``eating'' 14 twisted moduli in the closed-string sector. Moreover, the mass acquired at one loop by the last 2 blowing-up modes of $T^4/\Z_2$ should be computed in order to conclude whether the internal space is stabilized at the orbifold  point or not. 
The anomaly-free gauge symmetry in six dimensions may be written as $[SU(4)^4]_{\text{DD}} \times [U(2)/U(1)_{\rm diag}\times U(6)/U(1)_{\diag}]_{\text{NN}}$, where  $U(2)$ and $U(6)$ are spontaneously broken to $U(1)\times U(1)\equiv U(1)_{\rm diag}\times U(1)_\perp$ and $U(1)\times U(5)\equiv U(1)_{\rm diag}\times U(1)_\perp\times SU(5)$ when the Wilson-line background on $T^2$ is switched on. The anomaly-free gauge group in four dimensions is thus  
\begin{equation}
\left[SU(4)^4\right]_{\text{DD}} \times \left[U(1)_\perp\times \big(U(1)_\perp\times SU(5)\big)\right]_{\text{NN}}\,.
\end{equation}

As follows from the mass-term coefficients~(\ref{cot4}), all D3-brane postions along $T^4/\Z_2$ and $\tilde T^4/\Z_2$ turn out to be  rigid or massive. By taking into account the Green--Schwarz mechanism which stabilizes automatically 14 linear combinations of continuous locations along $\tilde T^2$, the mass-matrix of the 18 remaining positions can be found and leads to the conclusion that they are all massless except one which is massive. As explained in the previous subsection, all untwisted closed-string moduli are flat directions, including $M_{3/2}$ thanks to the vanishing of $\nF-\nB$.  The moduli in the ND+DN sectors are realized as strings ``stretched'' between the stack of 2 D3-branes T-dual to D9-branes located on the left side of \Fig{nfnb0_c} and any of the four stacks of 8 D3-branes T-dual to D5-branes. Indeed, they  share the same coordinates in $\tilde T^2$. In all four cases, we have in the notations of \Eqs{amplif} and~(\ref{resul}, 
\be
\begin{aligned}
&N_{i_0i'_0}=2\, , \quad  N_{i_0\hat \imath'_0}=0\, ,\quad D_{j_0i'_0}=8\, , \quad  D_{j_0\hat \imath'_0}=0\, ,\\
\Longrightarrow\quad &n_{i_0i_0'}-n_{i_0\hat \imath_0'}-1+d_{j_0i_0'}-d_{j_0\hat \imath_0'}-1 =3>0\, ,
\end{aligned}
\ee
which shows that all moduli fields in the ND+DN sector are massive. 

For completeness, let us review the counting of $\nB$ and $\nF$. In the NN and DD sectors, the bosonic degrees of freedom include those of an $\N=2$ vector multiplet in the adjoint representation of  the group~(\ref{u16c}), along with  those of hypermultiplets in antisymmetric $\oplus$ $\overline{\text{antisymmetric}}$ representations of all non-Abelian factors. In the ND+DN sector, we have the bosonic degrees of freedom of 4 hypermultiplets, all  transforming under a ``bifundamental'' representation of a $U(1)_{\rm NN}\times U(4)_{\rm DD}$ group.  Adding the 96 degrees of freedom arising from the closed-string sector, we obtain a total of $\nB=832$. On the contrary, $\nF$ contains only contributions from the ND+DN sector. The latter correspond to the fermionic degrees of freedom of hypermultiplets in the bifundamental representations of each pair of unitary group factors supported by stacks of D3-branes T-dual to D5-branes and stacks of D3- branes T-dual to D9-branes, provided they have distinct coordinates along $\tilde X^5$ but not $\tilde X^4$. Their total number is given by $\nF=4\times 16\times 13=832$, which equals $\nB$ as claimed before.


\paragraph{\em Positive one-loop potential:} 

The second configuration we are interested in lies in the component $(\cR,\tilde\cR)=(8,8)$ of the moduli space. It yields a positive potential satisfying  $\nF-\nB=32$ and is  depicted in \Fig{nfnbpos_e}.\footnote{It corresponds to the configuration shown in Fig.~6(e) of \Reff{ACP}, after a T-duality on $T^4/\Z_2$ \ie an interchange D5 $\leftrightarrow$ D9.} The D3-branes T-dual to the D5-branes are distributed in $T^4/\Z_2$ as 6 pairs and 2 stacks of 10. Similarly, the D3-branes T-dual to the D9-branes are located in  $\tilde T^4/\Z_2$ as 6 pairs and 2 stacks of 10. Because the precise positions in $\tilde T^2$ shown in \Fig{nfnbpos_e} do not involve the direction $\tilde X^4$, the configuration could be considered in five dimensions. 

Including the anomalous $U(1)$'s, the open-string gauge group  is
\be
\left[U(1)^6\times U(5)^2\right]_{\rm DD}\times\left[U(1)^6\times U(5)^2\right]_{\rm NN}\,,
\ee
both in four and six dimensions. It contains 16 unitary factors, which implies that all twisted sector moduli are massive, \ie that $T^4/\Z_2$ cannot be desingularized. Moreover, the anomaly-free gauge symmetry is reduced to 
\be
\left[SU(5)^2\right]_{\rm DD}\times \left[SU(5)^2\right]_{\rm NN}\,.
\ee

All  D3-brane positions in $T^4/\Z_2$ or $\tilde T^4/\Z_2$ turn out to be  rigid of massive at one loop, while the Green-Schwarz mechanism leaves $8$ massive and $8$ massless moduli associated with the locations in $\tilde T^2$. Moreover, all untwisted closed-string moduli are flat directions except the supersymmetry breaking scale $M_{3/2}$ which undergoes a runaway along a positive potential. There are moduli in the ND+DN sector arising from strings ``stretched'' between any stack of D3-branes T-dual to D5-branes and the pair of D3-branes T-dual to D9-branes located on the same edged of the box. They include 2 copies of scalars in ``bifundamental'' representations of $U(1)_{\rm NN}\times U(5)_{\rm DD}$ groups, and 6 copies  in ``bifundamental'' representations of $U(1)_{\rm NN}\times U(1)_{\rm DD}$ groups. In the former case the moduli are stabilized since we have
\be
\begin{aligned}
&N_{i_0i'_0}=2\, , \quad  N_{i_0\hat \imath'_0}=0\, ,\quad D_{j_0i'_0}=10\, , \quad  D_{j_0\hat \imath'_0}=0\, ,\\
\Longrightarrow\quad &n_{i_0i_0'}-n_{i_0\hat \imath_0'}-1+d_{j_0i_0'}-d_{j_0\hat \imath_0'}-1 =4>0\, ,
\end{aligned}
\ee
while in the latter case they remain massless since 
\be
\begin{aligned}
&N_{i_0i'_0}=2\, , \quad  N_{i_0\hat \imath'_0}=0\, ,\quad D_{j_0i'_0}=2\, , \quad  D_{j_0\hat \imath'_0}=0\, ,\\
\Longrightarrow\quad &n_{i_0i_0'}-n_{i_0\hat \imath_0'}-1+d_{j_0i_0'}-d_{j_0\hat \imath_0'}-1 =0\, .
\end{aligned}
\ee
Finally, the counting of the classically massless degrees of freedom can be done as in the brane configuration of \Fig{nfnb0_c} and leads to $\nF-\nB=32$.


\section{Conclusions}
\label{conclusion}

In this work, we have calculated the quadratic mass terms of the moduli fields arising in the ND+DN sector of the type~IIB orientifold model compactified on $T^2\times T^4/\Z_2$, when $\N=2$ supersymmetry is spontaneously broken \via the \SS mechanism implemented along one direction of $T^2$. Assuming the string coupling is weak, this is done at one loop by computing the two-point functions of ``boundary-changing vertex operators'' inserted on the boundaries of the annulus and M\"obius strip. The main difficulties of the derivation to which we have paid particular attention are the following:
\begin{itemize}
\item Using the stress-tensor method,  the correlators of ground-state boundary-changing fields and spin fields are found up to ``integration constants,'' which are functions of  the \Teich parameters of the double-cover tori. This leads to an ambiguity in the full amplitude of interest, which is lifted by taking the limit where the insertion points of the vertices coalesce. Indeed, the expression reduces in this case to contributions of the partition function that arise from states with specific Chan--Paton indices only. This very fact makes this step of the computation more involved than its counterpart for closed-string amplitudes of twisted-sector states  for which this identification is made with the entire partition function.
\item  The two-point function can be split into two parts referred to as ``external'' and ``internal.'' The former, which is dressed by a kinematic factor, can be used to derive the one-loop correction to the \K\"ahler metric and involves only correlation functions of ``ground-state boundary-changing fields.'' By contrast, the internal part which captures the mass correction also requires  correlators of ``excited boundary-changing fields''. These extra ingredients contain two contributions:\footnote{Denoted as $(1)$ and $(2)$ in the correlators $\langle \tau^u\tau^{\prime u}\rangle_{\rm qu}=\langle \tau^{\prime u}\tau^u\rangle_{\rm qu}$.} One arises from periodicity properties of the orbifold-background coordinates, and the other from pure local monodromy effects. Although the latter have often been neglected in favor of the former in the literature, both turn out to be of equal order of magnitude, as we have shown explicitly. 
\end{itemize}

The squared masses of all moduli fields have been derived at one loop and up to contributions that are suppressed\footnote{They can be exponentially or power-like suppressed.} when $M_{3/2}$ is lower than the other scales of the background. When the results are strictly positive, the corresponding scalars can be stabilized dynamically during the cosmological evolution of the universe~\cite{sta1, sta2, sta3, Bourliot:2009na, CatelinJullien:2007hw,        CoudarchetPartouche, CFP} in the regions in moduli space compatible with weak coupling and the assumption on $M_{3/2}$. However the potentially dominant contributions to the mass terms of moduli in the NN, DD or ND+DN open-string sectors can accidentally vanish for certain brane configurations.  In such cases, the issue of quartic interactions potentially inducing instabilities of the backgrounds arises. The fact that the untwisted closed-string moduli $(G+C)_{I'J'}$ and    $(G+C)_{IJ}$ are flat directions (up to exponentially suppressed corrections) seems to be a more severe difficulty. However, heterotic/type~I duality can be used to show that non-perturbative contributions of D1-branes, which are captured by a one-loop computation of the effective potential on the heterotic side, can stabilize some of these moduli~\cite{Estes:2011iw,CoudarchetPartouche,APP,PdV}.\footnote{See also \Reff{LP} for a stabilization of the K\"ahler and complex structure moduli of Calabi--Yau manifolds in type~IIB (type~IIA) thanks to D3-branes (D4- and D6-branes) contributions to the free energy, when $\N=2$ supersymmetry is effectively broken by thermal effects. } However, for large \SS direction \ie small  $M_{3/2}\equiv \Ms\sqrt{G^{55}}/2$ compared to $\Ms$, this mechanism is  ineffective for the components  $(G+C)_{5J}$ and $(G+C)_{5J'}$ (which include the degree of freedom of $M_{3/2}$ itself) for which extra physics should be invoked to yield their stabilization.


\section*{Acknowledgements} 

The authors would like to thank Ignatios Antoniadis and especially Emilian Dudas for useful inputs during the realization of this work. Steven Abel should also be warmly thanked for fruitful discussions and his participation to advanced stages of the project.  
 

\begin{appendices}
\makeatletter
\DeclareRobustCommand{\@seccntformat}[1]{%
  \def\temp@@a{#1}%
  \def\temp@@b{section}%
  \ifx\temp@@a\temp@@b
  \appendixname\ \thesection:\quad%
  \else
  \csname the#1\endcsname\quad%
  \fi
} 
\makeatother

\section{\bm Conventions of matrix actions on Chan--Paton indices }
\label{dico}
\renewcommand{\theequation}{A.\arabic{equation}}

 In \Reff{GimonPolchinski2}, the actions of the group elements $G\in\{1,g,\Omega,\Omega g\}$ on the Dirichlet or Neumann Chan--Paton indices $\alpha\in\{1,\dots,32\}$ are always represented by $32\times 32$ matrices. If needed, they define traces with an index $I$ (that would be denoted $i$ in our notations) labelling a fixed point of $T^4/\Z_2$ to indicate when they restrict to the matrix entries associated with the fixed point $I$ (see their Eq.~(2.22)). In our conventions, we  work directly with smaller matrices, one for each fixed point $ii'$ of $\tilde T^2\times T^4/\Z_2$ or $\tilde T^2\times \tilde T^4/\Z_2$, which are submatrices of those used in \Reff{GimonPolchinski2}. In this appendix, we would like to give a detailed correspondance between their notations and ours for the traces appearing in the open-string contributions to the partition function. 

Let us focus on the matrices acting on the Neumann Chan--Paton factors. In order to avoid any ambiguity, we first define the sets of indices $\cH_{ii'}$ associated with the fixed points $ii'$ that are used to generate the submatrices from the big ones. To this end, we label the fixed points in lexicographical order, $\left(11,12,13,14,21,\dots\right)$, and introduce a function $p(i,i')$ that gives the predecessor in this list,
\be
p(i,i')=\left\{\!\begin{array}{ll}
i,i'-1&\text{ if }i'\in\{2,3,4\}\\
i-1,4& \text{ if }i'=1 \phantom{{{L^L}^L}^L}
\end{array}
\right.\!\!.
\ee
The sets are then
\be
\cH_{11}=\left\{\!\begin{array}{l}
\varnothing~~~ \text{ if }N_{11}=0\\
\left\{1,\dots,\frac{N_{11}}{2}\right\}\cup\left\{17,\dots,16+\frac{N_{11}}{2}\right\}~~\text{ if }N_{11}\neq 0
\end{array}\right.\!\!,
\ee
and for $ii'\neq 11$,
\be
\cH_{ii'}=\left\{\!\begin{array}{l}
\varnothing~~~\text{ if }N_{ii'}=0\\
\left\{\frac{N_{p(i,i')}}{2}+1,\dots,\frac{N_{p(i,i')}}{2}+\frac{N_{ii'}}{2}\right\}\cup\left\{\frac{N_{p(i,i')}}{2}+17,\dots,16+\frac{N_{p(i,i')}}{2}+\frac{N_{ii'}}{2}\right\}~~\text{ if }N_{ii'}\neq 0
\end{array}\right.\!\!.
\ee
Our $N_{ii'}\times N_{ii'}$ matrices $\gamma_{{\rm N},G}^{ii'}$ are formed from $32\times 32$ matrices $\gamma_{{\rm N},G}$ as follows, 
\be
\gamma_{{\rm N},G}^{ii'}=\gamma_{{\rm N},G}|_{_{\cH_{ii'}}}\, ,
\ee
where the notation in the right-hand side means that we form submatrices by keeping the rows and columns $\alpha\in\cH_{ii'}$. 
The traces of $32\times 32$ matrices can then be expressed as 
\be
\begin{aligned}
\tr(\gamma_{{\rm N},G})&=\sum_{\alpha=1}^{32}(\gamma_{{\rm N},G})_{\alpha\alpha}=\sum_{i,i'}\sum_{\alpha\in\cH_{ii'}}(\gamma_{{\rm N},G})_{\alpha\alpha}=\sum_{i,i'}\tr(\gamma_{{\rm N},G}|_{_{\cH_{ii'}}})\\
&=\sum_{i,i'}\tr(\gamma_{{\rm N},G}^{ii'})\, ,
\end{aligned}
\ee
and similarly for the matrices associated with the  Dirichlet sector. 

Moreover, in order to justify the replacement~(\ref{repla}), let us define in our notations the $32\times 32$ matrix $\W_j$ that  is denoted $W_I$ and appears in Eq.~(2.22) of \Reff{GimonPolchinski2},
\be
\label{Wj}
\W_j=\prod_{I=6}^9\W_I^{2 a^I_j}\,,\quad\where\quad\W_I=I_2\otimes\begin{pmatrix}
e^{2i\pi a_1^I}I_{\frac{N_1}{2}}  &  & \empty_{\text{\Large{0}}} \\
 & \ddots & \\
\empty^{\text{\Large{0}}} & & e^{2i\pi a_{16}^I}I_{\frac{N_{16}}{2}}
\end{pmatrix},\quad N_i\equiv \sum_{i'}N_{ii'}\, . 
\ee
For $G=g$ we have to compute 
\be
\begin{aligned}
\tr(\W_j\gamma_{{\rm N},g})&=\sum_{\alpha=1}^{32}(\W_j\gamma_{{\rm N},g})_{\alpha\alpha}=\sum_{\alpha=1}^{32}\sum_{\beta=1}^{32}(\W_j)_{\alpha\beta}(\gamma_{{\rm N},g})_{\beta\alpha}=\sum_{\alpha=1}^{32}(\W_j)_{\alpha\alpha}(\gamma_{{\rm N},g})_{\alpha\alpha}\\
&=\sum_{i,i'}\sum_{\alpha\in\cH_{ii'}}(\W_j)_{\alpha\alpha}
(\gamma_{{\rm N},g})_{\alpha\alpha}=\sum_{i,i'}e^{4i\pi\vec{a}_i\cdot\vec{a}_j}\sum_{\alpha\in\cH_{ii'}}
(\gamma_{{\rm N},g})_{\alpha\alpha}\\
&=\sum_{i,i'}e^{4i\pi\vec{a}_i\cdot\vec{a}_j}\tr(\gamma_{{\rm N},g}^{ii'})\, .
\end{aligned}
\ee
In this derivation, we have used the fact that the matrix $\W_j$ is diagonal and that its components $\alpha\alpha$ for $\alpha\in\cH_{ii'}$ are $e^{4i\pi\vec{a}_i\cdot\vec{a}_j}$. 


\section{Closed-string sector partition function}
\label{ztk}
\renewcommand{\theequation}{B.\arabic{equation}}

In this appendix, we display the closed-string contributions to the one-loop partition function defined in \Eq{Zcl}. 

In order to write $Z_\T$, we introduce the lattices of zero modes of the bosonic coordinates associated with $T^4$ and $T^2$, 
\be
\begin{aligned}
&\Lambda_{\vec m, \vec n}^{(4,4)}(\tau)\, =q^{{1\over 4}P^{\rm L}_I G^{IJ}P^{\rm L}_J}\, \bar q^{{1\over 4}P^{\rm R}_I G^{IJ}P^{\rm R}_J}\,,  &&P^{\rm L}_I=m_I+G_{IJ}n_J \, , &&P^{\rm R}_I=m_I-G_{IJ}n_J\, , \\
&\Lambda_{\vec m', \vec n'}^{(2,2)}(\tau)=q^{{1\over 4}P^{\rm L}_{I'} G^{I'J'}P^{\rm L}_{J'}}\, \bar q^{{1\over 4}P^{\rm R}_{I'} G^{I'J'}P^{\rm R}_{J'}}\,,  &&P^{\rm L}_{I'}=m_{I'}+G_{I'J'}n_{J'} \, , &&P^{\rm R}_{I'}=m_{I'}-G_{I'J'}n_{J'}\, ,
\end{aligned}
\ee
where $\vec m$, $\vec n$ and $\vec m'$, $\vec n'$ are four-vectors and two-vectors whose components are integer momenta and winding numbers. 

Moreover, the worldsheet fermions generate an   $SO(8)$ affine  symmetry broken to $SO(4)\times SO(4)$ by the $\Z_2$-orbifold action. 
As a result, their contributions to the partition function take the forms of ordered pairs of characters of $SO(4)$. The latter can be expressed in terms of Jacobi modular forms and the Dedekind function, 
\be
\label{th}
\vartheta\big[{}^\alpha_\beta\big](z| \tau)=\sum_{k\in \Z} q^{{1\over 2}(k+\alpha)^2}e^{2i\pi(z+\beta)(k+\alpha)} \, , ~~\quad  \eta(\tau)= q^{1\over 24}\prod_{n=1}^{+\infty}(1-q^n)\, ,~~\quad q=e^{2i\pi \tau}\, .
\ee
Denoting as usual 
\be
\vartheta\big[{}^0_0\big](z| \tau)\equiv \vartheta_3(z| \tau)\, , ~~\vartheta\Big[{}^{{}^{\scriptstyle \phantom{.}\!0}}_{1\over 2}\Big](z| \tau)\equiv \vartheta_4(z| \tau)\, ,~~ \vartheta\Big[{}^{1\over 2}_{\phantom{\!|}_{\scriptstyle0}}\Big](z| \tau)\equiv \vartheta_2(z| \tau)\, ,~~ \vartheta\Big[{}^{1\over 2}_{1\over 2}\Big](z| \tau)\equiv \vartheta_1(z| \tau)\, ,
\ee 
the characters are given by~\cite{characters2, characters1, BLT}
\be
O_{4}=\frac{\vartheta_{3}^{2}+\vartheta_{4}^{2}}{2\eta^{2}}\, ,\quad~~ V_{4}=\frac{\vartheta_{3}^{2}-\vartheta_{4}^{2}}{2\eta^{2}}\, ,\quad ~~S_{4}=\frac{\vartheta_{2}^{2}+i^{-2}\vartheta_{1}^{2}}{2\eta^{2}}\, ,\quad ~~C_{4}=\frac{\vartheta_{2}^{2}-i^{-2}\vartheta_{1}^{2}}{2\eta^{2}}\, ,
\label{eq:CharacterDef}
\ee
where  it is understood that $\vartheta_n\equiv \vartheta_n(0|\tau)$.

Given these notations, the torus contribution to the partition function takes the following form, 
\begin{align}
\label{torus}
&Z_\T={1\over 4}\,{1\over \tau_2^2} \,\bigg\{\bigg[\left(\carq(V,O,O,V,+)+\carq(S,S,C,C,+)\right)\lattice\nonumber\\
&+\left(\carq(V,O,O,V,-)+\carq(S,S,C,C,-)\right)\tetad\nonumber\\
&+16\left(\carq(O,C,V,S,+)+\carq(S,O,C,V,+)\right)\tetaq\nonumber\\
&+16\left(\carq(O,C,V,S,-)+\carq(S,O,C,V,-)\right)\tetat\bigg]\sum_{\vec{m}',\vec{n}'}\LAMBDA(\vec{m}',{(n_{4},2n_{5})})\nonumber\\
&-\bigg[\left(\cars(V,O,O,V,+)\carsbar(S,S,C,C,+)+\cars(S,S,C,C,+)\carsbar(V,O,O,V,+)\right)\lattice\nonumber\\
&+\left(\cars(V,O,O,V,-)\carsbar(S,S,C,C,-)+\cars(S,S,C,C,-)\carsbar(V,O,O,V,-)\right)\tetad\nonumber\\
&+16\left(\cars(O,C,V,S,+)\carsbar(S,O,C,V,+)+\cars(S,O,C,V,+)\carsbar(O,C,V,S,+)\right)\tetaq\\
&+16\left(\cars(O,C,V,S,-)\carsbar(S,O,C,V,-)+\cars(S,O,C,V,-)\carsbar(O,C,V,S,-)\right)\tetat\bigg]\sum_{\vec{m}',\vec{n}'}\LAMBDA(\vec{m}'+\vec{a}'_{\text{S}},{(n_{4},2n_{5})})\nonumber\\
&+\bigg[\left(\carq(O,O,V,V,+)+\carq(C,S,S,C,+)\right)\lattice+\left(\carq(O,O,V,V,-)+\carq(S,C,C,S,-)\right)\tetad\nonumber\\
&+16\left(\carq(O,S,V,C,+)+\carq(S,V,C,O,+)\right)\tetaq\nonumber\\
&+16\left(\carq(O,S,V,C,-)+\carq(S,V,C,O,-)\right)\tetat\bigg]\sum_{\vec{m}',\vec{n}'} \LAMBDA(\vec{m}',{(n_{4},2n_{5}+1)})\nonumber\\
&-\bigg[\left(\cars(O,O,V,V,+)\carsbar(C,S,S,C,+)+\cars(C,S,S,C,+)\carsbar(O,O,V,V,+)\right)\lattice\nonumber\\
&+\left(\cars(O,O,V,V,-)\carsbar(S,C,C,S,-)+\cars(S,C,C,S,-)\carsbar(O,O,V,V,-)\right)\tetad\nonumber\\
&+16\left(\cars(O,S,V,C,+)\carsbar(S,V,C,O,+)+\cars(S,V,C,O,+)\carsbar(O,S,V,C,+)\right)\tetaq\nonumber\\
&+16\left(\cars(O,S,V,C,-)\carsbar(S,V,C,O,-)+\cars(S,V,C,O,-)\carsbar(O,S,V,C,-)\right)\tetat\bigg]\sum_{\vec{m}',\vec{n}'}\LAMBDA(\vec{m}'+\vec a_{\rm S}',{(n_{4},2n_{5}+1)})\bigg\},\nonumber
\end{align}
where the momenta $\vec m'$ of all fermionic degrees of freedom are shifted as shown in \Eq{shifts}.

The expression of the Klein-bottle contribution to the partition function involves only left-right symmetric states which are therefore bosonic. As a result it is identical to that found in the supersymmetric model of \Sect{BSGP},
\be\begin{aligned}
\label{klein}
Z_\K=&{1\over 4}\,{1\over \tau_2^2}\, \bigg\{\left(V_{4}O_{4}+O_{4}V_{4}\right)\latticek+32\left(O_{4}C_{4}+V_{4}S_{4}\right)\tetaqo\\
&-\left(S_{4}S_{4}+C_{4}C_{4}\right)\latticek-32\left(S_{4}O_{4}+C_{4}V_{4}\right)\tetaqo\bigg\}\sum_{\vec{m}'}\frac{P^{(2)}_{\vec{m}'}}{\eta^{4}}\,,
\end{aligned}
\ee
where all lattices are identical to those defined in the open-string sector, Eqs.~(\ref{latop}) and~(\ref{latop2}), and the characters and modular forms are evaluated at $2i\tau_{2}$. 


\section{\bm Leading behavior of $\varepsilon$ when $\alpha'\to 0$}
\label{lvar}
\renewcommand{\theequation}{C.\arabic{equation}}

In this appendix, we reconsider the small $\alpha'$ limit of the roots $U_1$ of  $\Omega(U)$ for arbitrary $|\Im Y_1|<\half\, \Im \taudc$, which are given in \Eq{change}.  Our goal is to find the leading behavior of~$\varepsilon$.

Using the full expansion of $\vartheta_1(z)$ given in \Eq{condi}, one obtains 
\be
 {\vartheta_1'\over \vartheta_1}(z)=\pi \cot (\pi z)+2i\pi\sum_{n\ge 1}\big[H(\qdc^n \,e^{-2i\pi z}) -H(\qdc^n \,e^{2i\pi z})\big]\, , \quad H(z)\equiv {z\over 1-z}\, , 
 \label{Hdef}
\ee
which can be used to rewrite the equation $\Omega(U_1)=0$ as follows, 
\begin{align}
\label{Heq}
i \sin(\pi Y_1) = 2 \sin(\pi U_1)\sin[\pi(Y_1-U_1)]\,\sum_{n\ge 1} &\Big[ H\big(\qdc^n \,e^{-2i\pi U_1}\big) -H\big(\qdc^n \,e^{2i\pi U_1}\big)\\
&\!\!\!\!\!+H\big(\qdc^n \,e^{-2i\pi (Y_1-U_1)}\big) -H\big(\qdc^n \,e^{2i\pi (Y_1-U_1)}\big)\Big]\, . \nonumber 
\end{align}
Applying the change of variable given in \Eq{change}, the above expression becomes 
\begin{align}
i\sin(\pi Y_1)=&\;\half \big(1-(-1)^m\, i\, e^{i\pi(\taudc+Y_1)}\,e^{2i\pi\varepsilon}\big)\big(1-(-1)^m\, i\, e^{i\pi(\taudc-Y_1)}\,e^{2i\pi\varepsilon}\big)\, e^{-2i\pi\varepsilon}\nonumber\espD\\
&\times \!\Bigg\{\!-{e^{-i\pi Y_1}\, e^{-2i\pi\varepsilon}\over 1+(-1)^m\, i\, e^{i\pi(\taudc-Y_1)}\, e^{-2i\pi\varepsilon}}\label{eqtot}\\
&~~~~~+2i\sum_{n\ge 1}  {\qdc^n\, e^{-i\pi Y_1} \big(\sin(2\pi\varepsilon)+(-1)^m \, \qdc^n\, e^{i\pi(\taudc-Y_1)}\big)\over \big(1+(-1)^m\, i\, \qdc^n\, e^{i\pi(\taudc-Y_1)}\, e^{-2i\pi\varepsilon}\big) \big(1-(-1)^m\, i\, \qdc^n\, e^{i\pi(\taudc-Y_1)}\, e^{2i\pi\varepsilon}\big)}\nonumber\\
&~~~~~-(Y_1\to -Y_1)\Bigg\}\, , \nonumber 
\end{align}
where the first term in the braces and its transformed under $Y_1\to -Y_1$ arise from the contributions $n=1$ of the first and fourth functions $H$ in \Eq{Heq}. 
\begin{itemize}
\item If $0<\Im Y_1<\half\, \Im \taudc$, and assuming $\varepsilon\to 0$ when  $\alpha'\to 0$, \Eq{eqtot} reads
\be
\begin{aligned}
(e^{2i\pi Y_1}-1)\, e^{-i\pi Y_1}=&\, \big(1-2i\pi\varepsilon-(-1)^m\, i\, e^{i\pi (\taudc-Y_1)}+\mbox{sub dom.}\big)\\
&\, \Big\{-e^{-i\pi Y_1}\big(1-2i\pi\varepsilon-(-1)^m\, i\, e^{i\pi (\taudc-Y_1)}+\mbox{sub dom.}\big)\\
&\, ~~+e^{i\pi Y_1}\big(1-2i\pi\varepsilon-(-1)^m\, i\, e^{i\pi (\taudc+Y_1)}+\mbox{sub dom.}\big)\Big\}\, ,
\end{aligned}
\ee
where all contributions $n\ge 1$ have been absorbed in ``subdominant'' contributions.   Simplifying this expression, one obtains
\be
0=4i\pi \varepsilon+2(-1)^m\, i\, e^{i\pi(\taudc-Y_1)}+\mbox{subdom.}\, , 
\ee
which leads to the asymptotics
\be
\varepsilon\underset{\alpha'\to 0}\sim-{(-1)^m\over 2\pi}\, e^{i\pi(\taudc-Y_1)}\underset{\alpha'\to 0}{\longrightarrow} 0\, .
\ee
Our assumption on the convergence of $\varepsilon$ being consistent, we have shown the existence of solutions $U_1$ with the above behavior. 
\item If $-\half\, \Im \taudc<\Im Y_1<0$, \Eq{eqtot} being invariant under $Y_1\to -Y_1$, one obtains immediately from the previous analysis that 
\be
\varepsilon\underset{\alpha'\to 0}\sim-{(-1)^m\over 2\pi}\, e^{i\pi(\taudc+Y_1)}\underset{\alpha'\to 0}{\longrightarrow} 0\, .
\ee
\end{itemize}

\end{appendices}



\end{document}